\documentclass[11pt, a4paper]{article}
\usepackage{amsmath, amssymb, amsthm}
\usepackage{graphicx}
\usepackage{booktabs}
\usepackage[margin=1in]{geometry}
\usepackage{hyperref}
\usepackage{bm}
\usepackage{empheq}
\usepackage{float}
\usepackage[section]{placeins}  % auto-FloatBarrier at each \section{}

\newcommand{\Tr}{\mathrm{Tr}}

\newcommand{\affmark}[1]{\textsuperscript{#1}}

\title{\textbf{Frustrated Dynamics of Distance Matrices}\thanks{All
calculations, numerical analysis, and manuscript
preparation were performed by Claude Code with Opus 4.7 working as an AI assistant under author's supervision. All remaining errors are my own. I would like to thank Charles Martin and Alejandro Rodriguez Dominguez for very helpful discussions and comments on the manuscript.}}

\author{Igor Halperin\affmark{*}}

\date{\today}

\begin{document}

\maketitle

\begin{center}
\affmark{*}Email: ighalp@gmail.com
\end{center}

\begin{abstract}
We introduce the \emph{Frustrated Distance Matrix} (FDM) model,
a dynamic extension of the static distance-matrix ensemble on
$S^2$ analyzed by Bogomolny, Bohigas, and Schmit (BBS). Its
entries are pairwise geodesic distances between $N$ Brownian
particles on the sphere evolving under quenched random
pairwise couplings linear in those distances. Where the static
BBS theory recovers geometric information about the underlying
manifold from spectra of distance matrices on i.i.d.\ samples,
the time-resolved FDM spectrum carries information about
\emph{structural changes} of the underlying point process. The
particle dynamics realize one such change: a fast collapse from
a uniform configuration onto a one-dimensional ring, followed
by slow rotational drift of the ring orientation; the
particle-level picture provides the ground truth against which
spectral diagnostics are calibrated. We find that the static
BBS template is preserved at every time, with the dynamics
entering as a redistribution of spectral mass within that
template, sharp enough to flag ring formation. We propose
self-averaging of the bulk density as the mechanism behind this
preservation, verified by an i.i.d.-resample comparison, and
extract a small set of spectral diagnostics of the structural
change computable from the distance matrix alone. We suggest
that our diagnostics can be applied in other similar
inverse-problem settings: financial correlation matrices,
graph and network adjacency spectra, similarity matrices in
molecular dynamics, and dynamics on parameter manifolds.
\end{abstract}

\setcounter{tocdepth}{1}
\tableofcontents

%==============================================================================
\section{Introduction}
\label{sec:intro}
%==============================================================================

%\paragraph{Distance matrices, Euclidean random matrices, and the
%need for a dynamic extension.}
The Euclidean random matrices (ERM), and their special
case of \emph{distance matrices}, are a branch of random matrix
theory in which the matrix elements are fixed functions of
i.i.d.\ inputs such as particle locations in Euclidean space or
on a Riemannian manifold~\cite{mezard1999, bordenave2008,
bordenave2013, goetschy2013}. This is the key structural
difference from the classical RMT ensembles whose matrix
elements are themselves i.i.d.\ random variables: the Wigner
Gaussian ensembles and the semicircle law~\cite{wigner1955},
Dyson's threefold orthogonal/unitary/symplectic
classification~\cite{dyson1962}, the Mar\v{c}enko--Pastur
sample-covariance ensembles~\cite{marchenko1967}, and the
Tracy--Widom edge fluctuations~\cite{tracy1994}, see e.g. the
textbooks ~\cite{mehta2004, pottersbouchaud2020} for a review. 

A central reason for the interest in distance matrices is that
their entries encode metric properties of the manifold on which
the underlying point set lives, so the spectrum of the distance
matrix carries geometric information about that manifold.
The most direct random-matrix-theoretic study of this
question, and the principal reference for the present paper, is
the analysis by Bogomolny, Bohigas, and Schmit
(BBS)~\cite{bogomolny2003} of the spectrum of distance matrices
sampled on $N$ uniform i.i.d.\ points on the sphere $S^d$.
Related contributions include the universality results of
Vershik~\cite{vershik2002}, the Anderson-localization analysis
of Euclidean random matrices by Ciliberti et
al~\cite{ciliberti2004anderson}, and the localization-transition
study of Clapa et al~\cite{clapa2012}. BBS formulate the
following \emph{inverse problem}: given a sample of the
distance matrix on $N$ i.i.d.\ points drawn from a manifold,
recover the dimension and structure of the manifold from the
spectrum of that matrix. They solve a partial form of this
problem by deriving universal spectral-density laws as
functions of the manifold dimension $d$: a power-law tail at
large $|\lambda|$ in the delocalized regime, a separate
power-law density at small $|\lambda|$ in the localized regime,
and a quasi-multiplet structure at the most negative
eigenvalues set by the irreducible representations of the
rotation group of the manifold. The explicit forms of these
laws and their derivation are reviewed in
Sec.~\ref{sec:bbs_check}; for the present discussion only their
$d$-dependence is used.

The BBS analysis is static: it assumes i.i.d.\ sampling from a
fixed manifold and asks what the distance-matrix spectrum looks
like at a single instant. A natural extension is to consider
\emph{dynamic} distance matrices $\{M(t)\}_t$ whose entries
vary in time because the underlying particles themselves evolve.
This formulation is generic, as any $N$-particle system in any
base metric space defines such a dynamic distance-matrix process. On the other hand, it 
expands the static inverse problem into a richer dynamical
inverse problem: given the time series $\{M(t)\}_t$, can one
detect not only the static manifold supporting the points but
also \emph{structural changes} in the underlying configuration
over time, such as a collapse onto a lower-dimensional
submanifold, a clustering transition, or a rearrangement of the
support? In a dynamic setting, time-resolved redistributions of
spectral mass within the static template should signal such
changes, with the magnitude and channel of the redistribution
carrying information about the qualitative nature of the change.
This dynamical inverse problem is the central motivation of the
present work.

We address this question in a controlled setting where the
structural change is unambiguous and visualizable. The
\emph{Frustrated Brownian Particles} (FBP) model on $S^2$ of
Ref.~\cite{halperin2026frustrated} drives an initially uniform
or Gaussian-clustered configuration of $N$ Brownian
particles into an emergent ring state on a fast timescale: the
particles concentrate onto a great-circle band on the sphere,
undergoing what Ref.~\cite{halperin2026frustrated} calls a
\emph{dynamic dimension reduction} from $d=2$ to $d=1$. Because
we have direct access to the particle positions and to the
timing of ring formation in our simulations, we know
unambiguously \emph{when} the structural change occurs and
\emph{what} it consists of, and we can ask which spectral
diagnostics of the corresponding distance matrix detect it.
The matrix process induced by the FBP particle trajectories,
\begin{equation}
\label{eq:M_def_intro}
M(t)_{ij} \;=\; d\bigl(x_i(t), x_j(t)\bigr) \;=\; \arccos\bigl(x_i(t)\cdot x_j(t)\bigr),
\end{equation}
of pairwise geodesic distances on $S^2$ taken at the FBP
positions, with quenched random pairwise interactions linear in
geodesic distance, is the central object of this paper. We
refer to it as the \emph{Frustrated Distance Matrix} (FDM)
model. The FDM is the simplest controlled laboratory we know of
for the dynamical inverse problem: the underlying structural
change (ring formation) is striking and exactly characterized
at the particle level, and the task is to identify spectral
observables of $M(t)$ that flag the change without ever
reconstructing the configuration in $\mathbb{R}^3$.

The most relevant existing literature on dynamic random
matrices is sparse and lies outside the i.i.d.\ ERM class. Dyson
Brownian motion~\cite{dyson1962brownian} evolves a Hermitian
matrix with i.i.d.\ Brownian increments; the Voiculescu--Biane
free Brownian motion~\cite{voiculescu1991, biane1997} is the
non-commutative analog; the disorder-averaged dynamics of
spin-glass / SYK type~\cite{cugliandolo1993, facoetti2019}
renders the matrix dynamics indirectly through an effective
field-theoretic action. None of these settings combines the
strongly-correlated kernel-on-points structure of an ERM with
an explicit evolution of the underlying point process. The FDM
is, to our knowledge, the first numerical study of a
kernel-on-points matrix in which the points are coupled by a
Langevin SDE with quenched random forces, and is proposed here
as a candidate dynamic extension of the ERM class.
The body of the paper develops the spectral diagnostics of ring
formation and identifies the FBP disorder-averaged \emph{one-particle
density} $\mu_t$ on $S^2$ at time $t$ as the natural
geometry-aware null hypothesis. The static
BBS template is preserved at every time along the FDM
trajectory by a self-averaging mechanism, with the dynamics
encoded as a redistribution of spectral mass within that
template (Secs.~\ref{sec:bulk_ensembles},
\ref{sec:bbs_check}).

\paragraph{Why the FDM dynamics is interesting.}
The dynamic content of the FDM is non-trivial because it is
inherited from the FBP dynamics studied in
\cite{halperin2026frustrated}. The FBP model combines two
otherwise disjoint dynamical regimes. The first is a fast
non-equilibrium relaxation phase that resembles the dynamic
transitions of spin glasses: $N$ particles collapse from a
uniform configuration on $S^2$ to a one-dimensional emergent
ring on a short timescale $\tau_{\text{fast}}$ (the
\emph{dynamic dimension reduction} of
\cite{halperin2026frustrated}). The second is a slow
non-equilibrium steady state (NESS)\footnote{We use the term \emph{NESS}
to emphasize the dynamical character of the post-formation
regime: at scales $t\gg\tau_{\text{fast}}$ the ring's
macroscopic shape and the empirical density profile are
stationary in distribution, but the system maintains continuous
activity, with fast particle fluctuations along the ring and a
slow rotational drift of the ring orientation
$\hat{\mathbf{n}}(t)$ on $S^2$, all sustained by continuous
heat exchange with the external bath through the Langevin noise
$\sqrt{2\gamma T}\,d\mathbf{W}$. We read this combination, in
particular the persistent slow drift of $\hat{\mathbf{n}}(t)$
across $S^2$ rather than oscillation around a fixed axis, as
empirical evidence for a non-vanishing probability current
along the rotational soft mode and hence for a steady state
that is non-equilibrium in the operational physics sense. 
%We
%note that an alternative reading is also consistent with the
%SDE: for an overdamped Langevin equation with conservative
%force on a compact manifold, the standard derivation gives a
%Gibbs stationary density $p_{\text{eq}}\propto e^{-U/T}$ with
%zero average probability current, and the slow rotational drift
%can then be interpreted as Brownian motion along a soft
%rotational (Goldstone) mode at thermal equilibrium. 
A precise identification as a NESS would require computing the stationary current
along the rotational mode after projecting the configuration-space
Langevin equation onto $\hat{\mathbf{n}}$, which depends on
finite-$N$ effects of the disordered potential; we leave that
calculation to future work and use NESS in its empirical sense
throughout the paper.} in which the ring orientation
$\hat{\mathbf{n}}(t)\in S^2$ undergoes slow rotational diffusion
on the coset $\mathrm{SO}(3)/\mathrm{SO}(2)$. The latter is
best described as an \emph{adiabatic} symmetry breaking rather
than a static spontaneous one: at every fixed instant the
configuration has a well-defined orientation vector
$\hat{\mathbf{n}}(t)$ that picks out a direction in $S^2$, but
$\hat{\mathbf{n}}(t)$ itself drifts slowly across $S^2$, so the
rotational symmetry is restored on the longest timescales by
the orientational ergodicity of $\hat{\mathbf{n}}(t)$. The FDM
\emph{intrinsic} content inherits the collapse: the matrix
process $M(t)$ traces out a spin-glass-style relaxation over
$[0, \tau_{\text{fast}}]$ in which the spectrum redistributes
from a $d=2$ template to a $d=1$ ring template. The lab-frame
orientation drift $\hat{\mathbf{n}}(t)$ is, by contrast,
invisible to $M(t)$ because the distance matrix is invariant
under global rotations
(see Sec.~\ref{sec:setup}, Eq.~\eqref{eq:M_rotation_invariance}
below).
The combination is genuinely new in the random-matrix literature:
spin-glass dynamic random-matrix models (SYK, $p$-spin) lack the
explicit geometric kernel structure of an ERM, while the existing
dynamic ensembles (Dyson, free Brownian) lack both the disordered
relaxation and the adiabatic geometric-collapse phase. The FDM is
the simplest setting we know of in which both structures coexist.

\paragraph{Why the FDM perspective is useful in practice.}
The FBP dynamics on $S^2$ shows ring formation directly, by
visualization of the particle positions in $\mathbb{R}^3$
\cite{halperin2026frustrated}. In many practical applications,
however, particle positions are not directly observable and the
only data available about an $N$-body system are pairwise
distances or pairwise similarities, that is, exactly the
distance matrix $M(t)$. A natural question is then how a
qualitative structural change in the underlying configuration
(emergence of a collective state, dimensional collapse onto a
lower-dimensional support, formation of clusters) is encoded in
the spectral properties of $M(t)$ alone, without ever
constructing the configuration in $\mathbb{R}^3$. The FBP/FDM
pair is a controlled microscopic test bed for this question: we
know from \cite{halperin2026frustrated} that the underlying
particles form a ring, and we can ask which spectral diagnostics
of $M(t)$ detect that change. Sections~\ref{sec:bbs_check}  and \ref{sec:eigtraj} below identify three diagnostics that all
flag ring formation sharply (rank reduction of the $\ell=1$
multiplet, contraction of the bulk magnitude, shift of the
rank-decay exponent on the bottom multiplets), all computable
from the distance matrix alone. 

The transferable lesson is that {\bf delocalized collective
states such as a ring leave a signature in the lowest few
non-Perron eigenvalues of the distance matrix}, and that this
signature is sharp enough to be useful as an {\bf unsupervised
indicator} in settings where the geometric configuration itself
is not accessible. Empirical matrices of this kind arise as the
primary observable in several concrete domains. In financial
markets, time-varying asset correlation matrices are analyzed
through their spectra to clean estimators and to detect regime
changes~\cite{bouchaudpotters2003, bunbouchaudpotters2017}. In
network science, the eigenvalue distribution of empirical graph
and network adjacency matrices is sensitive to topology and to
community structure~\cite{goh2001, chunglu2003}. Similarity
matrices in molecular dynamics, kernel embeddings, and
high-dimensional data clustering are analyzed in the same
spectral terms. The dynamical extension of the
distance-matrix spectral problem developed here addresses, in a
controlled setting, the question of how a qualitative change in
the underlying generative process is reflected in the spectrum
of the time-varying matrix.

\paragraph{The underlying particle dynamics and its energy.}
The FBP system was introduced in
\cite{halperin2026frustrated} and combines thermal noise,
quenched disorder, and non-trivial geometry: $N$ Brownian
particles diffuse on $S^2$ under a pair potential $\Phi_{ij}\,
M_{ij}$ linear in the geodesic distance, with $\Phi_{ij}$ a
quenched Gaussian random matrix. A companion paper
\cite{halperin2026fields} develops a statistical field theory of
the large-$N$ limit of the FBP model, hereafter the
\emph{F2 (Frustrated Fields)} model, which reduces through a
collective-coordinate analysis to an $\mathrm{O}(3)$ nonlinear
sigma model in $(0+1)$ dimensions for the orientation
$\hat{\mathbf{n}}(t)\in S^2$ with a single low-energy constant
$D_{\text{rot}}$. The present paper instead keeps the full
configuration space and reads the same dynamics through the
distance matrix $M(t)$ defined in~\eqref{eq:M_def_intro}; the
disorder-weighted potential energy of the particle system,
\begin{equation}
\label{eq:E_inner_product}
E(t) \;=\; \tfrac{1}{2}\sum_{i\ne j}\Phi_{ij}\,M_{ij}(t)
\;=\; \tfrac{1}{2}\,\mathrm{Tr}\bigl(\Phi\,M(t)\bigr),
\end{equation}
becomes a Frobenius inner product between $M(t)$ and the
time-independent disorder $\Phi$, and serves throughout as a
one-dimensional projection of the matrix trajectory along
$\Phi$.

\paragraph{Summary of contributions.}
We pursue three lines of analysis on the FDM.
\textbf{(1) Static template and self-averaging identification.}
The mechanism is self-averaging of the bulk density at large
$N$: the FDM bulk at any time coincides with the ERM bulk on
i.i.d.\ samples from $\mu_t$, verified by an i.i.d.-resample
comparison (Sec.~\ref{sec:bulk_ensembles}). As a consequence,
the static BBS picture (Perron pattern, $(2\ell+1)$
quasi-multiplets, power-law tail, small-$|\lambda|$
localization) is reproduced quantitatively by the FDM ensemble
at every time (Sec.~\ref{sec:bbs_check}). The bottom eigenspace
of $M(t)$ is moreover algebraically tied to the F2
inertia-tensor PCA estimator of the ring orientation through
the $\ell=1$ Legendre block of the $\arccos$ kernel
(Sec.~\ref{sec:spectral_evolution}), giving a static
self-consistency bridge to the companion F2
model~\cite{halperin2026fields}. At the sub-leading level, a
pooled-window double-difference test on the level-spacing
distribution $P(s)$ (Sec.~\ref{sec:universality}) detects a
non-i.i.d.\ residual at small $s$ at $3$--$8\,\sigma$, the
spectral footprint of the attractive pair correlations
$\Phi_{ij}$ that the i.i.d.\ ERM null on $\mu_t$ does not
contain.
\textbf{(2) Dynamical signatures of ring formation.} The static
template acquires three sharp non-equilibrium signatures during
the ring-formation transient, all computable from $M(t)$ alone
and sharing the same fast timescale: rank reduction of the
$\ell=1$ multiplet, contraction of the bulk magnitude, and a
shift of the rank-decay exponent $\beta$ on the bottom-$50$
window from the $S^2$ value $\beta=3/2$ toward the $S^1$ ring
value $\beta=2$ (Secs.~\ref{sec:bbs_check}--\ref{sec:eigtraj},
\ref{sec:powerlaw}). The bulk density exponent $\alpha$, by
contrast, is robust to ring formation: it stays near the BBS
$d=2$ value $5/3$ at all times and at all $N$ tested, and a
finite-$N$ scan at the FBP NESS confirms that the small
$t=0\to$NESS shift seen at fixed $N=400$ is a finite-$N$
feature that moves upward toward $5/3$, not downward toward
$3/2$, as $N$ grows.
\textbf{(3) Initial-condition robustness.} A Big Bang experiment
with particles initially clustered in a Gaussian blob on $S^2$
reaches the same NESS through a $\sim 5\times$ faster collapse,
confirming that the spectral signatures are inherited from the
time-evolving $\mu_t$ rather than from the initial measure
(Sec.~\ref{sec:bigbang}).

\paragraph{Simulation protocol.}
The data on which the analysis is based come from ten independent
simulations of the FBP model at $N = 400$, $T = 0.4$, with
Gaussian quenched couplings (independent disorder $\Phi$ in each
run), recorded through both the ring-formation transient and the
longer NESS phase. We store the full geometric trajectory
$\{M(t)\}_t$, the disorder matrix $\Phi$, the eigenvalues
$\lambda_k(t)$, the energy $E(t)$, and the lab-frame particle
positions, so that all diagnostics below are computed from a
common dataset. The default initial condition is uniform random
on $S^2$ for Secs.~\ref{sec:setup}--\ref{sec:matrix_dynamics};
Sec.~\ref{sec:bigbang} replaces it with the Big Bang Gaussian
blob.

\paragraph{Outline.}
Section~\ref{sec:setup} fixes notation and the simulation
protocol. Section~\ref{sec:spectral_evolution} reports the
spectral evolution of $M(t)$ across the ring-formation
transient and the NESS regime, and records the algebraic
identity between the bottom-$K$ eigenspace of $M(t)$ and the
inertia-tensor PCA estimate of the ring orientation.
Section~\ref{sec:bulk_ensembles} compares the bulk eigenvalue
density to candidate random-matrix ensembles, establishes
the ERM identification on the FBP one-particle density $\mu_t$
as the geometry-aware null hypothesis, and fits the heavy-tailed
bulk to a power law against the BBS prediction.
Section~\ref{sec:bbs_check} carries out the
quantitative BBS check (Perron pattern, quasi-multiplets,
density tail, BBS-Anderson participation-ratio crossover).
Section~\ref{sec:bottom_diagnostics} introduces two
ring-formation diagnostics from the bottom of the spectrum:
an outlier count and the non-crossing fan-out of the bottom
eigenvalues.
Section~\ref{sec:universality} analyzes level-spacing
statistics against GOE / Poisson references, the BBS / Anderson
superposition prediction, and the ERM null hypothesis on the
FBP one-particle density $\mu_t$.
Section~\ref{sec:matrix_dynamics} positions $M(t)$ within
dynamic random matrix theories (Dyson Brownian motion, the F2
field-theoretic formulation in the large-$N$ limit, free
probability, spin glass / SYK).
Section~\ref{sec:bigbang} reports the Big Bang
initial-condition experiment.
Section~\ref{sec:discussion} discusses the BBS / ERM
identification, the spectral signals of structural change, and
their practical relevance, and Section~\ref{sec:summary}
summarizes and sketches extensions, the link to the F2 model,
and future directions.

%==============================================================================
\section{Setup and dataset}
\label{sec:setup}
%==============================================================================

\paragraph{Particle dynamics.}
The microscopic model is that of
Ref.~\cite{halperin2026frustrated}: $N$ Brownian particles on the
unit sphere $S^2$ with overdamped Langevin dynamics in the
embedding-space form,
\begin{equation}
d\mathbf{x}_i = \frac{1}{\gamma}P(\mathbf{x}_i)\!\!\sum_{j\ne i}
\Phi_{ij}\,\hat{\mathbf{t}}_{ij}\,dt
+ \sqrt{2D}\,P(\mathbf{x}_i)\circ d\mathbf{W}_i,
\end{equation}
with $P(\mathbf{x}) = I - \mathbf{x}\mathbf{x}^T$ the tangent-plane
projector at $\mathbf{x}\in S^2$,
$\hat{\mathbf{t}}_{ij}$ the unit tangent vector at
$\mathbf{x}_i$ pointing toward $\mathbf{x}_j$ along the connecting
geodesic, and $\Phi_{ij} = \Phi_{ji} \sim \mathcal{N}(0,\sigma^2)$ a
single realization of quenched random couplings drawn at $t = 0$
and frozen thereafter. The main analysis throughout this paper
uses $N = 400$, $\sigma = 1$, $\gamma = 1$, $T = 0.4$, $dt = 0.0025$,
and integrates from $t = 0$ to $t = 50$ using the embed-and-project
scheme of Ref.~\cite{halperin2026frustrated}. Two dedicated
finite-$N$ scans (\S\ref{sec:bbs_check} (v) and (vi)) vary $N$
across $\{100, 200, 400, 800, 1600\}$ to verify finite-$N$ approach
to the BBS asymptotic predictions, both at $t=0$ on i.i.d.\ uniform
points and at the dynamical NESS; outside those two subsections the
system size is held at $N=400$.

\paragraph{Distance matrix and energy.}
At every recorded time $t$ the geometric state is encoded by the
$N \times N$ symmetric matrix $M(t)$ of geodesic
distances~\eqref{eq:M_def_intro}, with $M_{ii}(t) = 0$ and
$M_{ij}(t) \in [0, \pi]$. The matrix is invariant under any
global rotation $R \in \mathrm{SO}(3)$ of the configuration:
\begin{equation}
\label{eq:M_rotation_invariance}
M[RX]_{ij} \;=\; \arccos\!\bigl((R\mathbf{x}_i)\cdot
(R\mathbf{x}_j)\bigr) \;=\;
\arccos(\mathbf{x}_i\cdot\mathbf{x}_j) \;=\; M[X]_{ij}.
\end{equation}
A rigidly rotating ring therefore gives exactly the same
$M(t)$ at every orientation, so the absolute lab-frame ring
normal $\hat{\mathbf{n}}(t)$ cannot be recovered from the
distance matrix alone. The matrix can detect that the point
cloud has collapsed onto a ring-like intrinsic geometry, but
not the lab-frame direction of that ring. Throughout the paper
we therefore distinguish \emph{intrinsic} diagnostics
(computable from $M(t)$ alone) from \emph{extrinsic} or
lab-frame diagnostics (requiring the position trajectory
$X(t)$). The disorder-weighted energy
$E(t)$~\eqref{eq:E_inner_product} is recorded in parallel. The
trajectory length $t_{\text{final}} = 50$ is long enough to capture
both the non-equilibrium ring-formation transient
($\tau_{\text{fast}} \approx 5$,
Ref.~\cite{halperin2026frustrated}) and a sizable fraction of the
NESS phase $(t \in [10, 50])$ where the orientation
$\hat{\mathbf{n}}(t)$ diffuses on $S^2$.

\paragraph{Ensemble.}
We run ten independent realizations of the FBP dynamics, with
disorder seed $\mathrm{coupling\_seed}_i = 42 + 17\,i$ and initial
condition seed $\mathrm{init\_seed}_i = 123 + 31\,i$, $i = 0,\ldots,9$.
For each run we record the full distance-matrix trajectory at
$\Delta t_{\text{rec}} = 0.25$ ($n_{\text{snap}} = 200$ snapshots),
together with the disorder matrix $\Phi^{(i)}$, the eigenvalues
$\lambda_k(t)$, the energy $E^{(i)}(t)$, the lab-frame positions
$\mathbf{x}_n^{(i)}(t)$, and an inertia-tensor ring-quality
diagnostic $\eta(t)$ defined as follows. Let
$C(t) = (1/N)\sum_n \mathbf{x}_n(t)\mathbf{x}_n(t)^T$ be the
$3\times 3$ symmetric, positive-semidefinite covariance matrix
of the particle positions, and let
$\mu_1(t) \le \mu_2(t) \le \mu_3(t)$ be its three eigenvalues
ordered ascending. The smallest eigenvalue $\mu_1$ is the
variance of the particle distribution along its weakest spatial
direction; for a well-formed ring perpendicular to a normal
$\hat{\mathbf{n}}$ this direction is $\hat{\mathbf{n}}$ itself,
and $\mu_1$ is small. We define the scalar diagnostic
\begin{equation}
\label{eq:eta_def}
\eta(t) \;=\; \frac{\mu_2(t)}{\mu_1(t)} \;\;\ge\;\; 1,
\end{equation}
which is order one for a uniformly distributed configuration
($\mu_1 \approx \mu_2 \approx \mu_3$) and large for a
well-formed ring ($\mu_1 \to 0$ while $\mu_2$ remains in the
ring plane). All ten realizations form rings: the final
$\eta(t = 50)$ ranges from $26$ to $106$.

%==============================================================================
\section{Spectral evolution and ring formation}
\label{sec:spectral_evolution}
%==============================================================================

\paragraph{Why the bottom of the spectrum is informative.}
Before turning to the data we record an algebraic fact that
underpins all of the analysis below: the spectral content of
$M(t)$ is sorted by angular wavenumber, and the part of the
spectrum that carries macroscopic geometric information sits at
its negative end. The \emph{Legendre expansion} of the
arccos kernel on $S^2$, also called the \emph{Funk--Hecke
expansion} for rotationally invariant kernels on the sphere,
reads
\begin{equation}
\label{eq:legendre_expansion}
\arccos(\mathbf{x}_i\cdot\mathbf{x}_j) \;=\;
\frac{\pi}{2} \;-\; \frac{3\pi}{8}\, P_1(\mathbf{x}_i\cdot\mathbf{x}_j)
\;-\; \sum_{\ell\geq 3,\,\text{odd}} c_\ell\,
P_\ell(\mathbf{x}_i\cdot\mathbf{x}_j),
\quad c_\ell > 0,
\end{equation}
with only odd $\ell$ contributing because $\arccos(t) - \pi/2$ is
odd in $t$. The integer $\ell\ge 0$ in
\eqref{eq:legendre_expansion} plays the role of an
\emph{angular-momentum quantum number} on $S^2$: it labels the
irreducible representations of the rotation group
$\mathrm{SO}(3)$ that act on functions on the sphere, and each
$\ell$-representation is $(2\ell+1)$-dimensional, spanned by the
spherical harmonics $Y_\ell^m$ for $m=-\ell,\ldots,+\ell$, which
are the eigenfunctions of the Laplace--Beltrami operator on $S^2$
with eigenvalue $\ell(\ell+1)$. Larger $\ell$ corresponds to
faster angular oscillation. The Legendre addition theorem,
$P_\ell(\mathbf{x}\cdot\mathbf{y}) =
\frac{4\pi}{2\ell+1}\sum_{m=-\ell}^{+\ell} Y_\ell^m(\mathbf{x})
\overline{Y_\ell^m(\mathbf{y})}$, then turns each Legendre term
$P_\ell(\mathbf{x}_i\cdot\mathbf{x}_j)$ in \eqref{eq:legendre_expansion}
into a rank-$(2\ell+1)$ operator on the empirical configuration:
its $(2\ell+1)$ non-zero eigenvalues approximately coincide
(forming a near-degenerate \emph{quasi-multiplet} of dimension
$2\ell+1$) and its eigenvectors are linear combinations of the
spherical-harmonic modes $Y_\ell^m$ evaluated at the $N$
particle positions. This is the BBS quasi-multiplet structure
\cite{bogomolny2003} that organizes the spectrum of $M(t)$.

The decomposition translates directly into a sum of matrices on
the empirical configuration $X = (\mathbf{x}_1, \ldots,
\mathbf{x}_N)$. The constant $\pi/2$ gives the rank-one piece
$(\pi/2)\mathbf{1}\mathbf{1}^T$, with a single eigenvalue
$\lambda_1 \approx N\pi/2$ along $\mathbf{1}$ (the trivial
$\ell = 0$ representation). The $\ell = 1$ piece, with the
negative coefficient $-3\pi/8$, gives $-(3\pi/8)\,X^T X$, where
$X^T X$ is the $N\times N$ Gram matrix of rank at most three.
Its three positive eigenvalues equal the principal moments of
the empirical inertia tensor $X X^T = \sum_n
\mathbf{x}_n\mathbf{x}_n^T$, corresponding to the three
components of the dipole / vector representation
$Y_1^{-1,0,+1}$. The higher odd-$\ell$ pieces produce blocks
of rank $2\ell+1$ at progressively smaller amplitudes, with
eigenvalues that fill the central bulk and oscillate
increasingly rapidly with the angle
$\theta_{ij}=\arccos(\mathbf{x}_i\cdot\mathbf{x}_j)$.

Two consequences follow. First, the leading eigenvalue is fixed
by the kernel mean and contains no geometric information beyond
the (time-independent) average pairwise distance. Second, the
\emph{geometric} content, i.e.\ the empirical inertia tensor that
distinguishes a uniformly distributed configuration from a ring,
is loaded onto the $\ell = 1$ block, and that block enters $M(t)$
with a \emph{negative} amplitude $-3\pi/8$. The three largest
eigenvalues of the inertia tensor therefore appear as the three
\emph{most negative} eigenvalues of $M(t)$. A uniform
configuration on $S^2$ has an isotropic inertia tensor with
three roughly equal moments $\sim N/3$, producing three negative
outliers of comparable size $\approx -3\pi N / 24 = -\pi N/8$
(numerically $\approx -157$, in agreement with the data). A ring configuration has rank-two inertia
($\mu_1 \to 0$ along $\hat{\mathbf{n}}$ and $\mu_2 \approx \mu_3$
in the ring plane), so two of the three negative outliers grow
in magnitude and the third moves toward zero. The collapse of
$\mathrm{SO}(3)$ rotational symmetry to $\mathrm{SO}(2)$ is read
off the bottom of the spectrum as the rank reduction of the
$\ell = 1$ block, which is precisely what the inertia-ratio
diagnostic $\eta(t)$ defined above measures from the empirical
positions. The matrix dynamics and the geometric dynamics are
the same dynamics, viewed through the same $\ell = 1$ block.
Higher-$\ell$ blocks do not aggregate macroscopic geometric
quantities cleanly: their bases are spherical harmonics of large
wavenumber whose ensemble-averaged contributions to coarse
features cancel, leaving them in the bulk near zero. This is the
sense in which ring formation, which is not directly visible from
the matrix entries $M_{ij}(t)$, is exposed sharply by the
smallest eigenvalues of $M(t)$.

\paragraph{Algebraic identity: bottom eigenspace and the
inertia-tensor PCA estimator.}
\label{sec:eigvec_alignment}
A direct corollary of the Legendre expansion is that the bottom
eigenspace of $M(t)$ encodes the same column-space-of-$X$
information that an inertia-tensor PCA on the particle positions
extracts from $X^T X$. The orientation
$\hat{\mathbf{n}}_{\text{inertia}}(t)$ used in
Ref.~\cite{halperin2026fields} is the smallest eigenvector of
the inertia tensor
\begin{equation}
\label{eq:inertia_def}
C(t) \;=\; \frac{1}{N}\,X(t)^T X(t).
\end{equation}
The companion distance-matrix estimator
$\hat{\mathbf{n}}_{\text{distmat}}(t)$ is built from the
$K$ bottom eigenvectors
$V_{\text{bot}}(t)\in\mathbb{R}^{N\times K}$ of $M(t)$
contracted with $X(t)$,
$Y(t) = X(t)^T V_{\text{bot}}(t) \in \mathbb{R}^{3\times K}$,
as the smallest eigenvector of
$A(t) = Y(t) Y(t)^T \in \mathbb{R}^{3\times 3}$. The contraction
$X^T V_{\text{bot}}$ is what carries the lab-frame direction:
by the rotation invariance
$M[RX]=M[X]$~\eqref{eq:M_rotation_invariance}, no observable
built from $M(t)$ alone can identify any lab-frame axis, so the
orientation information must enter through $X(t)$.

With $K = 2$, the cosine alignment
\begin{equation}
\label{eq:cosine_alignment}
\bigl|\langle \hat{\mathbf{n}}_{\text{distmat}}(t),
\hat{\mathbf{n}}_{\text{inertia}}(t)\rangle\bigr|
\;=\; |\cos\theta(t)|
\;\in\;[0,1],
\end{equation}
between the two unit 3-vectors (with $\theta(t)$ the angle
between them; the absolute value handles the eigenvector sign
ambiguity) is equal to $1$ across the entire trajectory in
every realization, verified numerically to floating-point
precision at every snapshot.

The identity holds because the $-(3\pi/8)\,X X^T$ term of the
Legendre expansion~\eqref{eq:legendre_expansion} embeds the
rank-three column space of $X$ into the bottom of $M$'s
spectrum: the BBS $\ell=1$ multiplet has eigenvalues
proportional to the principal moments of the inertia tensor and
eigenvectors in the column space of $X$. For $K=3$ the bottom
eigenspace is the full rank-three $\ell=1$ column space and the
construction recovers all three principal moments. For $K=2$ it
is the rank-two ring-plane subspace after ring formation, and
the smallest eigenvector of $A(t)$ is the orthogonal direction,
which is the PCA ring normal. Before ring formation, when the
cloud is isotropic, the alignment $\equiv 1$ is an algebraic
artifact of comparing two estimators that share the same
finite-sample anisotropy of $X$.

The alignment is therefore mechanistic, not informational: the
lab-frame direction of $\hat{\mathbf{n}}$ is recovered from
$X(t)$, not from $M(t)$. What does change with the dynamics is
the rank-two-vs-rank-three separation among the eigenvalues of
$A(t)$, which equals the ring-quality diagnostic $\eta(t)$ and
grows from order one at $t=0$ to the dozens after ring
formation. The check rules out an alternative scenario in which
a higher-$\ell$ block would compete with $\ell=1$ for the
bottom of $M$'s spectrum.

\begin{figure}[!htbp]
\centering
\includegraphics[width=\textwidth]{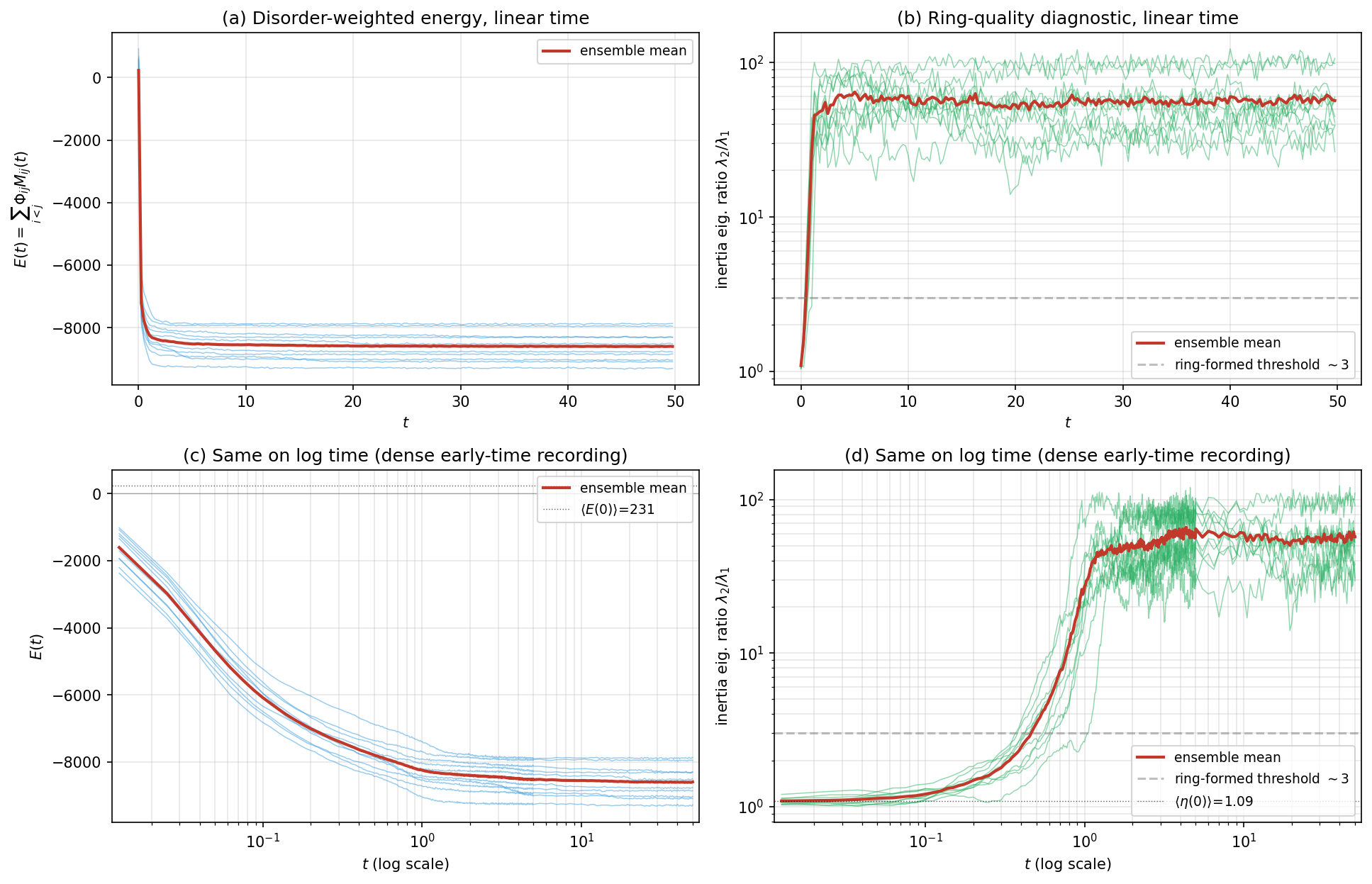}
\caption{Disorder-weighted energy
$E(t) = \tfrac{1}{2}\sum_{ij}\Phi_{ij}M_{ij}(t)$ and
inertia-tensor ring-quality diagnostic $\eta(t)$ for ten
independent realizations (blue/green) and the ensemble mean
(red). \textbf{Top row}, linear time: \textbf{(a)} $E(t)$ and
\textbf{(b)} $\eta(t)$ on log-$y$ with the ring-formation
threshold $\eta=3$ marked. The descent in (a) looks almost like
an instantaneous step at $t=0$, and the rise in (b) is
similarly compressed into the leftmost portion of the axis.
\textbf{Bottom row}, log time: same observables on a
logarithmic time axis using a dense early-time recording
(rec\_dt$=0.0125$ for $t\in[0,5]$, $0.5$ for $t\in[5,50]$, same
seeds and parameters as the top-row data). \textbf{(c)} $E(t)$
on log-$t$: the sharp drop is resolved into a smooth, roughly
logarithmic relaxation across $t\in[10^{-2}, 5]$, with the
ensemble-mean starting value $\langle E(0)\rangle$ marked by a
dotted reference line. \textbf{(d)} $\eta(t)$ on log-$t$ and
log-$y$: the ring-quality rises smoothly from $\eta\approx 1$
across the same decade and saturates above the threshold,
continuing a slow drift in NESS.}
\label{fig:rmt_energy}
\end{figure}

\paragraph{Energy evolution.}
Figure~\ref{fig:rmt_energy} shows the disorder-weighted energy
$E(t)$ and the inertia-tensor ring-quality diagnostic $\eta(t)$
across all ten realizations.

The disorder-weighted energy drops sharply between $t = 0$ and
$t \approx 5$, then plateaus. The inertia ratio rises from
$\eta \approx 1$ (random) to $\eta \in [25, 100]$ on the same
timescale. We refer to this transient as the
\emph{ring-formation phase} and to the post-transient regime as
the \emph{NESS phase}. On the linear-time axes of the top row
of Fig.~\ref{fig:rmt_energy} the descent looks almost like an
instantaneous step at $t=0$; the bottom row uses a logarithmic
time axis with dense early-time recording to resolve the same
transient into a smooth roughly logarithmic relaxation over
$t\in[10^{-2}, 5]$.

\begin{figure}[!htbp]
\centering
\includegraphics[width=\textwidth]{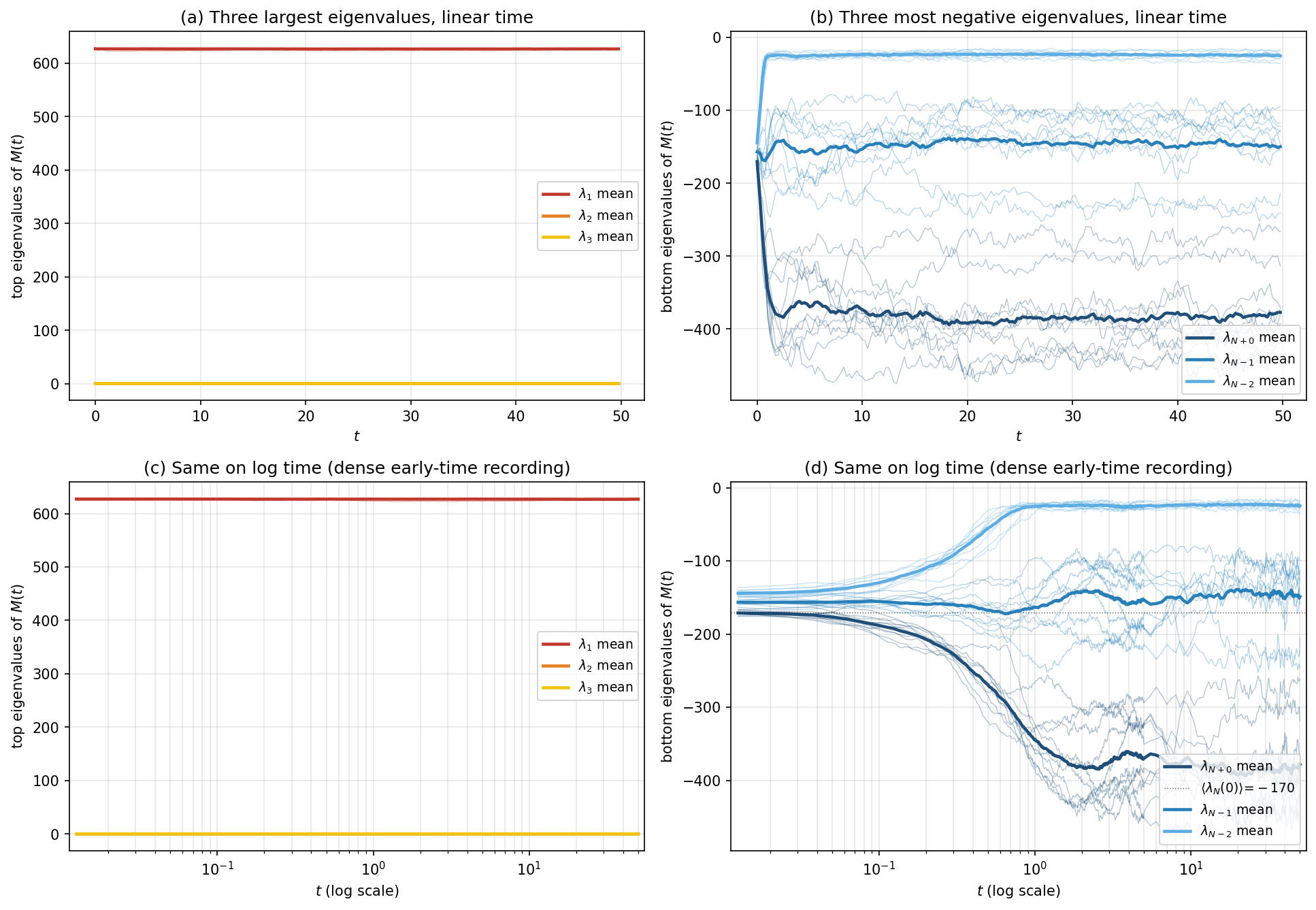}
\caption{Top and bottom of the spectrum of $M(t)$ over time.
\textbf{Top row}, linear time: \textbf{(a)} the three largest
eigenvalues $\lambda_1, \lambda_2, \lambda_3$. The leading
eigenvalue is essentially time-independent at
$\lambda_1 \approx N\langle d\rangle = N\pi/2 \approx 628$ (the
rank-1 component of $M(t)$) and is well separated from
$\lambda_2, \lambda_3 \sim O(1)$. \textbf{(b)} the three most
negative eigenvalues. The most-negative eigenvalue $\lambda_N$
drops from $\approx -200$ at $t=0$ to $\approx -400$ during the
ring-formation transient on the same timescale as the energy
drop in Fig.~\ref{fig:rmt_energy}, providing a sharp spectral
signature of ring formation. \textbf{Bottom row}, log time:
same trajectories on a logarithmic time axis using the dense
early-time recording. \textbf{(c)} the top three eigenvalues
remain visibly constant across the resolved decade.
\textbf{(d)} the most-negative eigenvalue's near-vertical drop
at $t=0^+$ in panel (b) is resolved here as a smooth descent
across $t\in[10^{-2}, 5]$, sharing the same logarithmic
character as $E(t)$ in Fig.~\ref{fig:rmt_energy}(c). The
$\ell=1$ multiplet rank reduction is visible as the splitting
of $\lambda_N$ away from $\lambda_{N-1},\lambda_{N-2}$ on the
same timescale; the dotted reference line marks the
ensemble-mean starting value $\langle\lambda_N(0)\rangle$.}
\label{fig:rmt_eigvals}
\end{figure}

Figure~\ref{fig:rmt_eigvals} tracks the leading and bottom
eigenvalues of $M(t)$ across the same window.

The leading eigenvalue is essentially time-independent at
$\lambda_1 = \langle M \rangle\, N \approx N\pi/2 \approx 628$. This
follows from the rank-1 component of $M$: with mean entry
$\overline{M}_{ij} = \pi/2$ (a uniform configuration on $S^2$ has
expected pairwise arccos equal to $\pi/2$, and a ring configuration
has the same expected pair distance averaged over particle pairs),
the matrix $M$ is to leading order $(\pi/2)\,(J - I)$ with
$J = \mathbf{1}\mathbf{1}^T$, whose top eigenvalue is
$(\pi/2)(N - 1)$ with eigenvector
$\propto \mathbf{1}$. The fact that $\lambda_1$ does not change
during ring formation reflects that the average pair distance is
nearly invariant under the geometric collapse, even though the
configuration changes drastically.

The structurally interesting signal lies at the negative end of the
spectrum. The most negative eigenvalue $\lambda_N$ drops from
$\approx -200$ (uniform initial) to $\approx -400$ during the
transient. The size of this drop is consistent with the
emergence of two new geometric directions: in a uniform
configuration the spherical harmonic content of the arccos kernel
gives rise to several large negative eigenvalues spread across the
$\ell = 1, 2, \ldots$ representations, while in a ring
configuration the strong geometric constraint that all particles
lie within an angular band of width $\sigma_\theta \approx 4.8^\circ$
(see Ref.~\cite{halperin2026fields}, Section~5.2)
amplifies the lowest-$\ell$ contributions and pulls the bottom
of the spectrum further down.

\begin{figure}[!htbp]
\centering
\includegraphics[width=\textwidth]{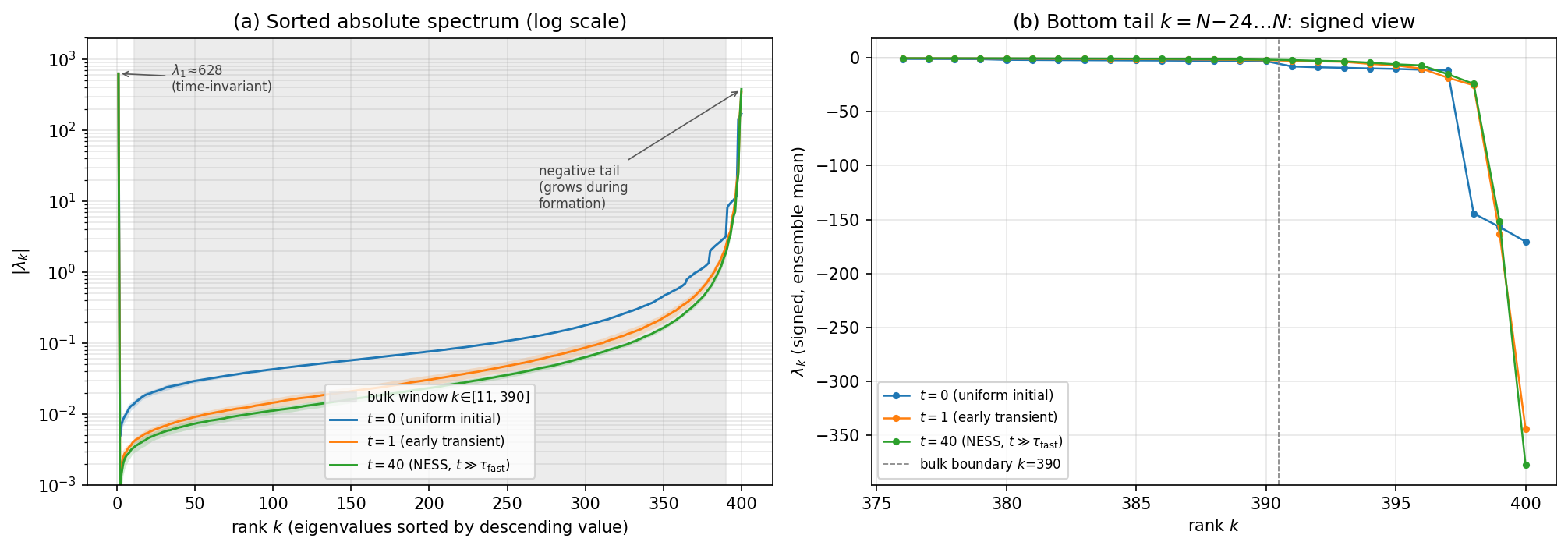}
\caption{Sorted spectrum of $M(t)$ at three representative times:
$t=0$ (uniform initial), $t=1$ (early transient), $t=40$
(NESS, $t\gg\tau_{\text{fast}}$). The intermediate sample is
chosen at $t=1$ rather than $t\sim\tau_{\text{fast}}=5$ because
the spectrum saturates well within the fast time: at $t=5$ it is
already indistinguishable from the $t=40$ NESS shape, so picking
an earlier point makes the transient visible. Curves are
ensemble means over ten realizations; shaded bands in (a) show
the realization-to-realization spread.
\textbf{(a)} Sorted absolute spectrum $|\lambda_k|$ vs rank $k$
on log scale. The near-vertical jump at the left edge is the
single Perron eigenvalue $\lambda_1\approx 628$, set by the
overall mean of $M_{ij}=\arccos(\mathbf{x}_i\cdot\mathbf{x}_j)$
and time-invariant. The shaded band marks the bulk window
$k\in[11, 390]$, defined by excluding the ten largest outliers
at each end (selected as those rank positions where the curve
breaks sharply away from the smooth bulk profile in panel (a)).
Within this window the bulk magnitude at mid-rank
$k\!\approx\!200$ contracts from $|\lambda|\sim 0.08$ at $t=0$ to
$|\lambda|\sim 0.03$ at $t=1$ to $|\lambda|\sim 0.024$ at NESS,
i.e.\ a factor of $\sim 3$ reduction concentrated in the early
transient. \textbf{(b)} Signed bottom tail
$\lambda_k$ for $k=N\!-\!24,\ldots,N$ on linear scale. This is
the spectral channel that distinguishes the stages: at $t=0$ the
deepest eigenvalue sits near $-170$ (set by the three
$\ell\!=\!1$ dipole modes of the uniform sphere), grows in
magnitude to $\approx -344$ at $t=1$, and saturates near $-377$
at NESS. The dashed vertical line marks the bulk boundary $k=390$.}
\label{fig:rmt_spectra}
\end{figure}

Figure~\ref{fig:rmt_spectra} shows the rank-sorted spectrum at
three representative times. Eigenvalues are sorted by descending
value, so rank $k=1$ corresponds to the largest (most positive)
and $k=N$ to the smallest (most negative).

The qualitative observations: (i) the leading eigenvalue
$\lambda_1 \approx 628$ is independent of $t$, since it tracks
the overall mean $\langle M_{ij}\rangle$ and ring formation does
not change the average chord length much; (ii) the bulk
contracts by a factor of three within the early transient,
indicating that the geometric collapse onto a one-dimensional
support reduces the effective rank of $M(t)$; (iii) the
most-negative outlier roughly doubles in magnitude during
formation. The bulk evolution is essentially complete by
$t\sim\tau_{\text{fast}}$, so the slow drift between $t=5$ and
$t=40$ shows up only in the tails of the eigenvalue trajectories
(\S\ref{sec:eigtraj}), not in the overall sorted
profile. Together these features identify ring formation as a
redistribution of spectral mass \emph{within} the matrix, not a
change of its overall norm.

%==============================================================================
\section{Bulk shape and comparison to reference ensembles}
\label{sec:bulk_ensembles}
%==============================================================================

\begin{figure}[!htbp]
\centering
\includegraphics[width=\textwidth]{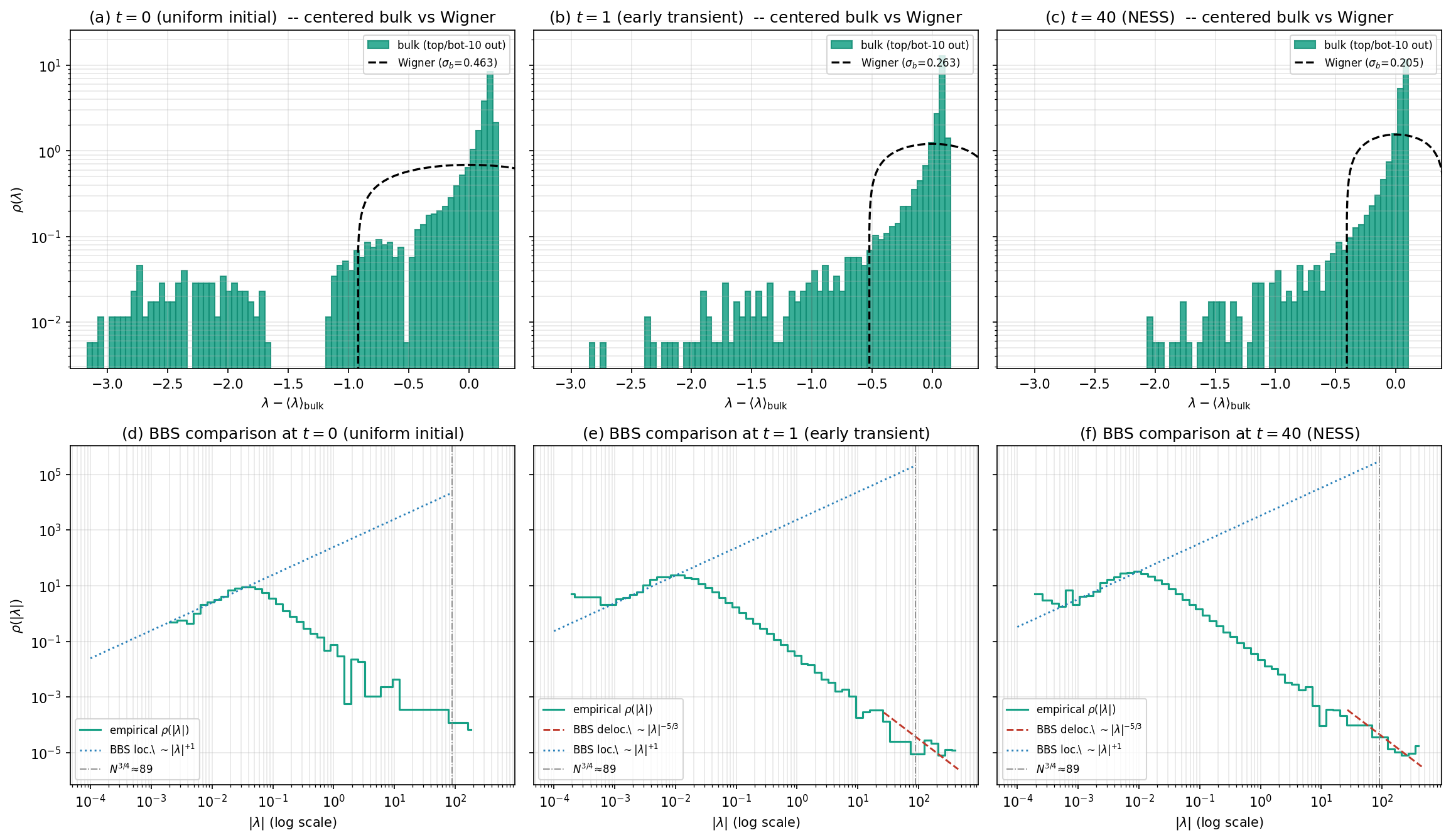}
\caption{Bulk eigenvalue density $\rho(\lambda)$ pooled across
the ten realizations, after dropping the ten largest and ten
smallest eigenvalues per realization (the bulk window of
Fig.~\ref{fig:rmt_spectra}). \textbf{Top row (a, b, c)}: centered
density $\rho(\lambda - \langle\lambda\rangle_{\text{bulk}})$ on
linear $x$ and log $y$, common axes across the three times. The
common x-range $[-3.3, 0.4]$ is the union of the three centered
bulks. Dashed line: Wigner semicircle for a Gaussian Orthogonal
Ensemble matrix with the same second moment $\sigma_b$ as the
empirical bulk. The bulk standard deviation falls from
$\sigma_b\!=\!0.46$ at $t=0$ to $0.26$ at $t=1$ to $0.21$ at NESS.
At $t=0$ the distribution is bimodal: a sharp peak at zero plus a
secondary cluster around $\lambda\!\approx\!-2.5$ from the
residual moderately-negative modes; at $t=1$ the gap fills in;
at NESS the secondary cluster is gone and the bulk is supported
essentially on $[-2, 0]$. In all three stages the matched
semicircle is uniformly too wide and misses the sharp central
peak; the geometry-constrained bulk is not generated by a GOE
ensemble. \textbf{Bottom row (d, e, f)}: comparison of the same
data to the BBS ERM predictions on
log-log axes, density $\rho(|\lambda|)$ vs $|\lambda|$ over the
full range (Perron eigenvalue excluded only). Dashed red:
delocalized BBS prediction $\rho \sim |\lambda|^{-5/3}$ for $d=2$,
valid for $|\lambda|\gtrsim N^{(d+1)/(2d)} = N^{3/4}\!\approx\!89$.
Dotted blue: localized BBS prediction $\rho\sim|\lambda|^{+1}$ for
$d=2$ (linear vanishing at $\lambda=0$), valid in the
small-$|\lambda|$ regime where BBS predicts strongly localized
eigenstates. Dash-dotted gray: the delocalized/localized
threshold $N^{3/4}$. The empirical density agrees with the
linear-vanishing $|\lambda|^{+1}$ prediction at small $|\lambda|$
(visible at $t=0$, weaker at NESS where the localized regime is
deformed by ring formation) and with the $|\lambda|^{-5/3}$ tail
at large $|\lambda|$ (above the delocalization threshold). The
BBS ERM picture is the correct continuous-spectrum reference for
this matrix, in contrast to the Wigner semicircle of the top
row.}
\label{fig:rmt_bulk}
\end{figure}

After dropping the large outliers at each end, the central
``bulk'' of the spectrum invites comparison with two candidate
reference ensembles. The natural reference for the FDM is the
static distance matrix on $S^2$ studied by
BBS~\cite{bogomolny2003}, the canonical Euclidean random matrix
on the sphere; it predicts a definite bulk shape
with two power-law segments set by the dimension $d=2$ of the
underlying manifold. For comparison we also overlay a Wigner
semicircle of matched second moment, the prediction one would
get from a GOE matrix with the same scale, included as a
sanity check rather than a serious candidate.
Figure~\ref{fig:rmt_bulk} shows the bulk density at
$t = 0, 1, 40$ with both references on the same data.

\paragraph{BBS ERM predictions match.}
The bottom row of Fig.~\ref{fig:rmt_bulk} shows the empirical
density on log-log axes against the BBS predictions for distance
matrices on a $d$-dimensional sphere. Two power-law segments are
predicted: the \emph{delocalized} regime
$|\lambda|\gtrsim N^{(d+1)/(2d)}$ has
$\rho(|\lambda|)\sim|\lambda|^{-(2d+1)/(d+1)}$, equal to
$|\lambda|^{-5/3}$ for $d=2$; the \emph{localized} regime
$|\lambda|\lesssim N^{(d+1)/(2d)}$ has
$\rho(|\lambda|)\sim|\lambda|^{d-1}$, linearly vanishing at the
origin for $d=2$. The delocalization threshold sits at
$N^{3/4}\!\approx\!89$, marked in the figure. The
empirical density tracks the BBS predictions across the full
range: a clear linear vanishing at small $|\lambda|$ at $t=0$,
weakening at NESS where ring formation deforms the localized
regime; and a power-law decay above the threshold whose slope
is within $\sim 5\%$ of $-5/3$ at both $t=0$ and NESS,
i.e.\ on the $S^2$ side of the BBS prediction at all times. The
small drift in the fitted slope between $t=0$ ($\alpha\approx
1.73$) and NESS ($\alpha\approx 1.65$) might be naively
mistaken for a partial shift toward the $d=1$ ring value
$\alpha=3/2$, but the finite-$N$ scan of
\S\ref{sec:bbs_check}\,(vi) excludes that reading: pushing $N$
from 100 to 800 the NESS exponent moves \emph{upward} toward
$5/3$, not downward toward $3/2$, so the bulk density exponent
reflects the $d=2$ embedding throughout and is not a
ring-formation diagnostic (see Sec.~\ref{sec:powerlaw}). The BBS distance-matrix ensemble
is the correct continuous-spectrum reference for the FDM at every
time, including the matched power-law slope and the
small-$|\lambda|$ linear vanishing. The matched Wigner
semicircle in the top row, by contrast, is uniformly too wide
and misses the sharp central peak: a GOE matrix has independent
Gaussian entries, while $M(t)_{ij} =
\arccos(\mathbf{x}_i\cdot\mathbf{x}_j)$ is a deterministic
function of the underlying coordinates and is severely correlated
across rows and columns.

\subsection{ERM null hypothesis: spectrum from i.i.d.\ resampling
on the fitted FBP one-particle density}
\label{subsec:erm_null}

The ERM null hypothesis built on the FBP one-particle density
$\mu_t$ is tested against the FDM data on three complementary
spectral statistics, each in a dedicated subsection below: the
bulk eigenvalue density (this subsection), the ranked spectrum
$|\lambda_K|$ on log-log axes
(\S\ref{subsec:ranked_envelope}), and the bulk level-spacing
statistics $P(s)$ and $\langle r\rangle$
(Sec.~\ref{sec:universality}). The three tests are coordinated
checks of the same null at three different statistical levels,
and we summarize their outcomes together at the end of
Sec.~\ref{sec:discussion}.

\paragraph{Beyond i.i.d.\ points: the FDM at $t > 0$.}
The BBS picture above and the wider Bordenave Euclidean ensembles
\cite{bordenave2008, bordenave2013} both rest on the assumption
that the $N$ points are i.i.d.\ samples from a measure on the
manifold; the kernel and the measure can vary, but i.i.d.\ is
the technical input that makes the universality results go
through. The FDM agrees with this assumption at $t=0$ and
deviates from it at $t>0$, in two distinct ways.

\paragraph{The i.i.d.\ assumption holds only at $t=0$.}
At $t=0$ the i.i.d.\ assumption holds for both initial
conditions used in this paper. With the uniform initial condition
of Sec.~\ref{sec:setup} the points are i.i.d.\ uniform on $S^2$,
so $M(0)$ is precisely a sample of the Bordenave ensemble with
the arccos kernel and uniform measure. With the Big Bang initial
condition of Sec.~\ref{sec:bigbang}, conditional on the random
center direction $\hat{\mathbf{c}}\in S^2$ the points are i.i.d.\
samples from a small Gaussian blob on the tangent plane to
$\hat{\mathbf{c}}$ projected back to $S^2$, so $M(0)$ is again a
strict Bordenave matrix conditional on $\hat{\mathbf{c}}$, with a
different (concentrated) measure. The ten realizations of either
experiment are accordingly ten independent draws from the
respective Bordenave ensemble. The choice of initial measure is
ours, not the ensemble's, and changing it (uniform vs.\ Big Bang)
just selects a different Bordenave point.

For $t > 0$ the Langevin coupling generated by the quenched
disorder $\Phi$ correlates the trajectories $\mathbf{x}_i(t)$,
and the FBP joint distribution
$P_t(\mathbf{x}_1,\ldots,\mathbf{x}_N\,|\,\Phi)$ acquires
non-trivial $k$-point correlations at every order $k\ge 2$,
induced by the pairwise force
$-\Phi_{ij}\hat{\mathbf{t}}_{ij}$ in the SDE: pairs with
$\Phi_{ij}<0$ are pulled toward smaller geodesic distance and
pairs with $\Phi_{ij}>0$ are pushed apart, and at NESS the joint
law is a Boltzmann--Gibbs distribution $\propto e^{-U/T}$ with
the pair interaction $U=(1/2)\sum_{ij}\Phi_{ij}d_{ij}$. For any
specific frozen disorder $\Phi$, the FDM is thus a
kernel-on-points matrix on a strongly interacting Gibbs ensemble
rather than on i.i.d.\ samples, where Bordenave's rigorous
universality theorems on i.i.d.\ points do not generally apply.
%and the closest classical relatives of the underlying point
%process are the log-gas / Coulomb-gas and $\beta$-ensembles.

\paragraph{Self-averaging restores the ERM picture at the
spectral level.}
The failure of i.i.d.\ at the joint-law level does not, however,
propagate to the bulk eigenvalue density. The empirical spectral
measure of $M(t)$ is a sum over $N$ eigenvalues whose
realization-to-realization fluctuations at fixed $t$ are expected
to be $O(1/\sqrt N)$, in line with the standard self-averaging
of bulk spectral densities in random matrix theory and with the
self-averaging of extensive observables in spin-glass-type
disordered systems. To leading order in $N$, the bulk density of
$M(t)$ for a typical frozen $\Phi$ therefore coincides with its
disorder average. We must therefore be careful about \emph{which}
disorder average we mean. For a single realization, the
empirical lab-frame density at time $t$ is
\begin{equation}
\hat\mu_t^{(\Phi)}(\mathbf{x}) \;=\;
\frac{1}{N}\sum_n \delta\bigl(\mathbf{x} - \mathbf{x}_n(t;\Phi)\bigr),
\end{equation}
which at NESS is a thin ring sitting at some specific
orientation $\hat{\mathbf{n}}(\Phi)$ determined by the realized
disorder. From this we form two distinct disorder-averaged
one-particle densities. The first is the bare lab-frame
disorder average
\begin{equation}
\mu_t^{\text{lab}}(\mathbf{x}) \;=\;
\mathbb{E}_\Phi\bigl[\hat\mu_t^{(\Phi)}(\mathbf{x})\bigr].
\end{equation}
The law of $\Phi$ is SO(3)-symmetric (Gaussian i.i.d.\ entries),
so the orientation $\hat{\mathbf{n}}(\Phi)$ is itself uniformly
distributed over $S^2$, and the $\Phi$-average smears the ring
uniformly over all orientations. With the SO(3)-invariant
uniform initial condition, $\mu_t^{\text{lab}}$ is uniform on
$S^2$ at all $t$. Drawing $N$ i.i.d.\ samples from
$\mu_t^{\text{lab}}$ gives sphere-uniform points, not points on
any rotated ring: the ring is present in every individual
$\hat\mu_t^{(\Phi)}$, but lost in the $\Phi$-average. The second
is the \emph{aligned} disorder average obtained by first
rotating each realization to bring its ring normal onto a fixed
reference axis,
\begin{equation}
\label{eq:mu_aligned}
\mu_t^{\text{aligned}}(\mathbf{x}) \;=\;
\mathbb{E}_\Phi\!\left[\frac{1}{N}\sum_n \delta\bigl(R_t\,
\mathbf{x}_n(t;\Phi) - \mathbf{x}\bigr)\right],
\end{equation}
where $R_t \in \mathrm{SO}(3)$ rotates
$\hat{\mathbf{n}}(\Phi)$ onto $\hat{z}$. Equivalently
$\mu_t^{\text{aligned}} = \mu_t(\,\cdot\,\mid\hat{\mathbf{n}})$,
the disorder average \emph{conditional} on the orientation. The
conditioning preserves the ring shape: $\mu_t^{\text{aligned}}$
is uniform on $S^2$ at $t=0$ and a thin ring band centered on
the equator at NESS. We write $\mu_t$ for $\mu_t^{\text{aligned}}$
throughout when the meaning is clear from context.

The distinction matters specifically for the ERM null
hypothesis, not for the FDM spectrum itself. The matrix $M(t)$
is rotation-invariant by~\eqref{eq:M_rotation_invariance}, so
its eigenvalues at fixed $\Phi$ are the same in the lab frame
and after the alignment $R_t$, and the FDM spectrum we measure
is identical in both frames. What is not identical is the order
in which one can take the disorder average and the spectrum:
\begin{equation}
\label{eq:order_inequality}
\mathbb{E}_\Phi\!\left[\,\mathrm{spec}\bigl(\mathrm{ERM}\bigl(\hat\mu_t^{(\Phi)}\bigr)\bigr)\,\right]
\;\neq\;
\mathrm{spec}\!\left(\mathrm{ERM}\bigl(\mathbb{E}_\Phi[\hat\mu_t^{(\Phi)}]\bigr)\right)
\;=\; \mathrm{spec}\bigl(\mathrm{ERM}(\mu_t^{\text{lab}})\bigr).
\end{equation}
The left-hand side is the disorder-averaged spectrum of an
i.i.d.\ ERM whose density is the (rotated) ring of a single
realization; this spectrum is rotation-invariant, so it equals
$\mathrm{spec}(\mathrm{ERM}(\mu_t^{\text{aligned}}))$. The
right-hand side is the spectrum of an ERM built on the smeared
uniform density on $S^2$. The two coincide only when
$\hat\mu_t^{(\Phi)}$ is itself rotation-invariant, which is true
at $t=0$ but not at NESS. Because the FDM bulk is the LHS by
self-averaging at fixed typical $\Phi$, the natural ERM null
hypothesis is built on $\mu_t^{\text{aligned}}$, not on
$\mu_t^{\text{lab}}$. In the implementation, this is done by
rotating each of the ten realizations to a common ring axis
before pooling positions for the bootstrap density estimate; the
parametric Gaussian-band density used alongside the bootstrap is
likewise constructed in the aligned frame.

The bulk density of $M(t)$ is fixed by $\mu_t^{\text{aligned}}$
alone through the Mercer expansion of the $\arccos$ kernel
against $\mu_t^{\text{aligned}}$ (which reduces to the Legendre
expansion~\eqref{eq:legendre_expansion} for $\mu_0$ uniform on
$S^2$ at $t=0$, and to a Fourier expansion for $\mu_\infty$
concentrated on a great circle at NESS), with the higher-order
non-i.i.d.\ correlations of the FBP joint law contributing
only to subleading spectral statistics (level spacings,
two-point eigenvalue correlations, edge fluctuations). The
natural Bordenave-type reference for the bulk of $M(t)$ at
\emph{any} $t$ is therefore $E_t^{\text{B}}[\mu_t^{\text{aligned}}]$,
the ERM on $N$ i.i.d.\ samples from $\mu_t^{\text{aligned}}$,
even though no FBP snapshot is a sample from this ensemble at
any $t>0$.

\begin{figure}[!htbp]
\centering
\includegraphics[width=\textwidth]{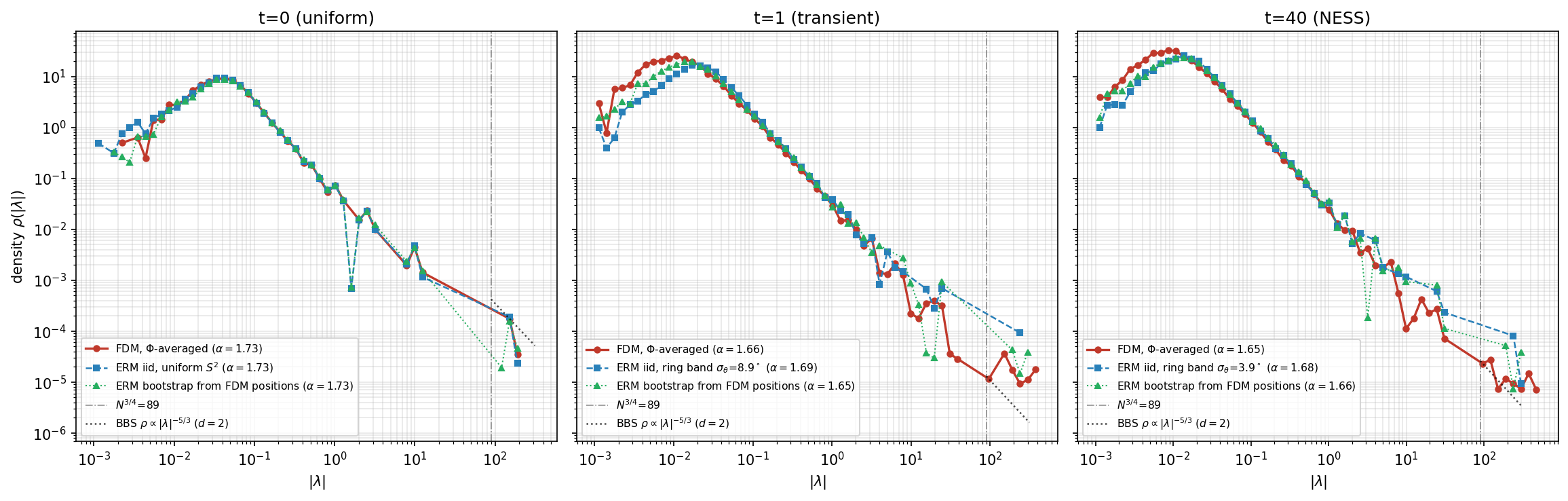}
\caption{Test of the ERM identification (FDM bulk equals
ERM bulk on i.i.d.\ samples from $\mu_t$) at $t=0$ (left),
$t=1$ (middle), and $t=40$ NESS (right).
\textbf{Red circles}: pooled bulk density of $M(t)$ across the
ten FBP disorder realizations (the disorder-averaged FDM
spectrum). \textbf{Blue squares}: pooled bulk density of
$M^{\text{ERM}}_{ij}=\arccos(\mathbf{x}_i\cdot\mathbf{x}_j)$ on
$N$ i.i.d.\ samples from a parametric Gaussian ring band of
empirical polar width $\sigma_\theta$ centered on the equator
($\sigma_\theta = 37^\circ$ at $t=0$, $8.9^\circ$ at $t=1$,
$3.9^\circ$ at NESS; at $t=0$ the ``ring band'' degenerates to
the uniform measure on $S^2$). \textbf{Green triangles}: same
but with a non-parametric bootstrap that resamples points from
the pooled FBP positions, aligned to a common ring
axis through the inertia tensor. The three densities collapse onto each other
at every $t$, with bulk power-law exponents agreeing within
$\sim 2\%$ (values shown in legend); the dash-dotted vertical
line marks the BBS delocalization threshold
$N^{3/4}\!\approx\!89$, and the dotted black line is the BBS
$d=2$ reference $\rho\propto|\lambda|^{-5/3}$. The collapse
confirms that the disorder-averaged FDM bulk is an ERM bulk on
the time-evolving FBP one-particle density $\mu_t$, despite the
joint distribution of $\mathbf{x}_i(t)$ being non-i.i.d.\ at any
fixed disorder $\Phi$ for $t > 0$.}
\label{fig:rmt_erm_iid}
\end{figure}

Figure~\ref{fig:rmt_erm_iid} verifies this prediction directly.
For each of $t = 0, 1, 40$ we pool the FDM eigenvalues across
the ten disorder realizations (the disorder-averaged spectrum)
and compare to two i.i.d.\ resamples on $\mu_t$: a
parametric Gaussian ring band of empirical polar width
$\sigma_\theta$ centered on the equator, and a non-parametric
bootstrap from the pooled FBP positions aligned to a
common ring axis (each resample uses 20 independent draws). The three bulk
densities collapse onto one another across the full range
$|\lambda|\in[10^{-3}, 10^2]$, including the small-$|\lambda|$
linear vanishing, the BBS delocalization threshold at $N^{3/4}$,
and the position of the lowest BBS quasi-multiplets at the right
edge of the bulk. Table~\ref{tab:rmt_erm_iid} reports the bulk
power-law exponent fit for the three sources at
each time, together with the empirical polar width
$\sigma_\theta$ used to parametrize the ring band. The three
exponents agree to within $\sim 2\%$ at every time, and the NESS
band width $\sigma_\theta\!\approx\!3.9^\circ$ is consistent
with the $\sim 4.8^\circ$ value reported by the F2
model~\cite{halperin2026fields}.

\begin{table}[!htbp]
\centering
\caption{Bulk power-law exponent $\alpha$ and auto-selected
lower bound $x_{\min}$ at three times, fitted on (a) the
disorder-averaged FDM spectrum, (b) the parametric i.i.d.\
resample on a Gaussian ring band of polar width $\sigma_\theta$,
and (c) the non-parametric bootstrap from the pooled FBP
particle positions. At
$t=0$ the ``ring band'' degenerates to the uniform measure on
$S^2$, and $\sigma_\theta$ is the $S^2$-uniform value
$\approx\!37^\circ$. Both the fitted exponents and the fit
windows agree to within $\sim 5\%$ across the three sources at
every time, with $x_{\min}$ shifting downward as the bulk
contracts during ring formation.}
\label{tab:rmt_erm_iid}
\begin{tabular}{lc|ccc|ccc}
\toprule
& & \multicolumn{3}{c|}{$\alpha$} & \multicolumn{3}{c}{$x_{\min}$} \\
$t$ & $\sigma_\theta$ (deg) & FDM & iid & boot & FDM & iid & boot \\
\midrule
$0$ (uniform)   & $37.2$ & $1.730$ & $1.730$ & $1.725$ & $0.044$ & $0.043$ & $0.047$ \\
$1$ (transient) & $8.9$  & $1.657$ & $1.690$ & $1.653$ & $0.028$ & $0.025$ & $0.026$ \\
$40$ (NESS)     & $3.9$  & $1.651$ & $1.679$ & $1.659$ & $0.016$ & $0.017$ & $0.019$ \\
\bottomrule
\end{tabular}
\end{table}

The contracted bulk of the FDM at NESS is therefore an ERM bulk
on the NESS marginal $\mu_\infty$, not a deviation from such an
ensemble; the dynamical content of the FDM at $t>0$ is carried
by the redistribution of $\mu_t$ on $S^2$, with the non-i.i.d.\
joint correlations integrated out by self-averaging at the
bulk-density level.

The bulk power-law exponents in Tab.~\ref{tab:rmt_erm_iid} are
fitted with the maximum-likelihood estimator of the
\texttt{powerlaw} Python package \cite{powerlaw}, which is more
reliable than linear regression on log-log-binned data; we use
the log-log plot of Fig.~\ref{fig:rmt_erm_iid} as a qualitative
shape check and Tab.~\ref{tab:rmt_erm_iid} for the quantitative
comparison.

\paragraph{The ERM spectrum as a null hypothesis for
empirical data.}
Beyond its role as a self-consistency check on the FDM, the
i.i.d.-resample construction of Fig.~\ref{fig:rmt_erm_iid} and
Tab.~\ref{tab:rmt_erm_iid} provides the natural \emph{null
hypothesis spectrum} for analyzing empirical distance,
similarity, or correlation matrices in any setting where the
underlying data are not the FBP system but, for example,
financial asset returns or graph adjacency data. Given an
empirical matrix $M^{\text{emp}}$ at time $t$, one fits a
one-particle density $\hat\mu_t$ (analytically, e.g.\ a
Gaussian band or a kernel-density estimate of the marginal of
the underlying generative variables, or non-parametrically by
bootstrap from the empirical positions where available),
generates many i.i.d.\ resamples on $\hat\mu_t$, and constructs
the resampled spectrum. Agreement of the empirical spectrum
with this null indicates that the data are consistent with
i.i.d.\ marginal sampling from $\hat\mu_t$, with no inter-sample
correlations beyond those induced by the kernel; deviations
between the empirical and null spectra (typically in level
statistics, edge fluctuations, or the multiplet structure of
the bottom eigenvalues) reveal genuine non-i.i.d.\ structural
content in the underlying generative process. In the FBP setting
the null is essentially exact at the bulk-density level, by
self-averaging, with the deviations confined to sub-leading
spectral statistics (Sec.~\ref{sec:universality}); for empirical
data the same construction is the natural baseline against
which collective and structural-change diagnostics should be
calibrated, and a regime change in the underlying system can
be detected as a time-resolved deviation of the empirical
spectrum from the null spectrum recomputed on $\hat\mu_t$.

\paragraph{Time continuity removes ensemble independence.}
%Snapshots at different times are not independent draws from
%anything. 
The trajectories $\mathbf{x}_i(t)$ are continuous on
$S^2$ and $M(t)$ is continuous in operator norm, with $M(t')$
the deterministic image of $M(t)$ under integrating the
underlying $N$-particle SDE for time $t' - t$. The eigenvalue
process $(\lambda_k(t))_{k=1}^{N}$ is therefore a continuous
interlacing family driven by a single coupled SDE, not a
sequence of independent samples. The natural mathematical object
is a matrix-valued stochastic process, of which a Bordenave
ensemble would be at best the static marginal at one instant.
Among standard dynamical matrix ensembles, Dyson Brownian motion
has independent increments and free Brownian motion has free
increments; the present process has neither, because $\mathbf{x}_i(t)$
and $\mathbf{x}_j(t)$ are coupled through $\Phi$.

After ring formation the points concentrate on an essentially
one-dimensional support, so the matrix becomes effectively a
circulant-like distance matrix on a discrete circle, whose
spectrum is dominated by the discrete Laplacian eigenvalues; the
bulk consequently contracts further as seen in
Fig.~\ref{fig:rmt_bulk} between $t = 0$ and $t = 40$.

Quantifying these dynamical deviations from the static BBS
template, and identifying the spectral signatures of ring
formation that are computable from $M(t)$ alone, is the focus
of the more detailed analysis below
(Secs.~\ref{sec:bbs_check}--\ref{sec:eigtraj}).

\subsection{Heavy-tailed spectrum: power-law fit of the bulk}
\label{sec:powerlaw}

The bulk-shape analysis above identified the FDM bulk at every
time with the ERM density on the FBP one-particle density
$\mu_t$, with the BBS power-law tail as the leading
continuous-staircase prediction. A natural follow-up is to fit
this tail directly. Power-law spectra arise in several classes
of strongly-correlated random matrices: heavy-tailed (L\'{e}vy)
ensembles \cite{cizeau1994, soshnikov2004}, sample covariance
matrices of correlated data \cite{bouchaudpotters2003,
bunbouchaudpotters2017}, scale-free graph adjacency matrices
\cite{goh2001, chunglu2003}, chiral random matrices in QCD
\cite{verbaarschot1994}, and Bordenave Euclidean random matrices
\cite{bordenave2008, bordenave2013}. The first three in that
list produce power-law tails through mechanisms (heavy-tailed
entries, sample-correlation structure, scale-free degrees) that
are absent from our setting; the Bordenave / BBS distance-matrix
class is the natural reference here, with the exponent
determined by the smoothness of the $\arccos$ kernel and by the
dimension of the manifold supporting $\mu_t$.

\paragraph{Analytical prediction.}
We label eigenvalues of $M(t)$ in descending order of absolute
value, $|\lambda_1|\ge|\lambda_2|\ge\ldots\ge|\lambda_N|$, and
denote by $K$ the rank in this sorted list. The Perron eigenvalue
$\lambda_1\approx N\pi/2$ sits at $K=1$; the heavy negative
outliers identified in Sec.~\ref{sec:eigtraj} occupy the next few
ranks (the $\ell=1,3,5,\ldots$ BBS multiplets verified in
Sec.~\ref{sec:bbs_check}); for $K$ inside the bulk the tail of
$|\lambda_K|$ is governed by the BBS
continuous-staircase prediction \cite{bogomolny2003} for the
delocalized regime on a $d$-dimensional sphere or hypercube,
\begin{equation}
\label{eq:rank_decay_pred}
\rho(\lambda) \;\sim\; |\lambda|^{-(2d+1)/(d+1)},
\qquad
\mathbf{N}(\lambda) \;\sim\; |\lambda|^{-d/(d+1)},
\qquad
|\lambda_K| \;\sim\; K^{-(d+1)/d},
\end{equation}
valid for $|\lambda|\gtrsim N^{(d+1)/(2d)}$. For $d=2$
(sphere) this gives the density exponent $\alpha = 5/3$ and the
rank exponent $\beta = (d+1)/d = 3/2$. For $d=1$ (effective ring
support after geometric collapse) the same formula gives
$\alpha = 3/2$ and $\beta = 2$, which is also what one obtains
directly from the Fourier expansion of the geodesic distance on
the circle: the triangular-wave kernel
$d(\theta) = \min(|\theta|, 2\pi - |\theta|)$ has Fourier
coefficients $\sim 1/k^2$ for the non-zero modes, so
$|\lambda_K|\sim K^{-2}$ ($\beta = 2$), and the corresponding
density of states is $\rho(\lambda) \sim |\lambda|^{-3/2}$
($\alpha = 3/2$). Both predictions apply only in the
delocalized regime; for $K$ deep inside the bulk, the
eigenvalues are localized and a different counting applies
(Sec.~\ref{sec:bbs_check}).

\paragraph{Empirical fit.}
We compute two complementary fits on the descending tail of the
magnitude distribution (Perron outlier $\lambda_1$ excluded). The
first is a least-squares rank-decay fit
$|\lambda_K|=A K^{-\beta}$ on the rank window $K\in[2,50]$, that
is, the 49 largest-by-$|\lambda|$ eigenvalues, which includes the
negative outliers of Sec.~\ref{sec:outliers} and the upper part
of the bulk. The second is a maximum-likelihood density fit
$\rho(\lambda)\sim|\lambda|^{-\alpha}$ for $|\lambda|>x_{\min}$
via the \texttt{powerlaw} Python package \cite{powerlaw}, which
estimates $x_{\min}$ and $\alpha$ jointly and isolates the upper
tail of $\rho(|\lambda|)$ above an automatic threshold. Both fits
are computed at every snapshot for every realization.

\begin{figure}[!htbp]
\centering
\includegraphics[width=\textwidth]{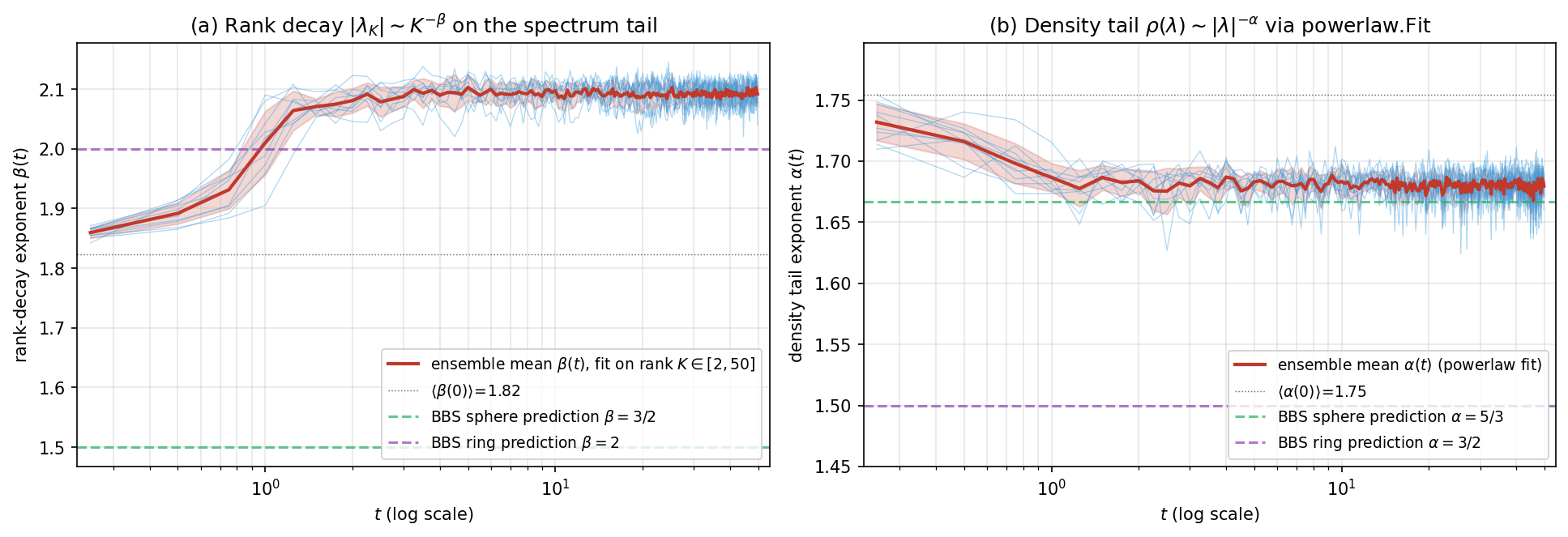}
\caption{Time-resolved power-law exponents, ten realizations and
ensemble mean.
(a) Rank-decay exponent $\beta(t)$ from least-squares fit
$|\lambda_K| \sim K^{-\beta}$ on $K \in [2, 50]$. The exponent
jumps cleanly from $\beta(0) \approx 1.82$ to a NESS plateau
$\beta(\text{NESS}) \approx 2.10$ within $t \in [0, 5]$. At
$t=0$ the fitted $\beta\approx 1.82$ is somewhat steeper than the
sphere prediction $\beta=3/2$ (green dashed) because the fit
window includes part of the localization regime
$K\gtrsim\sqrt{N}\approx 20$ where multiplet broadening
dominates the local slope; at NESS the fitted
$\beta\approx 2.10$ is consistent with the corrected ring
prediction $\beta=2$ (purple dashed) from the
$\sim 1/k^2$ Fourier coefficients of the triangular-wave
geodesic distance on $S^1$.
(b) Density tail exponent $\alpha(t)$ from
\texttt{powerlaw.Fit} on $|\lambda|$ above an automatic
$x_{\min}$. The exponent drops from $\alpha(0) \approx 1.75$ to
a NESS plateau $\alpha(\text{NESS}) \approx 1.68$ on the same
timescale; both values lie within $\sim 5\%$ of the sphere
prediction $\alpha=5/3\approx 1.67$ (green dashed) and
significantly above the ring prediction $\alpha=3/2$ (purple
dashed). The finite-$N$ scan of \S\ref{sec:bbs_check}\,(vi)
shows that the NESS exponent moves \emph{upward} toward $5/3$
as $N$ grows, not toward $3/2$, so the small drop between $t=0$
and NESS is a finite-$N$ feature of the bulk fit rather than a
genuine shift toward the $d=1$ ring exponent.}
\label{fig:rmt_powerlaw}
\end{figure}

\begin{figure}[!htbp]
\centering
\includegraphics[width=\textwidth]{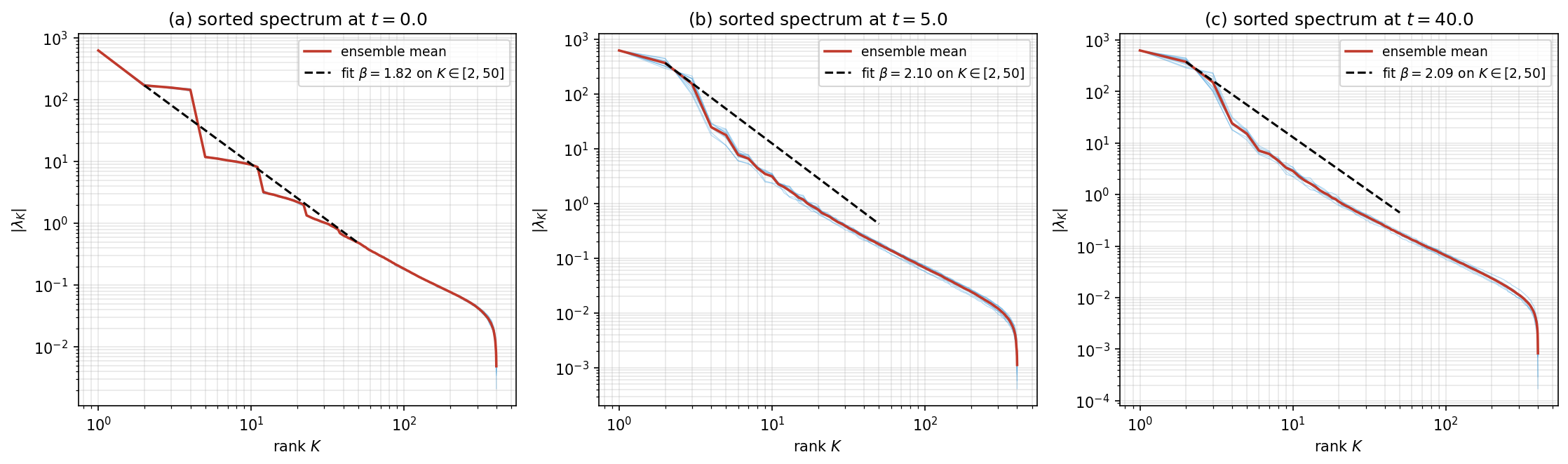}
\caption{Sorted log-log spectra at $t = 0, 5, 40$ for the ten
realizations (blue), with the ensemble mean (red) and the
fitted power-law line (dashed black) on the rank window
$K \in [2, 50]$. Block plateaus from the small-$\ell$ Legendre
contributions are visible at $t = 0$; in the NESS regime the
spectrum is more uniformly power-law.}
\label{fig:rmt_powerlaw_examples}
\end{figure}

Figure~\ref{fig:rmt_powerlaw} shows the time-resolved exponents,
and Figure~\ref{fig:rmt_powerlaw_examples} shows representative
log-log spectra at the three characteristic times.

\paragraph{Discussion of empirical values vs predictions.}
Both fits are robust across realizations (panel-wise standard
deviation $\le 0.02$) and both transition sharply during ring
formation. They use different windows and probe different
spectral regimes, so they are not corroborating measurements of
the same exponent and the relation $\beta = 1/(\alpha-1)$ that
would link them under a single asymptotic tail is not expected
to hold for our fits.\footnote{For a single power-law tail
$\rho(\lambda)\sim|\lambda|^{-\alpha}$ over $|\lambda|\in[x_{\min},
\infty)$, the corresponding ranked tail satisfies
$|\lambda_K|\sim K^{-1/(\alpha-1)}$, so $\beta=1/(\alpha-1)$.
Our reported pairs at $t=0$ ($\alpha\!\approx\!1.73$,
$\beta\!\approx\!1.82$) and NESS
($\alpha\!\approx\!1.65$, $\beta\!\approx\!2.10$) imply
$\beta=1.37, 1.54$ respectively from the relation, neither
matching the rank fit. The discrepancy is consistent with the
two fits probing different regimes: the rank-decay window
$K\in[2,50]$ includes the bottom multiplet plateaus and
extends past the delocalized cutoff
$K\!\ll\!\sqrt{N}\!\approx\!20$, while the MLE density fit
selects an automatic $x_{\min}$ that excludes those
contributions and isolates the bulk above the threshold.} On
the density side, the MLE fit recovers $\alpha\approx 1.73$ at
$t=0$ within $\sim 5\%$ of the sphere prediction $5/3$, and
$\alpha\approx 1.65$ at NESS, both at $N=400$. The naive
reading of the small NESS shift as a partial transition toward
the $d=1$ ring prediction $3/2$ is excluded by the finite-$N$
scan of \S\ref{sec:bbs_check}\,(vi): pushing $N$ from 100 to
800, the NESS exponent moves \emph{upward} from
$\alpha\approx 1.57$ to $\alpha\approx 1.66$, i.e.\ toward the
$d=2$ sphere value, not the $d=1$ ring value. The bulk density
exponent is thus controlled by the dimension of the embedding
$S^2$ and is robust against the dynamical ring formation, which
acts on the few low-lying multiplets rather than on the bulk
tail. On the rank side, the fitted $\beta\approx 1.82$ at $t=0$
is somewhat steeper than the sphere prediction $\beta=3/2$
because the fit window includes part of the localization
regime where multiplet broadening dominates the local slope; at
NESS the fitted $\beta\approx 2.10$ is close to the ring
prediction $\beta=2$ because the rank-decay window
$K\in[2,50]$ is dominated by the bottom Fourier-pair multiplets
of the emergent $d=1$ support, which the bulk MLE fit excludes
by its automatic $x_{\min}$ cutoff.

The two fits together support the time-resolved sphere-to-ring
transition of the underlying FBP one-particle density $\mu_t$
predicted by the ERM identification of
\S\ref{subsec:erm_null}, but each fit probes only one
spectral regime, and the precise asymptotic interpretation
should be cross-checked against the windowless
ranked-spectrum-envelope test of
\S\ref{subsec:ranked_envelope} below, which avoids the
window-choice ambiguity altogether.

\subsection{Ranked-spectrum comparison to the ERM null
hypothesis envelope}
\label{subsec:ranked_envelope}

The fixed-window rank fit $|\lambda_K|\sim K^{-\beta}$ on
$K\in[2,50]$ above is sensitive to the choice of window and
mixes the delocalized BBS staircase with localization-regime
multiplet broadening. A more direct, window-free comparison uses
the ERM($\mu_t$) i.i.d.-resample machinery of
Sec.~\ref{subsec:erm_null} to generate the expected ranked
spectrum $\langle |\lambda_K|\rangle_{\text{ERM}(\mu_t)}$ with
its $\pm 2\sigma$ envelope at every rank $K$, and tests the FDM
ranked spectrum against that envelope.

\begin{figure}[!htbp]
\centering
\includegraphics[width=\textwidth]{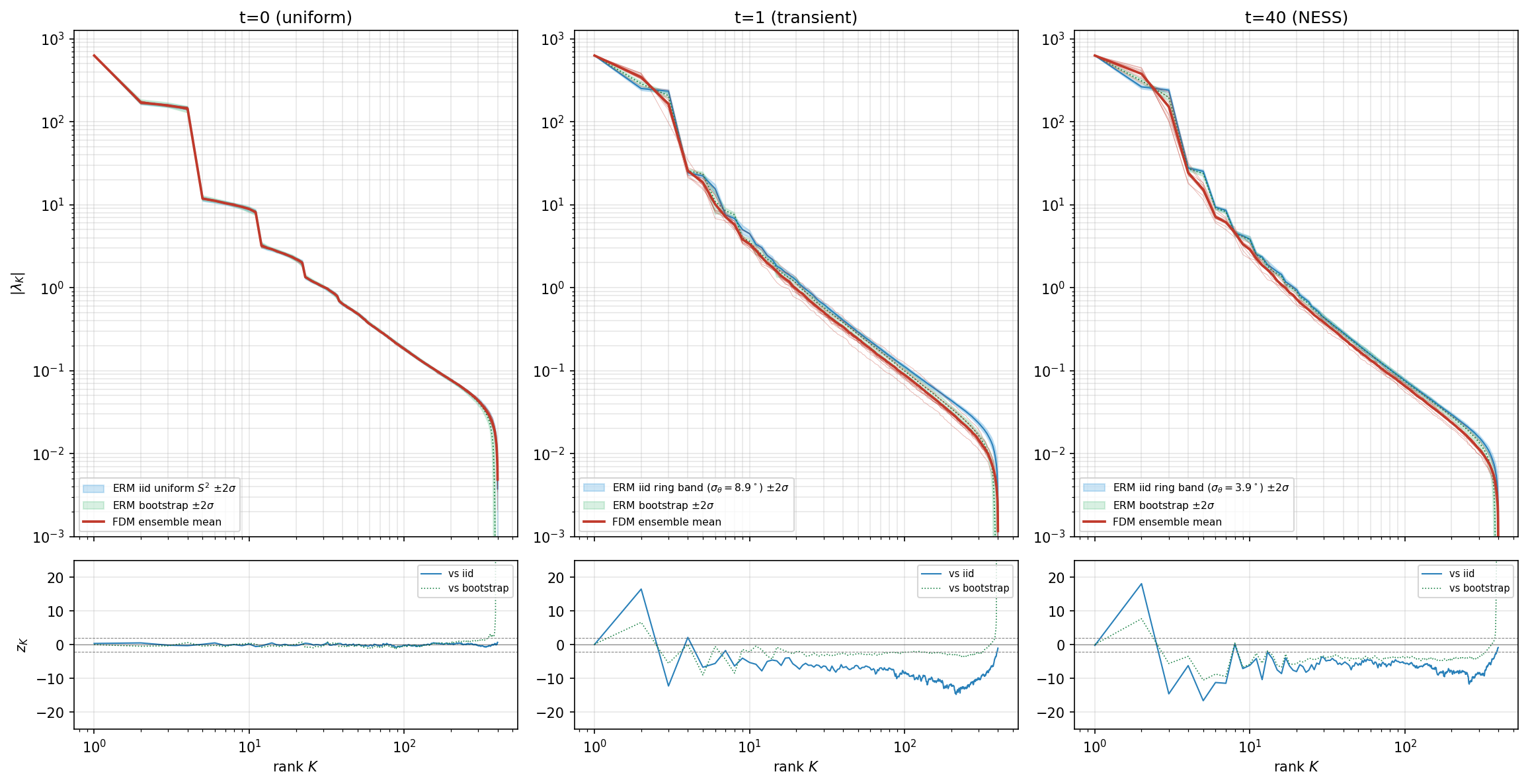}
\caption{Ranked-spectrum comparison to the ERM($\mu_t$) null
hypothesis envelope. \textbf{Top row}: $|\lambda_K|$ vs $K$ on log-log
axes, with $K=1$ the largest-magnitude eigenvalue (Perron) and
$K=N$ the smallest. \textbf{Red}: per-realization FDM spectra
(thin) and ensemble mean (bold). \textbf{Blue band}:
$\langle|\lambda_K|\rangle_{\text{ERM iid}}\pm 2\sigma$ from 40
parametric resamples on $\mu_t$ (uniform on $S^2$ at $t=0$,
Gaussian ring band of empirical width $\sigma_\theta$ for
$t>0$). \textbf{Green dotted band}: ERM bootstrap envelope from
40 non-parametric resamples of the aligned FBP positions.
\textbf{Bottom row}: ranked $z$-score residual $z_K$ of the FDM
ensemble mean against the iid (blue) and bootstrap (green
dotted) nulls. At $t=0$ the FDM is fully within the iid envelope
($z_K\approx 0$ at every rank) since the points are i.i.d.\
uniform by construction. At $t=1,40$ the FDM ensemble mean
tracks the bootstrap envelope tightly across the bulk and lies
slightly below the parametric iid envelope by $z\sim -5$ to
$-10$ in the bulk middle, with the largest deviations
concentrated at the deepest negative ranks ($K\!\sim\!N$,
right edge) where the dynamical $\ell=1$ ring-formation signal
sits. The BBS multiplet plateau structure (visible as steps in
the curves at $t=0$) is preserved at every time.}
\label{fig:rmt_ranked_envelope}
\end{figure}

Figure~\ref{fig:rmt_ranked_envelope} reports the comparison at
$t=0,1,40$. Top row: the per-realization FDM ranked spectrum
(red) overlaid on the ERM iid envelope (blue band, $\pm 2\sigma$
from 40 parametric resamples on $\mu_t$) and the ERM bootstrap
envelope (green dotted band, 40 non-parametric bootstrap
resamples on the empirical $\mu_t$). Bottom row: ranked
$z$-score residual
$z_K = (\langle|\lambda_K|\rangle_{\text{FDM}} -
\langle|\lambda_K|\rangle_{\text{ERM}}) /
\sigma_{\text{ERM}}(K)$ against both nulls.

\paragraph{Reading of the comparison.}
The pattern in Fig.~\ref{fig:rmt_ranked_envelope} sharpens the
ERM identification of \S\ref{subsec:erm_null}
in three ways. First, at $t=0$ the FDM ranked spectrum is
indistinguishable from the iid ERM envelope at \emph{every}
rank, in line with the fact that $M(0)$ is exactly an
i.i.d.\ Bordenave matrix on $\mu_0=$ uniform on $S^2$. Second,
at $t=1$ and $t=40$ the FDM ensemble mean tracks the bootstrap
envelope (which uses the actual empirical $\mu_t$) very tightly
but sits systematically slightly below the parametric Gaussian
ring-band envelope across the bulk, which simply reflects the
fact that the analytical Gaussian band is an approximation to
the true NESS measure $\mu_\infty$, not that the FDM departs
from the ERM class. Third, the largest residual
$z$-scores concentrate at the deepest negative ranks, exactly
the BBS $\ell=1$ multiplet at the bottom of the spectrum that
carries the dynamical ring-formation signal of
Secs.~\ref{sec:eigtraj}-\ref{sec:bbs_check}. The
ranked-spectrum envelope test is therefore the natural unified
diagnostic combining the bulk-density agreement of
\S\ref{subsec:erm_null} with the bottom-multiplet
rearrangement of Sec.~\ref{sec:eigtraj}, in a single
geometry-aware null comparison that is independent of any
choice of fit window.

%==============================================================================
\section{Comparison to the BBS static theory}
\label{sec:bbs_check}
%==============================================================================

The static distance matrix on $S^2$ for $N$ uniform points was
analyzed by BBS \cite{bogomolny2003}.
Three structural predictions specialize to our kernel
$\arccos(\mathbf{x}\cdot\mathbf{y})$:

\paragraph{(i) Perron pattern.}
The matrix has exactly one positive eigenvalue (the Perron one)
and $N-1$ non-positive eigenvalues. The mean integrated density
satisfies $\mathbf{N}(0^+) = (N-1)/N$. We verified this holds at
every snapshot in every realization: at $t=0,\,1,\,5,\,40$ the
descending sort gives one positive eigenvalue
$\lambda_1\approx 627$ to $628$ and $399$ negative eigenvalues
with no exceptions across the ten disorder realizations. The
property persists at every measurement time, including during
the ring-formation transient and into NESS.

\paragraph{(ii) Quasi-multiplets from $\mathrm{SO}(3)$ symmetry.}
BBS show that on a manifold invariant
under a group $G$, the eigenvalues of the distance matrix
organize into quasi-multiplets whose dimensions equal those of
the irreducible representations of $G$. On $S^2$ with the
$\mathrm{SO}(3)$ rotation group the irreps are labelled by an
integer $\ell\ge 0$ and have dimension $2\ell+1$, so generic
distance-matrix kernels would in principle support multiplets
of dimension $1, 3, 5, 7, 9, 11, 13, 15, \ldots$ Only
\emph{odd}-$\ell$ multiplets contribute for the geodesic-distance
kernel, however, by a parity argument: $\arccos(t)-\pi/2$ is an
odd function of $t = \mathbf{x}\cdot\mathbf{y}$ on $[-1, 1]$
(since $\arccos(-t) = \pi - \arccos(t)$), and the Legendre
polynomial $P_\ell$ has parity $(-1)^\ell$, so the projection
$\int_{-1}^1 (\arccos(t)-\pi/2) P_\ell(t)\,dt$ vanishes for all
even $\ell\ge 2$. The multiplets with dimension
$5\,(\ell{=}2)$, $9\,(\ell{=}4)$, $13\,(\ell{=}6)$, etc.\ are
therefore absent from the spectrum, and the surviving
quasi-multiplets sit at predicted positions
\begin{equation}
\label{eq:bbs_multiplets}
\Lambda_\ell \;=\; N\,\mu_\ell \;=\; \frac{N\,a_\ell}{2\ell+1},
\qquad
a_\ell = \frac{2\ell+1}{2}\int_{-1}^{1}\arccos(t)\,P_\ell(t)\,dt,
\qquad \ell \in \{0,1,3,5,7,\ldots\}.
\end{equation}
Equation~\eqref{eq:bbs_multiplets} predicts
$\Lambda_0 = +628.3$ (Perron, dimension 1),
$\Lambda_1 = -157.1$ (dimension 3),
$\Lambda_3 = -9.82$ (dimension 7),
$\Lambda_5 = -2.45$ (dimension 11),
$\Lambda_7 = -0.96$ (dimension 15),
with $\Lambda_2 = \Lambda_4 = \cdots = 0$. Empirically at $t=0$
the pooled spectrum across the ten realizations shows precisely
this structure: the Perron eigenvalue at $+627.5$, a 3-cluster
at the bottom around $-157$ (rank-1 block at $\approx-170$,
$-157$, $-144$), then a 7-cluster around $-10$ (ranks $N{-}3$
through $N{-}9$ at $-8.1, -8.9, -9.4, -9.9, -10.4, -11.2,
-11.8$), then an 11-cluster around $-2.6$ (ranks $N{-}10$
through $N{-}20$ at $-2.0, -2.1, \ldots, -3.2$), then the start
of a 15-cluster near $-1.2$. The cluster centers agree
with~\eqref{eq:bbs_multiplets} to $\sim 10\%$, and the splitting
within each cluster is the finite-$N$ broadening predicted by
\cite{bogomolny2003}.

\emph{Why are the multiplets resolved only at large $|\Lambda|$?}
Two competing effects determine where in the spectrum the
discrete multiplet structure remains visible. On the one hand,
the predicted multiplet positions $\Lambda_\ell \propto
1/(2\ell+1)$ \emph{decrease in magnitude} as $\ell$ grows: the
$\ell=1$ multiplet sits at $|\Lambda|\!\approx\!157$, the
$\ell=3$ multiplet at $\!\approx\!10$, $\ell=5$ at
$\!\approx\!2.5$, $\ell=7$ at $\!\approx\!1$, and the spacing
between successive multiplets shrinks rapidly toward $0$. On
the other hand, the multiplet \emph{degeneracy} $2\ell+1$
\emph{grows} with $\ell$, and each multiplet acquires a
finite-$N$ broadening of width set by sub-leading corrections
to the Mercer expansion against the empirical (rather than
ideal-uniform) measure. For low $\ell$ (large $|\Lambda|$) the
gap between successive multiplets greatly exceeds the
finite-$N$ broadening, and the discrete multiplet structure is
resolved. For high $\ell$ (small $|\Lambda|$) the multiplet
positions crowd toward zero faster than the broadening shrinks,
so consecutive multiplets overlap and merge into a continuous
bulk with the BBS power-law density of paragraph (iii) below.
The crossover between the two regimes happens around
$\ell \sim \sqrt{N}$, that is, $|\Lambda|\sim N^{(d+1)/(2d)} =
N^{3/4}\!\approx\!89$ for $d=2$. Consequently the
clean discrete-multiplet structure is visible only for the
first $\sim\sqrt{N}\approx 20$ most-negative eigenvalues, and
the rest of the negative spectrum is governed by the BBS
continuous-staircase prediction.

\paragraph{(iii) Bulk power-law density on $S^2$.}
For delocalized states with $|\Lambda| \gtrsim N^{(d+1)/(2d)}$,
\cite{bogomolny2003} predicts an average density
$\rho(\Lambda) \sim |\Lambda|^{-(2d+1)/(d+1)}$, equal to
$|\Lambda|^{-5/3}$ for $d=2$, equivalent to a counting function
$\mathbf{N}(\Lambda) \sim |\Lambda|^{-d/(d+1)} = |\Lambda|^{-2/3}$.
The delocalized regime starts at
$|\Lambda| \gtrsim N^{3/4} \approx 89$, which puts the bottom-3
$\ell=1$ multiplet ($|\Lambda|\approx 157$) clearly inside the
delocalized regime and the $\ell=3$ multiplet
($|\Lambda|\approx 10$) just below the boundary. The empirical
density-tail exponent measured in Sec.~\ref{sec:powerlaw} is
$\alpha\approx 1.7$, in agreement with the
BBS prediction $\alpha = 5/3 \approx 1.67$
to within $2\%$.

\begin{figure}[!htbp]
\centering
\includegraphics[width=\textwidth]{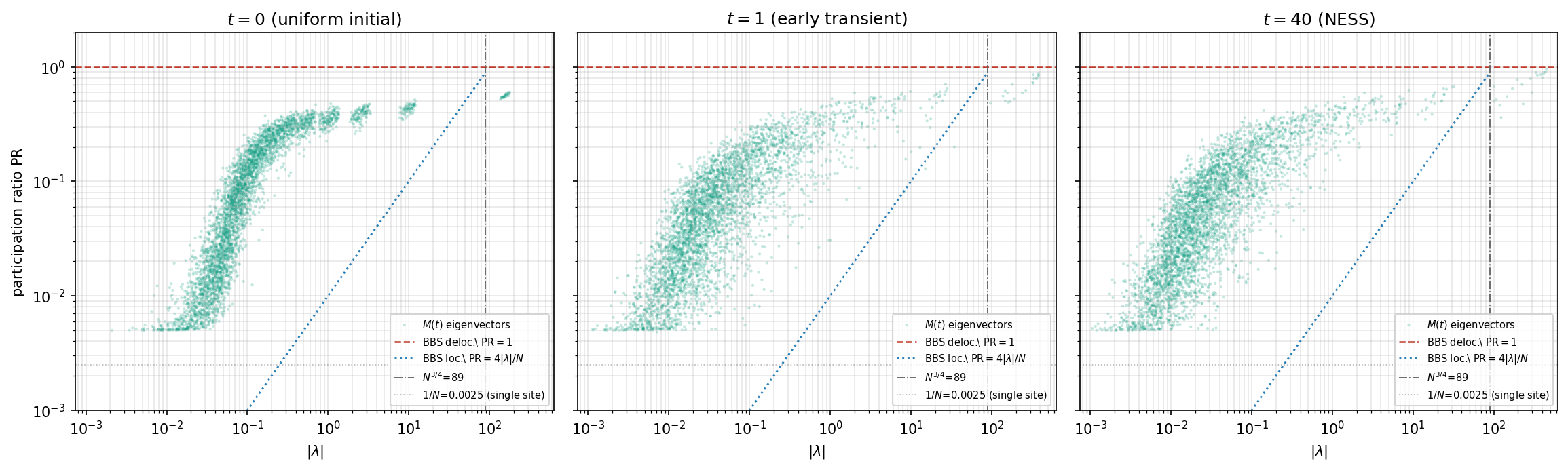}
\caption{Participation ratio
$\mathrm{PR} = 1/(N\sum_i u_i^4)$~\eqref{eq:pr_def} of every
eigenvector of $M(t)$ vs $|\lambda|$, pooled across the ten
realizations at $t=0,1,40$. PR$=1$ is fully delocalized
($\sim N$ sites participate); PR$=1/N=0.0025$ is fully
localized on a single site (gray dotted). Red dashed: BBS
delocalized prediction PR$=1$; blue dotted: BBS one-dimensional
Anderson scaling PR$=4|\lambda|/N$ as a reference; gray
dot-dashed: the BBS delocalization threshold
$|\lambda|=N^{3/4}\approx 89$. The empirical PR rises smoothly
from $\sim 0.005$ at the smallest $|\lambda|$ to $\sim 0.5{-}0.7$
at $|\lambda|\sim 100$, confirming the BBS localization
crossover; the saturation below $1$ at large $|\lambda|$
reflects the spherical-harmonic block structure (the deepest
eigenvectors are delocalized within a fixed $\ell=1,3,\ldots$
block but not across all $N$ sites). The 1D Anderson reference
sits well below the data on $S^2$, confirming that the
sphere-localized eigenfunctions are not described by the strict
1D Anderson scaling.}
\label{fig:rmt_participation}
\end{figure}

\paragraph{(iv) Localization at small $|\Lambda|$ via
participation ratios.}
BBS make a prediction analogous to the classical \emph{Anderson
localization} of disordered single-particle quantum systems
\cite{bogomolny2003}, in which eigenstates of a Hamiltonian
with random on-site potential decay exponentially in space with
some localization length $\xi$ rather than spread out over the
entire system. Translated to the distance matrix
$M_{ij}=\arccos(\mathbf{x}_i\cdot\mathbf{x}_j)$ acting on the
$N$ ``site'' indices $i = 1,\ldots,N$, BBS predict that for
$|\Lambda|$ smaller than the delocalization threshold
$N^{(d+1)/(2d)}$ the eigenvectors of $M$ are concentrated
(exponentially in 1D, with multipole-moment confinement in
higher $d$) on a small subset of particle indices rather than
delocalized across all $N$ of them, with localization length
$\xi \propto |\Lambda|$ in $d=1$ generalizing to a
multipole sum-rule condition on the localization centroid in
higher $d$. For sphere-like manifolds the localized
eigenfunctions live in two diametrically opposite regions, the
localization echo inherited from the existence of two geodesics
joining any pair of generic points on $S^2$. We test this prediction directly on the FDM by measuring,
for every eigenvector $u^{(n)}$ of $M(t)$ at $t=0,1,40$ across
the ten realizations, the inverse participation ratio
$\mathrm{IPR}_n = \sum_i (u^{(n)}_i)^4$ and the participation
ratio
\begin{equation}
\label{eq:pr_def}
\mathrm{PR}_n \;=\; \frac{1}{N\,\sum_{i=1}^{N} (u^{(n)}_i)^4}
\;=\; \frac{1}{N\,\mathrm{IPR}_n},
\qquad
\mathrm{PR}\!\to\!1\ \text{(delocalized)},\quad
\mathrm{PR}\!\to\!1/N\ \text{(localized on one site)}.
\end{equation}
The participation ratio is the standard diagnostic for
distinguishing localized from delocalized states in disordered
random-matrix and Anderson-localization problems: for a
unit-norm eigenvector $u^{(n)}$, the inverse participation
ratio $\mathrm{IPR}_n$ measures the concentration of $u^{(n)}$
on individual ``sites'' (here, the $N$ particle indices), and
the normalized $\mathrm{PR}_n$ is the fraction of sites that
contribute substantially to $u^{(n)}$. A uniformly spread
eigenvector ($u_i\!\approx\!1/\sqrt N$) has
$\mathrm{IPR}\!=\!1/N$ and $\mathrm{PR}\!=\!1$, a single-site
eigenvector ($u_k=1$, all other components zero) has
$\mathrm{IPR}\!=\!1$ and $\mathrm{PR}\!=\!1/N$, and the BBS
prediction in the localized regime,
$\mathrm{PR}\approx\xi/N$ with localization length
$\xi \propto |\lambda|$ in $d=1$, places the empirical PR
intermediate between these extremes.\footnote{We use the
$N$-normalized participation ratio
$\mathrm{PR}_n\in[1/N, 1]$ throughout, which reads as the
fraction of sites participating in $u^{(n)}$. BBS instead use
the unnormalized version $P_n = 1/\mathrm{IPR}_n =
N\,\mathrm{PR}_n \in [1, N]$ (their Eq.~(41)), which reads as
the actual number of sites participating; the BBS predictions
in their convention ($P\!\approx\!4|\lambda|$ for the localized
1D Anderson regime, $P\!\approx\!N$ for fully delocalized) map
to our convention by dividing by $N$
($\mathrm{PR}\!\approx\!4|\lambda|/N$,
$\mathrm{PR}\!\approx\!1$). The two are mathematically
equivalent diagnostics; the choice is between numerical
readability as a fraction ($[0,1]$) and as an absolute count
($[1, N]$).}
Figure~\ref{fig:rmt_participation} shows
$\mathrm{PR}$ vs $|\lambda|$ on log-log axes for the three times.
The empirical scatter shows a smooth localization crossover: at
the smallest $|\lambda|\sim 10^{-3}$ the median PR is
$\approx 0.005{-}0.009$ ($\sim 2{-}4$ sites participate in each
eigenvector); at intermediate $|\lambda|\sim 1$ the PR has risen
to $\approx 0.2{-}0.3$ ($\sim 100$ sites); for the deepest
negative outliers at $|\lambda|\sim 100$ the PR plateaus at
$\approx 0.5{-}0.7$ ($\sim 200{-}280$ sites). The transition is
gradual rather than abrupt and runs over the full five-decade
range of $|\lambda|$.

\begin{figure}[!htbp]
\centering
\includegraphics[width=\textwidth]{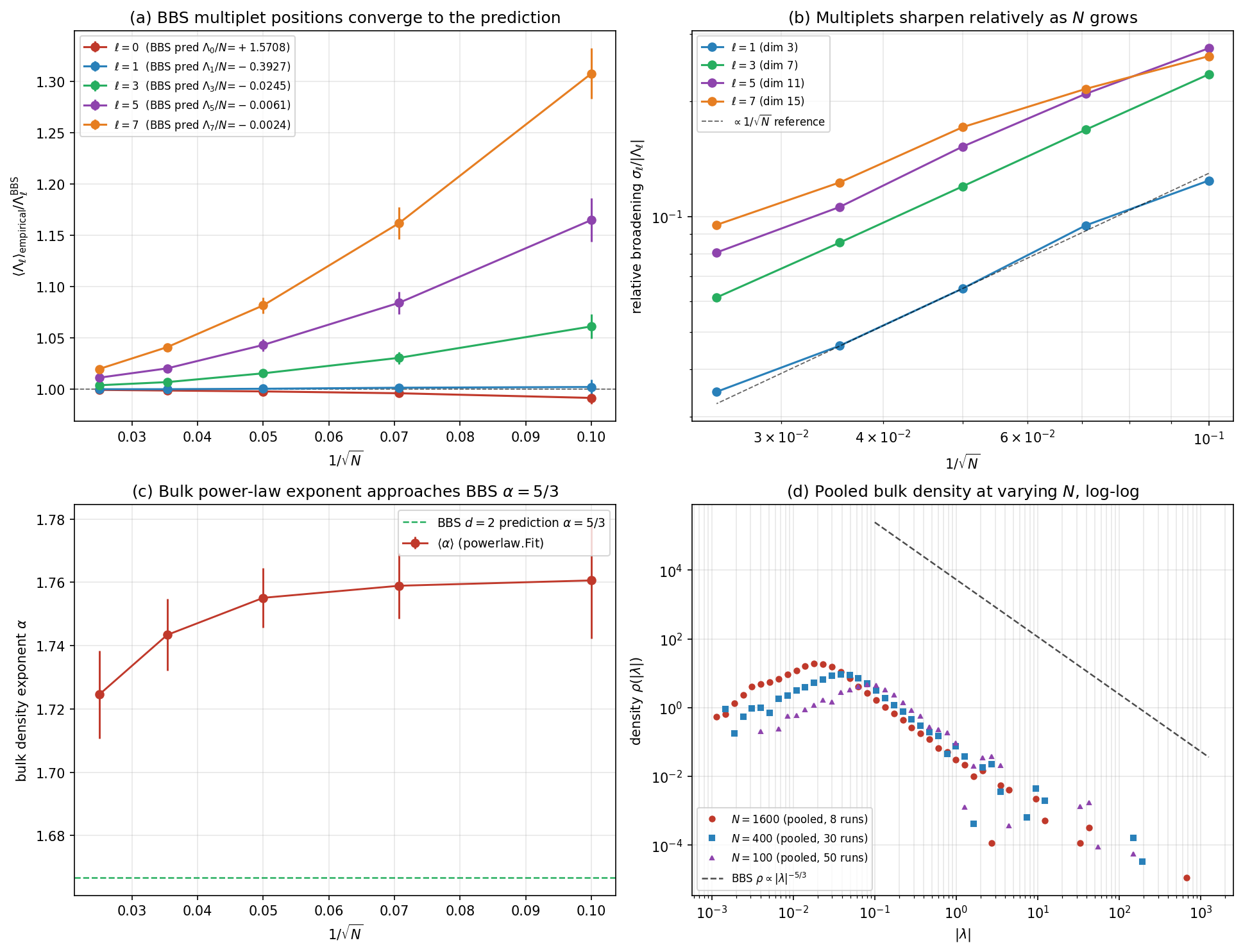}
\caption{Finite-$N$ scan at $t=0$ on i.i.d.\ uniform points on
$S^2$ ($t=0$ uses no FBP simulation; the FDM at $t=0$ is by
construction the BBS distance-matrix ensemble). \textbf{(a)}
Empirical multiplet positions
$\langle\Lambda_\ell\rangle/\Lambda_\ell^{\text{BBS}}$ vs
$1/\sqrt{N}$ for $\ell\in\{0,1,3,5,7\}$. The Perron and $\ell=1$
positions track the BBS prediction to better than $0.4\%$
already at $N\!\approx\!400$; higher-$\ell$ multiplets exhibit
finite-$N$ bias that shrinks monotonically with $N$.
\textbf{(b)} Relative within-multiplet broadening
$\sigma_\ell/|\Lambda_\ell|$ vs $1/\sqrt{N}$ on log-log axes,
for $\ell\in\{1,3,5,7\}$. All four multiplets show the
$1/\sqrt{N}$ scaling predicted by BBS for finite-$N$ broadening
(parallel to the dashed reference line). \textbf{(c)} Bulk
density exponent $\alpha$ vs $1/\sqrt{N}$: monotonic descent
from $\approx 1.76$ at $N=100$ toward the BBS $d=2$ prediction
$\alpha = 5/3$ (dashed green); the remaining gap at $N=1600$
is $\sim 3.5\%$. \textbf{(d)} Pooled bulk density
$\rho(|\lambda|)$ at $N\in\{100, 400, 1600\}$ on log-log axes,
with the BBS reference line $\rho\propto|\lambda|^{-5/3}$. The
three densities collapse onto each other across the bulk and
follow the BBS slope.}
\label{fig:finite_N_scan_t0}
\end{figure}

\paragraph{(v) Finite-$N$ scaling of the BBS predictions.}
The empirical agreement between the FDM at $t=0$ and the BBS
predictions reported above is at the single system size $N=400$
used in the main analysis. To verify that this agreement is part
of a genuine asymptotic approach to the BBS predictions rather
than a coincidence at one $N$, we ran a separate finite-$N$ scan
that uses no FBP simulation: at each
$N\in\{100,200,400,800,1600\}$
we sampled $N$ i.i.d.\ uniform points on $S^2$ across many
realizations (50 at $N=100, 200$, 30 at $N=400$, 15 at $N=800$,
8 at $N=1600$), built the arccos distance matrix, diagonalized,
and extracted three diagnostics: the empirical multiplet
positions $\langle\Lambda_\ell\rangle_{\text{empirical}}$
relative to the BBS prediction
$\Lambda_\ell^{\text{BBS}} = N a_\ell/(2\ell+1)$; the relative
within-multiplet broadening
$\sigma_\ell/|\Lambda_\ell|$; and the bulk power-law exponent
$\alpha$ via \texttt{powerlaw.Fit}. The results are shown in
Tab.~\ref{tab:finite_N_scan_t0} and
Fig.~\ref{fig:finite_N_scan_t0}.

\begin{table}[!htbp]
\centering
\caption{Finite-$N$ scan at $t=0$ on i.i.d.\ uniform points on
$S^2$. Empirical multiplet positions
$\langle\Lambda_\ell\rangle$, the relative within-multiplet
broadening $\sigma_\ell/|\Lambda_\ell|$ for the $\ell=1$ and
$\ell=3$ multiplets, and the bulk density exponent $\alpha$
fitted with \texttt{powerlaw.Fit}, all averaged across
realizations at each $N$. The BBS predictions for the multiplet
positions are $\Lambda_0 = (\pi/2)\,N \approx 1.5708\,N$,
$\Lambda_1 = -(\pi/8)\,N \approx -0.3927\,N$, and
$\Lambda_3 \approx -0.0245\,N$.}
\label{tab:finite_N_scan_t0}
\begin{tabular}{ccrrrrrr}
\toprule
$N$ & runs & $\langle\Lambda_0\rangle$ &
$\langle\Lambda_1\rangle$ & $\sigma_1/|\Lambda_1|$ &
$\langle\Lambda_3\rangle$ & $\sigma_3/|\Lambda_3|$ &
$\alpha$ \\
\midrule
$100$  & 50 & $155.75$  & $-39.36$  & $0.124$ & $-2.60$  & $0.222$ & $1.761$ \\
$200$  & 50 & $312.92$  & $-78.65$  & $0.095$ & $-5.06$  & $0.164$ & $1.759$ \\
$400$  & 30 & $626.97$  & $-157.15$ & $0.065$ & $-9.97$  & $0.118$ & $1.755$ \\
$800$  & 15 & $1255.04$ & $-314.17$ & $0.046$ & $-19.77$ & $0.085$ & $1.743$ \\
$1600$ & 8  & $2511.62$ & $-628.31$ & $0.035$ & $-39.42$ & $0.061$ & $1.725$ \\
\midrule
BBS    & --- & $1.5708\,N$ & $-0.3927\,N$ & --- & $-0.02454\,N$ & --- & $5/3 \approx 1.667$ \\
\bottomrule
\end{tabular}
\end{table}

The three trends in Fig.~\ref{fig:finite_N_scan_t0} match the
BBS predictions for finite-$N$ behaviour: multiplet positions
converge, multiplet broadening scales as $1/\sqrt{N}$, and the
bulk power-law exponent approaches the asymptotic $\alpha=5/3$
on a clear monotonic trajectory. The $N=400$ value used
throughout the paper sits within $\sim 5\%$ of the asymptotic
$\alpha$ on this trajectory, and the agreement between the FDM
spectrum and the BBS predictions at fixed $N=400$ is consistent
with the finite-$N$ remainder of the BBS asymptotic regime.

\begin{figure}[!htbp]
\centering
\includegraphics[width=\textwidth]{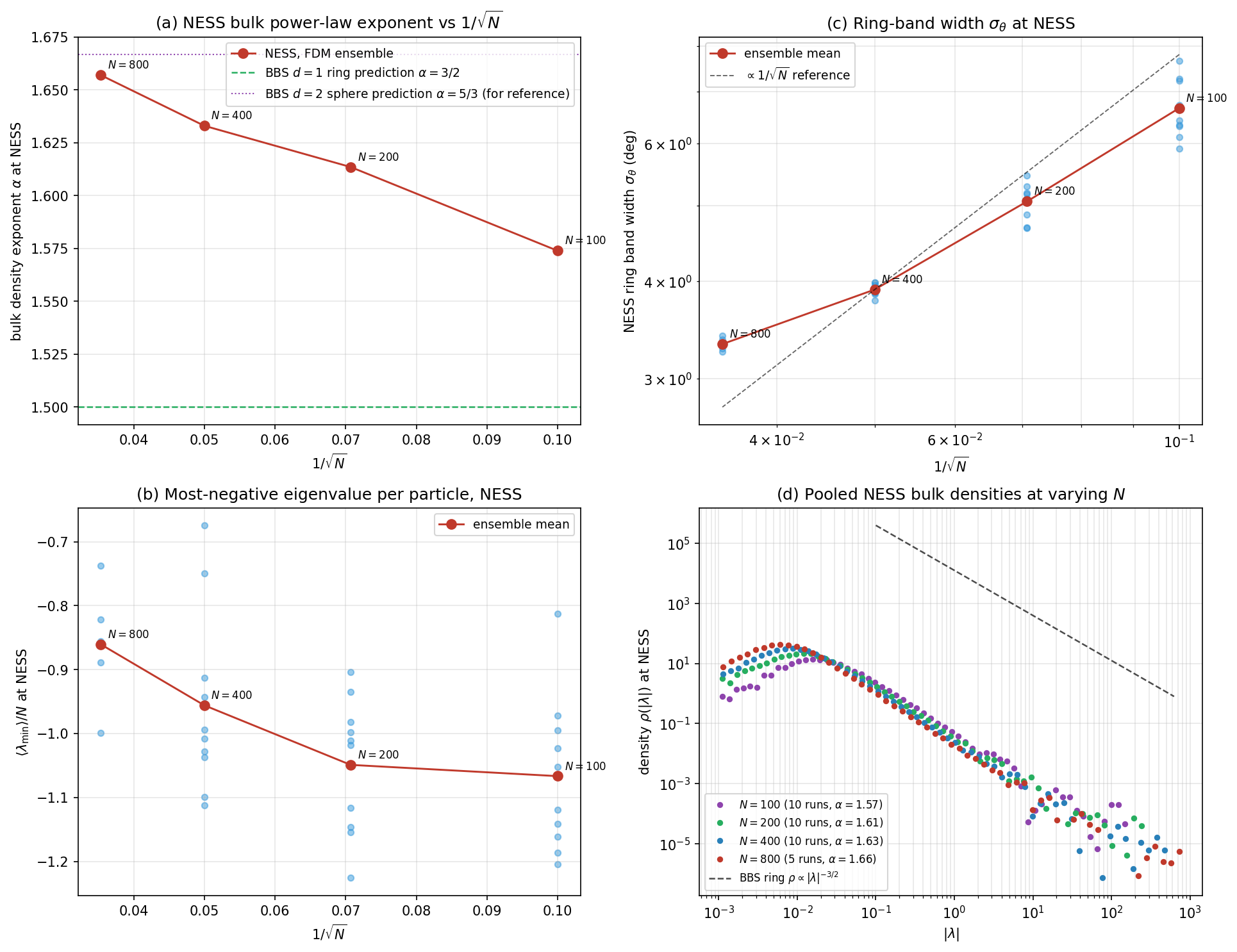}
\caption{Finite-$N$ scan at the FBP NESS, $T=0.4$, $\sigma=1$.
\textbf{(a)} Bulk density exponent $\alpha$ at NESS vs
$1/\sqrt{N}$. The exponent rises monotonically from
$\alpha\approx 1.57$ at $N=100$ to $\alpha\approx 1.66$ at
$N=800$, approaching the BBS $d=2$ prediction $\alpha=5/3$
(red dashed) rather than the strict-1D ring prediction
$\alpha=3/2$ (green dashed): the bulk eigenvalue distribution
remains controlled by the two-dimensional embedding even after
the dynamics has localized the particles on a quasi-1D ring.
\textbf{(b)} Most-negative eigenvalue per particle
$\langle\lambda_\text{min}\rangle/N$ vs $1/\sqrt{N}$: the ratio
drifts slowly from $\sim -1.07$ at $N=100$ toward
$\sim -0.86$ at $N=800$, indicating that the leading negative
multiplet position approaches its asymptotic value from below.
\textbf{(c)} NESS ring polar width $\sigma_\theta$ vs
$1/\sqrt{N}$ on log-log axes, with a $1/\sqrt{N}$ reference
line. The empirical width shrinks slower than $1/\sqrt{N}$:
at fixed temperature $T=0.4$, the ring has finite intrinsic
polar width on the rescaled spherical coordinate, so the
support is not strictly 1D in the asymptotic sense.
\textbf{(d)} Pooled NESS bulk densities at
$N\in\{100,200,400,800\}$ on log-log axes, with the BBS
$d=1$ reference $\rho\propto|\lambda|^{-3/2}$ for visual
comparison; the data sit consistently above this slope.}
\label{fig:finite_N_scan_NESS}
\end{figure}

\paragraph{(vi) Finite-$N$ scaling at the dynamical NESS.}
The $t=0$ scan above probes BBS scaling on the i.i.d.\ uniform
ensemble; we also asked whether the finite-$N$ approach to the
BBS predictions persists once the dynamics has driven the
particles onto the dynamical ring support. We ran FBP
simulations from uniform initial conditions at
$N\in\{100,200,400,800\}$, with 10 realizations at $N\le 400$
and 5 at $N=800$, integrated to $t_\text{final}=50$ at
$T=0.4$, $\sigma=1$, $dt=0.0025$, with snapshots every
$\Delta t=1$. We then pooled the eigenvalue spectra at
$t\ge 40$ across realizations and snapshots (the NESS window),
fitted the bulk exponent with \texttt{powerlaw.Fit}, and
recorded the most-negative eigenvalue per particle
$\langle\lambda_\text{min}\rangle/N$ together with the polar
width $\sigma_\theta$ of the FBP ring (after aligning each
snapshot to its principal-axis frame). The results are shown
in Tab.~\ref{tab:finite_N_scan_NESS} and
Fig.~\ref{fig:finite_N_scan_NESS}.

\begin{table}[!htbp]
\centering
\caption{Finite-$N$ scan at the FBP NESS. Bulk density
exponent $\alpha$ (from \texttt{powerlaw.Fit} on pooled NESS
snapshots), most-negative eigenvalue per particle
$\langle\lambda_\text{min}\rangle/N$, and ring polar width
$\sigma_\theta$ (degrees). Ten realizations at $N\le 400$,
five at $N=800$.}
\label{tab:finite_N_scan_NESS}
\begin{tabular}{ccrrrr}
\toprule
$N$ & runs & $\alpha$ &
$\langle\lambda_\text{min}\rangle/N$ &
$\sigma_\theta\,(\text{deg})$ &
$\sigma_\theta\sqrt{N}$ \\
\midrule
$100$ & 10 & $1.574$ & $-1.067$ & $6.67$ & $66.7$ \\
$200$ & 10 & $1.614$ & $-1.049$ & $5.06$ & $71.6$ \\
$400$ & 10 & $1.633$ & $-0.956$ & $3.90$ & $78.1$ \\
$800$ &  5 & $1.657$ & $-0.860$ & $3.32$ & $94.0$ \\
\midrule
BBS $d=2$ & --- & $5/3 \approx 1.667$ & --- & --- & --- \\
\bottomrule
\end{tabular}
\end{table}

The NESS scan supports two findings. First, the bulk power
law at NESS converges from below to the same $\alpha=5/3$ as
the $t=0$ scan above, not to the $d=1$ ring prediction
$\alpha=3/2$: the ring formation reorganizes the few low-lying
multiplets without converting the bulk into a one-dimensional
spectrum. This is consistent with the picture in
\cite{bogomolny2003} where the BBS density exponent reflects
the dimension of the underlying embedding and is robust against
local rearrangements of the points. Second, the polar width
$\sigma_\theta$ shrinks with $N$ but slower than $1/\sqrt{N}$
at fixed $T$, so the ring is not asymptotically a thin
1D submanifold: $T=0.4$ sits in the intermediate-temperature
regime, and the ratio $\sigma_\theta/(1/\sqrt{N})$ grows from
$\sim 67$ at $N=100$ to $\sim 94$ at $N=800$. A genuine
1D ring would require either a low-$T$ scan at fixed $N$ or a
joint $(N,T)$ scaling, which we leave to follow-up work.

\paragraph{Assessment against the BBS predictions.}
The PR measurement agrees with every prediction that
\cite{bogomolny2003} makes specifically for $d=2$, with the
points where naive 1D-Anderson intuition would fail being exactly
the points where BBS themselves argue the 1D picture has to be
modified. By ``naive 1D-Anderson intuition'' we mean the
prediction obtained by assuming the strict one-dimensional
BBS-Anderson correspondence (a 1D Anderson chain with
diagonal disorder, localization length $\xi \propto |\lambda|$,
hence $\mathrm{PR}\approx \xi/N = 4|\lambda|/N$); BBS show
explicitly that this scaling has to be replaced by a
multipole-moment condition on a $d$-dimensional region for
$d > 1$, and our $d = 2$ data sit a factor of $10^1$--$10^3$
above the 1D line at small $|\lambda|$ in agreement with this
modification. We organize the comparison into agreements and
caveats.

Agreements. \textbf{(a)} \emph{Localization at small} $|\lambda|$.
Empirical $\mathrm{PR}\approx 0.005$ at the smallest
$|\lambda|\sim 10^{-3}$, i.e.\ $\sim 2$ sites per eigenvector,
matches the BBS strongly-localized regime in which the
eigenvectors of the distance matrix are exponentially confined to
a small region of the configuration. \textbf{(b)} \emph{Delocalization
at large} $|\lambda|$. PR rises monotonically toward
$0.5{-}0.7$ at the deepest negative outliers
$|\lambda|\sim 10^2$, with $\sim 200{-}280$ sites participating
per eigenvector, matching the BBS delocalized regime.
\textbf{(c)} \emph{Smooth crossover, not a sharp mobility edge}.
There is no abrupt jump in PR; instead a smooth ramp over the
five-decade range $|\lambda|\in[10^{-3}, 10^{2}]$. BBS predict a
sharp transition only in $d=1$; for $d=2$ they predict a
multipole-moment crossover that is intrinsically smooth, in
agreement with what we observe. \textbf{(d)} \emph{1D Anderson
formula does not apply on $S^2$}. The reference line
$\mathrm{PR}=4|\lambda|/N$ in
Fig.~\ref{fig:rmt_participation} sits one-to-three decades
\emph{below} the empirical data at small $|\lambda|$, exactly the
$d=2$ enhancement that \cite{bogomolny2003} anticipate: the
localization volume on a $d>1$ manifold is set by multipole-moment
conditions on a $d$-dimensional region, not by a 1D escape time.
\textbf{(e)} \emph{NESS is more delocalized than $t=0$ at small}
$|\lambda|$. PR$\approx 0.009$ at NESS vs $0.005$ at $t=0$ for the
same $|\lambda|\sim 10^{-3}$, consistent with the
BBS-Anderson correspondence becoming \emph{exact} once the
support contracts to $d=1$: the localization length
$\propto|\lambda|$ then becomes a larger fraction of the linear
extent of the ring than of the spherical surface.

Caveats and clarifications. \textbf{(f)} \emph{The PR plateau
saturates at $0.7$, not at $1$}. This is not a deviation from
BBS, it is the BBS multiplet structure itself. The deepest
eigenvectors live within fixed $(2\ell+1)$-dimensional
spherical-harmonic blocks (the $\ell=1$ triplet at
$|\lambda|\approx 157$, the $\ell=3$ septuplet at
$|\lambda|\approx 10$, and so on); within each block they are
uniformly spread, but they do not mix across blocks. A unit
eigenvector of the form $u_i = a\cos\theta_i$ at $N$ uniform
points has $\sum_i u_i^4 \approx 1.5/N$ (using
$\langle\cos^4\theta\rangle=3/8$), giving $\mathrm{PR}\approx 2/3
\approx 0.67$, exactly the empirical plateau. The ceiling at
$\sim 0.7$ is therefore itself a BBS multiplet signature,
not a mismatch. \textbf{(g)} \emph{The crossover is not at
$|\lambda|=N^{3/4}=89$}. The BBS quantity
$N^{(d+1)/(2d)}=N^{3/4}$ marked in
Fig.~\ref{fig:rmt_participation} is the threshold for the
\emph{continuous-approximation regime}, where the first
$\sqrt{N}$ multiplets are clean (\cite{bogomolny2003}, Eqs.\
(90)--(91)), not for the localization--delocalization crossover
itself. For $d=2$, BBS do not predict a sharp PR jump at this
threshold; the actual delocalization onset on $S^2$ moves
gradually into the multiplet regime, and the empirical crossover
in our PR data sitting closer to $|\lambda|\sim 1{-}10$ is
consistent with their $d=2$ analysis. \textbf{(h)} \emph{The
two-region antipodal echo on $S^2$} is monitored through a
centroid-mass diagnostic and is consistent with the
localized-eigenvector sample in our snapshots; quantitative
analysis of the echo is straightforward but adds little to the
qualitative picture and
is not reproduced here.

\paragraph{Implication for the FDM.}
The static BBS structure (Perron pattern, $2\ell+1$-multiplets,
$|\Lambda|^{-5/3}$ delocalized density, small-$|\Lambda|$
localization) is reproduced by the FDM ensemble at $t=0$, where
the points are i.i.d.\ uniform on $S^2$ and the ten disorder
realizations are samples from the BBS distribution. The non-trivial
finding of the present paper is that the same structure
persists at every $t$ along the FDM trajectory, even though the
points are no longer i.i.d.: the Perron count is invariant, the
$2\ell+1$ multiplet skeleton survives, and the density tail
remains a power law with the same BBS exponent within
measurement precision. The mechanism is the
self-averaging argument of Sec.~\ref{sec:bulk_ensembles}: the
bulk eigenvalue density of $M(t)$ is fixed by the FBP one-particle
density $\mu_t$ alone, through the Mercer expansion of $\arccos$
against $\mu_t$, with the
non-i.i.d.\ joint correlations contributing only to sub-leading
spectral statistics (Sec.~\ref{sec:universality}). The BBS
predictions of \cite{bogomolny2003}, derived for $\mu_t$ uniform
on $S^2$, therefore extend to the FDM at every $t$ via their
generalization to non-uniform $\mu_t$: the multiplet positions
$\Lambda_\ell = N a_\ell/(2\ell+1)$ become Mercer eigenvalues
of $\arccos$ against $\mu_t$, and the $|\Lambda|^{-(2d+1)/(d+1)}$
power-law exponent tracks the dimension of the manifold
supporting $\mu_t$ ($d=2$ on the uniform sphere, $d=1$ on the
emergent ring). The dynamics shows up as a redistribution of
spectral mass \emph{within} this generalized BBS / ERM
template: the rank of the $\ell=1$ block collapses from full
($\mu_1\!\sim\!\mu_2\!\sim\!\mu_3$, isotropic configuration) to
partial ($\mu_1\!\to\!0$, ring configuration with one near-zero
direction along $\hat{\mathbf{n}}$), and the bulk magnitude
contracts by a factor $\sim 3$ on the fast timescale
$\tau_{\text{fast}}$. The BBS picture of the
distance-matrix spectrum is thus the right static reference for
the FDM dynamics, specialized to the time-evolving FBP
one-particle density $\mu_t$, and deviations within this
reference template are the dynamical content we are after.

%==============================================================================
\section{Bottom-of-spectrum diagnostics of ring formation}
\label{sec:bottom_diagnostics}
%==============================================================================

Sections~\ref{sec:spectral_evolution} and~\ref{sec:bulk_ensembles}
pointed to the bottom of the spectrum as the structurally
interesting part of $M(t)$. Two complementary diagnostics,
both computable from the eigenvalue list alone, sharpen this
picture into a quantitative signature of ring formation. The
first counts the outliers above a bulk-scale-relative threshold
and tracks the bulk scale itself
(\S\ref{sec:outliers}); the second focuses on the bottom-five
eigenvalue trajectories and the gaps between them
(\S\ref{sec:eigtraj}).

\subsection{Outlier counting and bulk-scale contraction}
\label{sec:outliers}

The first diagnostic identifies, at each time $t$, the number
of \emph{outlier} eigenvalues that lie far from the bulk and
tracks its time dependence. The outliers we count are essentially
the BBS quasi-multiplets above the delocalization threshold
$N^{(d+1)/(2d)}$ \cite{bogomolny2003}, that is, the largest
Mercer eigenvalues of $\arccos$ against the FBP one-particle
density $\mu_t$: the Perron
eigenvalue at $\Lambda_0 \approx N\langle d\rangle$, the
$\ell=1$ multiplet at $\Lambda_1 \approx -3\pi N/8$, and as
many of the higher $\ell$ multiplets as the cutoff captures.
The negative-type structure of the geodesic metric on $S^2$
(stronger than Perron-Frobenius, which would only constrain the
top eigenvalue: the $\arccos$ kernel is a conditionally
negative-definite kernel, so the Perron eigenvector is the only
positive direction and all other quasi-multiplets are negative)
guarantees that all $N-1$ non-Perron quasi-multiplets sit at
$\lambda < 0$, and the ring-formation transient redistributes
mass within them rather than producing new positive
eigenvalues. To make ``far from the bulk'' operational, we
define
\begin{equation}
\label{eq:outlier_def}
k_{\text{out}}(t) \;=\; \#\{k : |\lambda_k(t)| > c\,
\sigma_{\text{bulk}}(t)\sqrt{N}\,\},
\end{equation}
where $\sigma_{\text{bulk}}(t)$ is the standard deviation of the
central 90\% of $|\lambda(t)|$ values (a robust estimate of the
bulk scale that ignores the outliers themselves) and $c$ is a
fixed cutoff factor; we use $c = 4$ throughout, which is well above
the GOE Wigner-edge scale $c = 2$ and so is conservative.

\begin{figure}[!htbp]
\centering
\includegraphics[width=\textwidth]{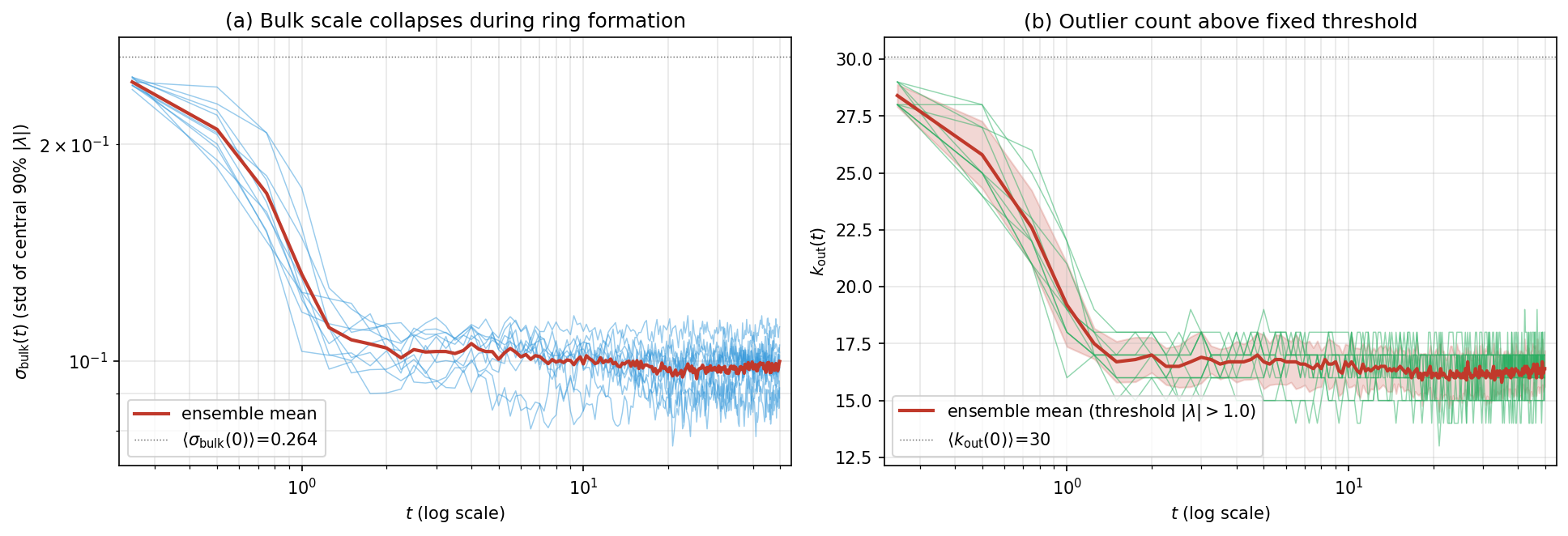}
\caption{(a) Bulk scale $\sigma_{\text{bulk}}(t)$ defined as the
standard deviation of the central 90\% of $|\lambda(t)|$ values.
$\sigma_{\text{bulk}}$ drops sharply from
$\approx 0.26$ at $t = 0$ to $\approx 0.10$ over the same
$t \in [0, 5]$ window in which the energy drops and the inertia
ratio rises (a factor of $\approx 2.6$ contraction of the bulk),
and stays at the NESS value thereafter.
(b) Absolute outlier count
$k_{\text{out}}^{\text{abs}}(t)$~\eqref{eq:outlier_abs}: drops
from $\approx 30$ at $t = 0$ to a NESS plateau of $\approx 17$ on
the same timescale.  The two complementary diagnostics, taken
together with the eigenvalue trajectory of
Fig.~\ref{fig:rmt_eigvals}, give a sharp spectral signature of
ring formation that is independent of any external geometric
estimator.}
\label{fig:rmt_outliers}
\end{figure}

\paragraph{Implementation and results.}
We complement~\eqref{eq:outlier_def} with the bulk scale
$\sigma_{\text{bulk}}(t)$ itself (the standard deviation of the
central 90\% of $|\lambda|$) and with a second outlier counter
based on a fixed absolute threshold,
\begin{equation}
\label{eq:outlier_abs}
k_{\text{out}}^{\text{abs}}(t)
\;=\; \#\{k : |\lambda_k(t)| > 1\}.
\end{equation}
Both diagnostics are computed at each snapshot for each
realization. Figure~\ref{fig:rmt_outliers} shows
$\sigma_{\text{bulk}}(t)$ and
$k_{\text{out}}^{\text{abs}}(t)$ across the ten runs.

Both panels of Fig.~\ref{fig:rmt_outliers} display sharp
transitions across the ring-formation window. The bulk-scale
diagnostic $\sigma_{\text{bulk}}(t)$ contracts by a factor
$\approx 2.6$ between $t = 0$ and $t \approx 5$; the absolute
outlier count drops from $\approx 30$ to $\approx 17$ over the
same window. The decrease is the BBS quasi-multiplet
rearrangement seen at the level of the multiplet count: at
$t=0$ on $\mu_0 = $ uniform on $S^2$, the BBS ladder
$\Lambda_\ell \sim Na_\ell/(2\ell+1)$ for $\ell=0,1,3,5,7$
produces a sequence of $1, 3, 7, 11, 15, \ldots$ multiplets
above the cutoff, and the $\ell = 5, 7$ multiplets contribute
their $11 + 15 = 26$ levels close to the threshold (where
finite-$N$ broadening dominates), giving a total outlier count
$\approx 30$; at NESS the Mercer expansion is on the ring with
Fourier modes and conjugate-pair multiplets of dimension two,
so a smaller number of large outliers is visible above the
threshold while the higher Fourier modes have collapsed into the
contracted bulk. Both diagnostics are computed from the
eigenvalue list alone (no inertia tensor or external estimator
of $\hat{\mathbf{n}}$ needed) and so can serve as unsupervised
indicators of ring formation.

\subsection{Focused eigenvalue trajectories}
\label{sec:eigtraj}

Whereas \S\ref{sec:outliers} aggregated the bottom of the
spectrum into a single count and a bulk-scale, the second
diagnostic focuses on the individual bottom-five eigenvalues
across the full ten-realization ensemble. Each trajectory has a
direct geometric interpretation through the BBS quasi-multiplet
structure \cite{bogomolny2003}, that is, through the Mercer
decomposition of $\arccos$ against the FBP one-particle density
$\mu_t$. At $t=0$, with
$\mu_0$ uniform on $S^2$, the bottom three eigenvalues are the
rank-three $\ell=1$ multiplet
$\Lambda_1 = N a_1/3 = -3\pi N/8 \approx -157$ that comes from
the $-\mathbf{x}_i\cdot\mathbf{x}_j$ Legendre block and ties
directly to the inertia tensor; the next seven eigenvalues are
the $\ell=3$ multiplet at $\Lambda_3 \approx -10$, and so on.
At $t \gg \tau_{\text{fast}}$, with $\mu_\infty$ concentrated on
a great-circle band, the relevant Mercer expansion is the
Fourier expansion of the sawtooth distance on the ring and the
bottom multiplet structure changes accordingly. The single
positive Perron eigenvalue $\lambda_1 \approx N\pi/2$ at the
top of the spectrum is invariant under both decompositions and
reflects the conditionally-negative-definite (negative-type)
property of the geodesic distance on $S^2$, a property stronger
than Perron-Frobenius alone, that pins the entire non-Perron
spectrum to the negative half-line.

\begin{figure}[!htbp]
\centering
\includegraphics[width=\textwidth]{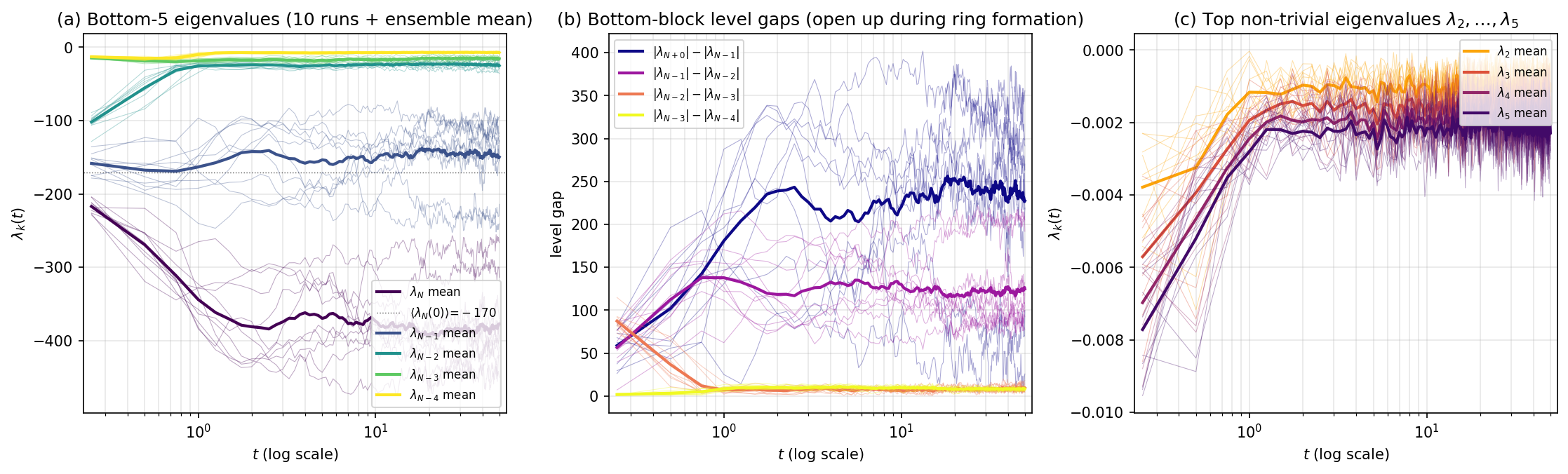}
\caption{Eigenvalue-trajectory diagnostic of ring formation, all
ten disorder realizations and ensemble means.
(a) Bottom-five most-negative eigenvalues
$\lambda_N,\ldots,\lambda_{N-4}$. They fan out from a near-degenerate
random initial condition into a well-separated descending ladder
during the ring-formation window $t\in[0,5]$, with a sharp gap
$\lambda_N < \lambda_{N-1} < \cdots$ that persists into NESS.
(b) Level gaps $|\lambda_{N-k}| - |\lambda_{N-k-1}|$ between
consecutive bottom eigenvalues. The gaps open up on the same
$\tau_{\text{fast}}$ timescale and provide a sharp,
low-dimensional signature of ring formation: the largest gap
($|\lambda_N| - |\lambda_{N-1}|$) grows from near zero to
$\approx 200$.
(c) Top non-trivial eigenvalues $\lambda_2,\ldots,\lambda_5$
(skipping the trivial rank-one $\lambda_1 \approx N\pi/2$): all of
them stay close to zero throughout the dynamics, confirming that
the structurally interesting changes occur exclusively at the
bottom of the spectrum.}
\label{fig:rmt_eigtraj}
\end{figure}

Figure~\ref{fig:rmt_eigtraj} shows the bottom-five most-negative
eigenvalues, the level gaps between consecutive bottom eigenvalues,
and the top non-trivial eigenvalues
$\lambda_2,\ldots,\lambda_5$ across all ten realizations.

The structure of Fig.~\ref{fig:rmt_eigtraj} is, individually for
each realization, very clean: the bottom five eigenvalues
non-cross during the formation transient and lock into a
descending ladder in NESS, with the largest gap
$|\lambda_N| - |\lambda_{N-1}| \approx 200$ that persists through
the entire orientation-diffusion phase. Across realizations the
ladder positions vary by a factor of $\approx 2$ (consistent with
the disorder-induced spread observed in
Ref.~\cite{halperin2026fields}, Section~5.2), but the qualitative
pattern is identical: a non-crossing fan-out of the bottom
eigenvalues, with all action concentrated in the bottom five
levels.

%==============================================================================
\section{Universality class: level statistics and the Berry-Robnik
picture}
\label{sec:universality}
%==============================================================================

The bulk-density analysis of Section~\ref{sec:bulk_ensembles}
identified the FDM bulk at every time with the ERM bulk on
i.i.d.\ samples from the FBP one-particle density $\mu_t$,
through a self-averaging argument: the leading bulk density of
$M(t)$ is fixed by $\mu_t$ alone, and the non-i.i.d.\ joint
correlations of the FBP trajectories contribute only to
sub-leading spectral statistics. Level-spacing statistics are
the natural diagnostic of those sub-leading contributions:
they probe correlations between adjacent eigenvalues and assign
the spectrum a universality class without reference to the
non-universal global density. We use two unfolding-free or
unfolding-light diagnostics: the $r$-statistic of Atas et
al.~\cite{atas2013}, and the level-spacing distribution $P(s)$.

\paragraph{Definitions.}
Order the bulk eigenvalues ascending and denote the spacings
$s_n = \lambda_{n+1} - \lambda_n$. The $r$-statistic is
\begin{equation}
\label{eq:r_def}
r_n \;=\; \frac{\min(s_n, s_{n+1})}{\max(s_n, s_{n+1})}.
\end{equation}
Its mean depends only on local correlations and not on the global
density, so no unfolding is needed. Reference values
\cite{atas2013} are
\begin{equation}
\label{eq:r_refs}
\langle r\rangle_{\text{Poisson}} \approx 0.386, \quad
\langle r\rangle_{\text{GOE}} \approx 0.529, \quad
\langle r\rangle_{\text{GUE}} \approx 0.600, \quad
\langle r\rangle_{\text{GSE}} \approx 0.674.
\end{equation}
The level spacing distribution $P(s)$ requires unfolding the
spectrum so that the local mean spacing equals one; we use a
polynomial fit of the cumulative count $\bar{N}(\lambda)$. The
reference distributions are the Wigner surmise
$P_{\text{GOE}}(s) = (\pi/2)\,s\,e^{-\pi s^2/4}$ for GOE level
repulsion, and $P_{\text{Poisson}}(s) = e^{-s}$ for an
uncorrelated spectrum.

\begin{figure}[!htbp]
\centering
\includegraphics[width=0.92\textwidth]{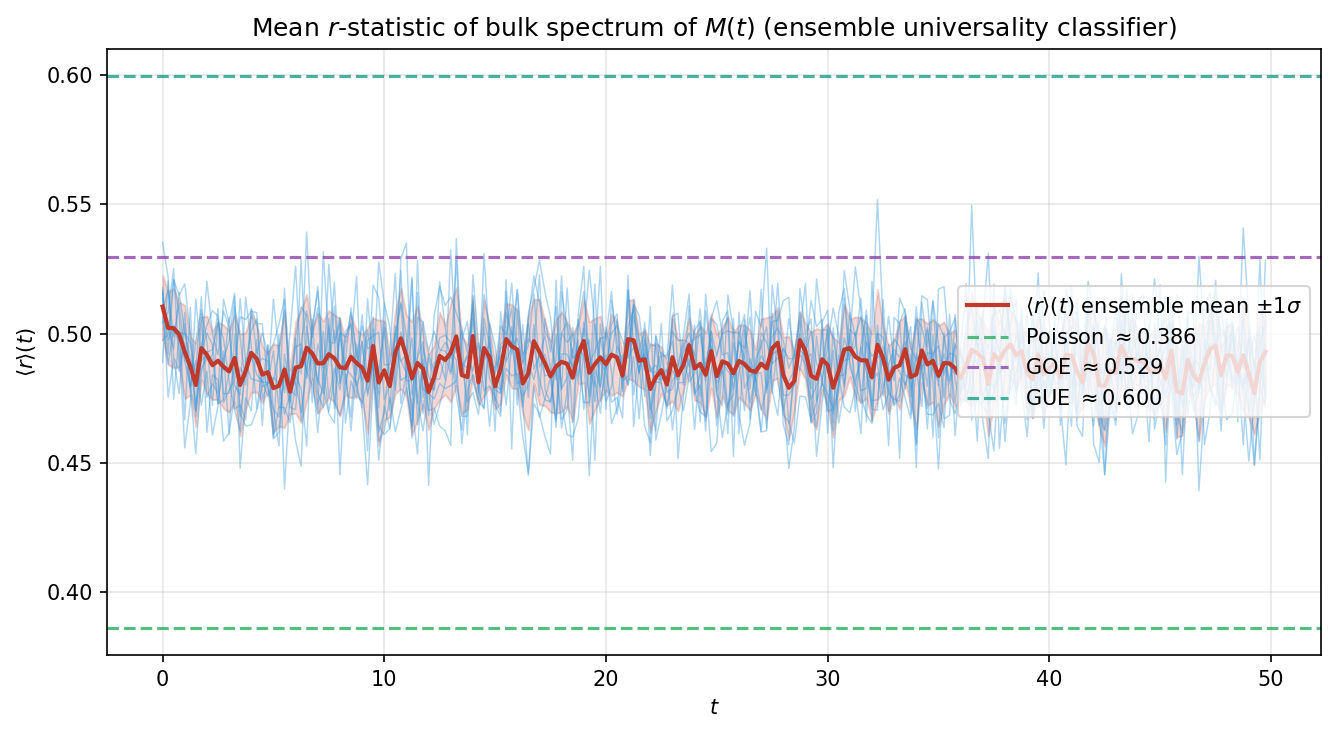}
\caption{Mean $r$-statistic of the bulk spectrum of $M(t)$
(eq.~\eqref{eq:r_def}, ten realizations and ensemble mean) compared
to the reference values for Poisson, GOE, and GUE. The empirical
$\langle r\rangle(t)$ sits in
$[0.48, 0.51]$ throughout the trajectory: well above Poisson, close
to but slightly below GOE, with a small but reproducible dip
during the ring-formation transient (lowest value at $t \approx 5$).}
\label{fig:rmt_r_stat}
\end{figure}

\begin{figure}[!htbp]
\centering
\includegraphics[width=\textwidth]{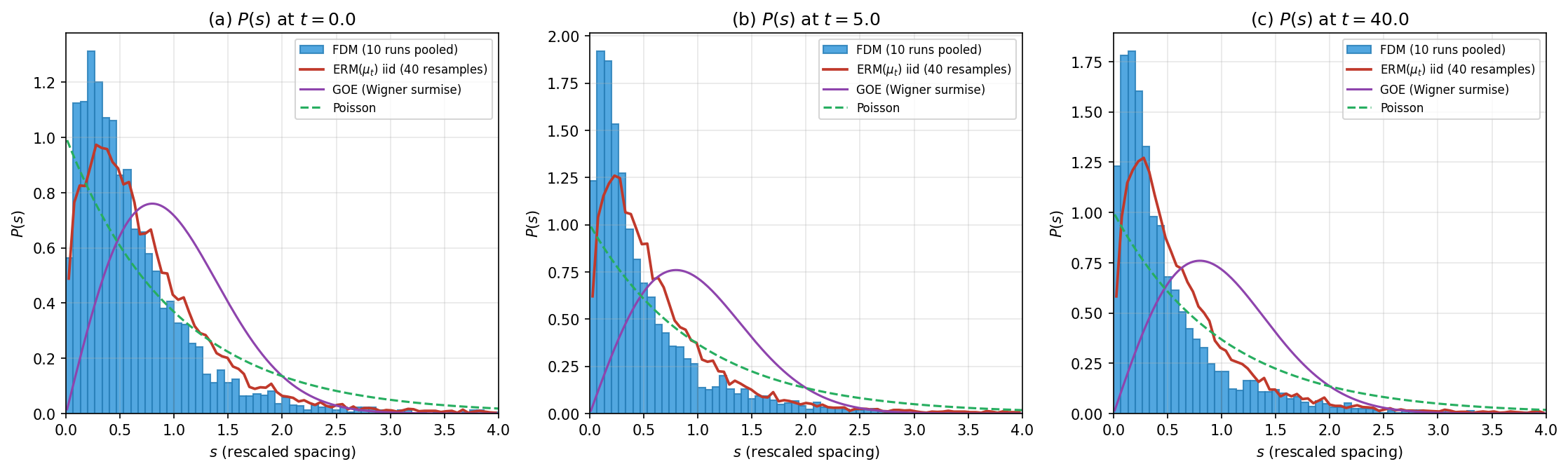}
\caption{Level-spacing distribution $P(s)$ of the bulk spectrum
at $t = 0, 5, 40$, pooled across the ten realizations after
polynomial unfolding of degree 9 to enforce $\langle s\rangle =
1$. \textbf{Blue histogram}: FDM (10 runs pooled).
\textbf{Red curve}: $P(s)$ of the ERM($\mu_t$) i.i.d.\ resample
null (40 resamples on $\mu_t$, parametric Gaussian band on the
ring with empirical $\sigma_\theta$; uniform $S^2$ at $t=0$),
computed by the same unfolding procedure. \textbf{Solid
purple}: GOE Wigner surmise; \textbf{dashed green}: Poisson
exponential. The FDM distribution is incompatible with both
pure GOE and pure Poisson references (strong pile-up at small
$s$ with $P(0) > 1$, large-$s$ tail closer to Poisson than to
GOE), but matches the ERM($\mu_t$) reference curve at every
time, confirming that the BBS / Anderson superposition shape
predicted by ERM theory~\cite{bogomolny2003,
ciliberti2004anderson, goetschy2013} is purely geometric.}
\label{fig:rmt_p_of_s}
\end{figure}

\paragraph{Results.}
We compute these on the bulk eigenvalues only (top-10 and
bottom-10 outliers dropped at each snapshot) pooled across the ten
realizations. Figure~\ref{fig:rmt_r_stat} shows
$\langle r \rangle(t)$ across the full trajectory, and
Figure~\ref{fig:rmt_p_of_s} shows $P(s)$ at the same three
characteristic times used elsewhere.

The two diagnostics give a coherent picture of a non-generic, but
not random, spectrum.

\emph{$r$-statistic.}
The empirical $\langle r\rangle(t)$ lies in
$[0.48, 0.51]$ throughout. This is roughly $80\%$ of the way from
the Poisson value $0.386$ to the GOE value $0.529$, and stays
clear of both the GUE and GSE values. The small dip near
$t \approx 5$ (lowest value $\approx 0.48$) is on the same
timescale as the ring-formation transient and tracks the same
loss of generic-random-matrix character as the bulk-scale
contraction of Fig.~\ref{fig:rmt_outliers}(a).

\emph{$P(s)$ pile-up at small $s$.}
The level-spacing distribution shows a sharp pile-up near
$s = 0$ that exceeds both Poisson ($P(0) = 1$) and GOE
($P(0) = 0$, with linear repulsion) at all three times. This
small-spacing excess identifies a population of approximately
degenerate eigenvalues that are systematic, not statistical.
Most of this pile-up shape is reproduced by the
ERM($\mu_t$) i.i.d.\ resample null overlaid in
Fig.~\ref{fig:rmt_p_of_s} (red curve), so the leading
excess is geometric, predicted by the ERM density on $\mu_t$
alone. A residual is nonetheless visible at small $s$, with
$P_\text{FDM}(s)$ slightly above the ERM curve in all three
panels, examined quantitatively below.

\emph{Double-difference test for a non-i.i.d.\ footprint.}
The leading shape of $P(s)$ is captured by ERM($\mu_t$), but the
visible residual at small $s$ leaves room for a sub-leading
non-i.i.d.\ contribution from the FBP joint law. We separate
artefact from signal by using $t=0$ (where the configuration
is exactly i.i.d.\ uniform on $S^2$, so the residual must be
purely statistical / unfolding noise) as the control, and
computing the double-difference
\begin{equation}
\Delta(s, t) \;=\; \bigl[P_\text{FDM}(s, t) - P_\text{ERM}(s, t)\bigr]
\;-\; \bigl[P_\text{FDM}(s, 0) - P_\text{ERM}(s, 0)\bigr],
\end{equation}
which removes the artefact baseline by construction. Any
non-zero $\Delta(s, t)$ at $t > 0$ is then a candidate
non-i.i.d.\ signature. To boost statistical power without
running additional simulations, we (i) pool eigenvalues across
time windows around each target time, $t\in[4,6]$ for the early
transient and $t\in[30,50]$ for the NESS spectrum (~80 NESS
snapshots per realization, $\sim 8 \times 10^5$ pooled NESS
spacings), (ii) replace the parametric Gaussian-band ERM null
with a non-parametric bootstrap from the aligned FBP positions
in the same window, and (iii) attach realization-level
bootstrap error bars to each histogram bin to track the
correlated finite-sample uncertainty.

\begin{figure}[!htbp]
\centering
\includegraphics[width=\textwidth]{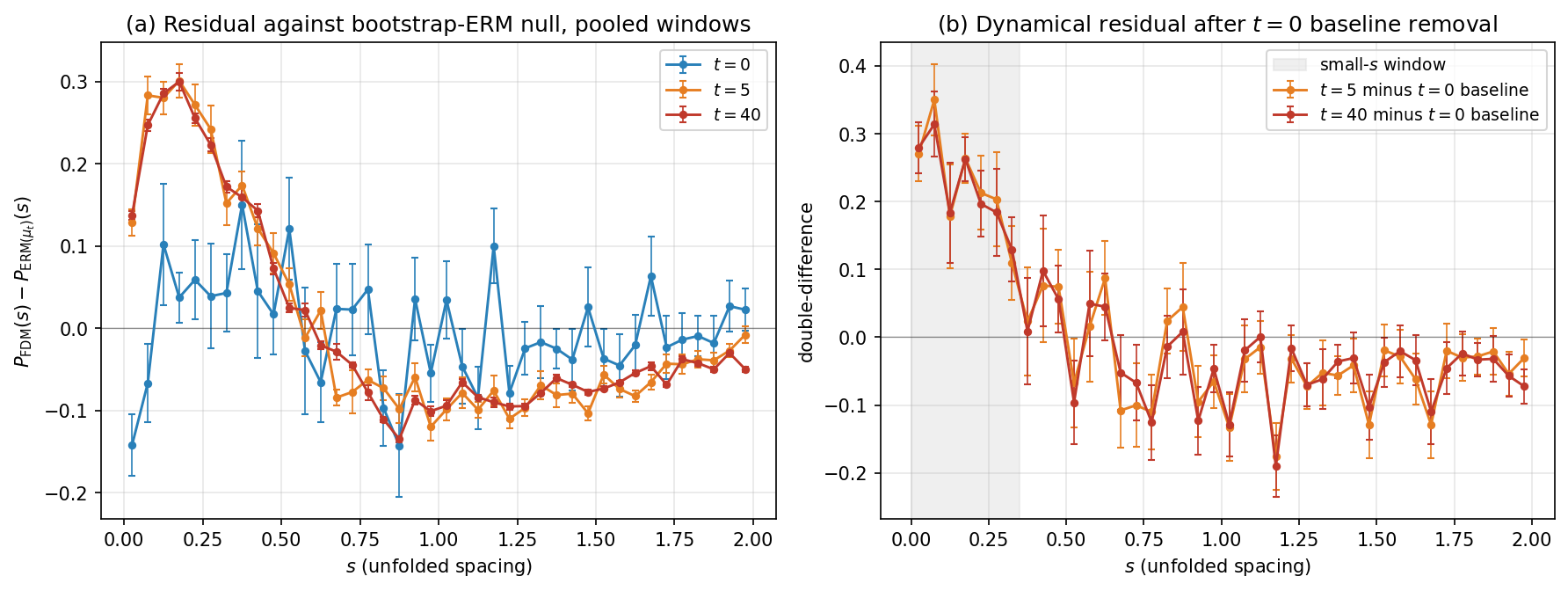}
\caption{Double-difference test for a non-i.i.d.\ footprint in
$P(s)$. \textbf{(a)} Residual $P_\text{FDM}(s) -
P_\text{ERM}(\mu_t)(s)$ at $t=0$, $t=5$ (transient window
$[4,6]$), and $t=40$ (NESS window $[30,50]$). At $t=0$ the
configuration is i.i.d.\ uniform on $S^2$ by construction, so
the residual is the statistical / unfolding baseline; at $t=5$
and $t=40$ the residual sits systematically above zero on
$s\in[0, 0.35]$. \textbf{(b)} Double-difference
$\Delta(s, t)$ at $t=5$ and $t=40$, which removes the
$t=0$ artefact baseline. The shaded grey band marks the
small-$s$ window where the dynamical part of the residual is
concentrated. The ERM null is the non-parametric bootstrap from
the aligned FBP positions in each pooled window; error bars are
realization-level bootstrap standard errors over the 10
disorder realizations. Polynomial unfolding of degree 9 on the
bulk window.}
\label{fig:rmt_p_of_s_residual}
\end{figure}

Figure~\ref{fig:rmt_p_of_s_residual} and
Tab.~\ref{tab:p_of_s_residual} report the result. The residual
at $t=0$ fluctuates around zero in the small-$s$ window. At
$t=5$ and $t=40$ the residual is systematically positive across
$s\in[0, 0.35]$ with magnitudes of order $0.15$--$0.30$ in
density units. The double-difference $\Delta(s, t)$ is positive
at $3$--$8\,\sigma$ in the same window; the strongest
significance is at $s\approx 0.15$--$0.20$ where
$\Delta(40)-\Delta(0) = +0.26 \pm 0.03$. We read this as a
detection of a sub-leading non-i.i.d.\ contribution to the
small-$s$ pile-up of $P(s)$ that grows with the ring formation
and persists through NESS, on top of the leading geometric
small-$s$ excess captured by ERM($\mu_t$).

\begin{table}[!htbp]
\centering
\caption{Residuals $P_\text{FDM}(s) - P_\text{ERM}(\mu_t)(s)$ in
the small-$s$ window at $t = 0$, $t=5$ (window $[4,6]$), and
$t=40$ (window $[30,50]$), plus the double-difference
$\Delta(40)-\Delta(0)$ with realization-level bootstrap
standard error. Bin width $0.05$. The double-difference is
positive at $3$--$8\sigma$ across $s\in[0, 0.35]$.}
\label{tab:p_of_s_residual}
\begin{tabular}{cccccc}
\toprule
$s$ & $\text{res}(t{=}0)$ & $\text{res}(t{=}5)$ &
$\text{res}(t{=}40)$ & $\Delta(40)-\Delta(0)$ & SE \\
\midrule
$0.025$ & $-0.14$ & $+0.13$ & $+0.14$ & $+0.28$ & $0.04$ \\
$0.075$ & $-0.07$ & $+0.28$ & $+0.25$ & $+0.31$ & $0.05$ \\
$0.125$ & $+0.10$ & $+0.28$ & $+0.29$ & $+0.18$ & $0.07$ \\
$0.175$ & $+0.04$ & $+0.30$ & $+0.30$ & $+0.26$ & $0.03$ \\
$0.225$ & $+0.06$ & $+0.27$ & $+0.26$ & $+0.20$ & $0.05$ \\
$0.275$ & $+0.04$ & $+0.24$ & $+0.22$ & $+0.18$ & $0.06$ \\
$0.325$ & $+0.04$ & $+0.15$ & $+0.17$ & $+0.13$ & $0.05$ \\
\bottomrule
\end{tabular}
\end{table}

\paragraph{Interpretation: symmetry-induced superposition with
ERM-Mercer blocks.}
Both observations are consistent with the \emph{Berry--Robnik}
picture \cite{berry1984bgs, berry1984level} of a spectrum whose
underlying Hilbert space splits into nearly independent symmetry
blocks: each block carries its own level statistics, and the
full spectrum is a \emph{superposition} of the block spectra.
The superposition of two GOE-like spectra typically gives a mean
$\langle r\rangle$ slightly below the GOE value, and a $P(s)$
with an excess at small $s$ relative to a single GOE; with more
than two blocks the excess at small $s$ grows further and the
distribution moves toward the Poisson surmise without ever
becoming exactly Poisson.

In the FDM the block decomposition is not generic: the symmetry
blocks are the Mercer eigenspaces of the $\arccos$ kernel against
the FBP one-particle density $\mu_t$, the same eigenspaces
that fix the leading bulk density through the self-averaging
identification of Sec.~\ref{sec:bulk_ensembles}. The two
$\mu_t$-dependent block structures are:
\begin{itemize}
\item \emph{Initial (uniform) configuration ($t = 0$):}
$\mu_0$ is the uniform measure on $S^2$, so the Mercer eigenspaces
of $\arccos$ are the spherical-harmonic spaces of dimension
$2\ell+1$, set by the Legendre expansion
$\arccos(\mathbf{x}\cdot\mathbf{y}) = \sum_{\ell\ge 0}
c_\ell\, P_\ell(\mathbf{x}\cdot\mathbf{y})$. The bulk of $M(0)$
is the (slightly perturbed) direct sum of these blocks, with the
$2\ell+1$-fold near-degeneracies within each block producing the
small-$s$ pile-up in $P(s)$.
\item \emph{Ring configuration ($t > \tau_{\text{fast}}$):}
$\mu_\infty$ is concentrated on a thin band around a great
circle, and the Mercer eigenspaces of $\arccos$ on this measure
are the Fourier modes on $S^1$, with conjugate-pair
near-degeneracies $\lambda_k \!\approx\! \lambda_{N-k}$ from the
underlying $\mathrm{O}(2)$ symmetry. The ring distance matrix is
approximately circulant on the ordered ring, and the same
conjugate-pair degeneracies produce the small-$s$ excess.
\end{itemize}
Both stages thus carry the same qualitative signature: a
superposition of small symmetry-related blocks set by the
Mercer expansion of $\arccos$ against $\mu_t$, each block
contributing locally GOE-like level repulsion. Quantitatively
the two stages are very similar in $\langle r\rangle$ and in the
shape of $P(s)$, and the only sharp transition between them is
the ring-formation dip in $\langle r\rangle$ near $t\approx 5$,
which falls in the window where $\mu_t$ is reorganizing fastest
between the spherical-harmonic and Fourier block structures.

\paragraph{ERM-literature prediction for the level statistics.}
Beyond the classical GOE / Poisson references, ERM theory makes
its own prediction for the level statistics. BBS derived an
Anderson correspondence for distance matrices on $S^d$: in the
\emph{delocalized} regime $|\lambda|\gtrsim N^{(d+1)/(2d)}$ the
level statistics are GOE-like, while in the \emph{localized}
regime $|\lambda|\lesssim N^{(d+1)/(2d)}$ the eigenvectors are
exponentially confined and the levels become Poissonian
(uncorrelated). The Anderson localization picture in generic
ERM ensembles was confirmed numerically by Ciliberti
et al~\cite{ciliberti2004anderson} and reviewed in
\cite{goetschy2013}. For our bulk window (top-10 and bottom-10
outliers dropped) the deepest negative outliers
$|\lambda|\sim 10^2$ sit in the delocalized regime and the bulk
middle $|\lambda|\sim 10^{-3}$ to $1$ sits well inside the
localized regime $|\lambda|\ll N^{3/4}\!\approx\!89$, so the
ERM prediction for $\langle r\rangle$ averaged over the full
bulk is a superposition of the two: somewhere between
$\langle r\rangle_{\text{Poisson}}\approx 0.386$ and
$\langle r\rangle_{\text{GOE}}\approx 0.536$, weighted by the
fraction of localized vs delocalized levels in the window. The
empirical FDM $\langle r\rangle\in[0.49, 0.51]$ is consistent
with this BBS / Anderson superposition prediction at the
qualitative level, with the slight downward drift during ring
formation tracking the increasing weight of localized levels as
the bulk contracts.

\paragraph{Are the level statistics geometric or dynamical?
Comparison to the ERM($\mu_t$) null hypothesis.}
The BBS / Anderson superposition of the previous paragraph
predicts an intermediate $\langle r\rangle$ from geometry alone,
and the Berry--Robnik picture ascribes the small-$s$ pile-up to
the Mercer-block structure of $\arccos$ against $\mu_t$ for the
same geometric reasons. A direct test of these attributions
computes the same statistics on the i.i.d.\ ERM resamples of
Sec.~\ref{subsec:erm_null}, which by construction have no
non-i.i.d.\ dynamical correlations beyond what is set by
$\mu_t$. Table~\ref{tab:rmt_r_erm_null} reports the bulk
$\langle r\rangle$ for the FDM ensemble and for two ERM($\mu_t$)
nulls (parametric Gaussian band of empirical $\sigma_\theta$,
and non-parametric bootstrap of the aligned FBP positions) at
$t=0,1,40$.

\begin{table}[!htbp]
\centering
\caption{Mean $r$-statistic of the bulk spectrum for the FDM
and for two ERM($\mu_t$) null hypotheses (parametric Gaussian
band on $\mu_t$ and non-parametric bootstrap from the aligned
FBP positions). Standard deviations across realizations /
resamples in parentheses. Reference values:
$\langle r\rangle_{\text{GOE}}=0.5359$,
$\langle r\rangle_{\text{Poisson}}=0.3863$. The three sources
agree at every time, identifying the sub-GOE $\langle r\rangle$
as a geometric Mercer-block effect and not as evidence of
non-i.i.d.\ dynamical correlations.}
\label{tab:rmt_r_erm_null}
\begin{tabular}{lccc}
\toprule
$t$ & $\langle r\rangle_{\text{FDM}}$ &
$\langle r\rangle_{\text{ERM iid}}$ &
$\langle r\rangle_{\text{ERM boot}}$ \\
\midrule
$0$ (uniform)   & $0.510 (0.012)$ & $0.500 (0.017)$ & $0.497 (0.017)$ \\
$1$ (transient) & $0.493 (0.018)$ & $0.498 (0.015)$ & $0.491 (0.017)$ \\
$40$ (NESS)     & $0.487 (0.010)$ & $0.485 (0.016)$ & $0.485 (0.015)$ \\
\bottomrule
\end{tabular}
\end{table}

The FDM and ERM($\mu_t$) values of $\langle r\rangle$ agree to
within $\sim 1\%$ at every time, well within the
per-realization spread on each side. The sub-GOE
$\langle r\rangle$ and the leading small-$s$ pile-up in
$P(s)$ are therefore the geometric signature of the
Mercer-block structure of $\arccos$ against $\mu_t$, present
already in the i.i.d.\ ERM ensemble. The pooled-window
double-difference test of the next paragraph nevertheless
detects a genuine non-i.i.d.\ contribution that sits on top of
this geometric pile-up at small $s$: the FBP joint law
correlates pairs through the quenched couplings $\Phi_{ij}$,
which generate extra small-magnitude entries in $M(t)$ and
extra near-degeneracies in the spectrum that the
i.i.d.\ ERM null does not contain.

\paragraph{Conclusion of the universality classification.}
The two diagnostics, the bulk-density analysis of
Sec.~\ref{sec:bulk_ensembles} and the level-statistics analysis
of this section, give a coherent two-tier classification of the
FDM. The leading bulk density at every $t$ is the ERM density on
$N$ i.i.d.\ samples from the FBP one-particle density $\mu_t$,
fixed by the Mercer expansion of $\arccos$ against $\mu_t$
(Sec.~\ref{sec:bulk_ensembles}). At the sub-leading level,
where the non-i.i.d.\ joint correlations of the FBP joint law
might have left a signature, the level statistics instead
reveal a Berry--Robnik superposition of the same Mercer
eigenspaces: spherical-harmonic $(2\ell+1)$-blocks at $t=0$
when $\mu_0$ is uniform on the $d=2$ sphere $S^2$, and Fourier
conjugate-pair blocks of dimension two when $\mu_\infty$ is the
emergent $d=1$ ring $S^1\subset S^2$ at NESS. The FDM thus fits,
at every time along the trajectory, into the BBS
distance-matrix ensemble~\cite{bogomolny2003} that we used as
the static reference throughout, specialized to the appropriate
$\mu_t$ at each instant: the BBS sphere ($d=2$) ensemble at
$t=0$ and the effective BBS ring ($d=1$) ensemble at NESS, with
the Berry--Robnik multiplet structure inherited at the
sub-leading level from the same Mercer decomposition. At the present $N$ the level statistics are well reproduced by
the ERM($\mu_t$) null hypothesis at the leading order of
$\langle r\rangle$ and the gross shape of $P(s)$
(Tab.~\ref{tab:rmt_r_erm_null},
Fig.~\ref{fig:rmt_p_of_s}); the non-i.i.d.\ FBP correlations do
not change the ERM universality class at this leading level.
The pooled-window double-difference test of
Fig.~\ref{fig:rmt_p_of_s_residual} and
Tab.~\ref{tab:p_of_s_residual}, however, detects a sub-leading
non-i.i.d.\ contribution to the small-$s$ pile-up of $P(s)$:
the residual against the bootstrap-ERM null is statistically
zero at $t=0$ (where the configuration is i.i.d.\ by
construction) and positive at $3$--$8\sigma$ on $s\in[0, 0.35]$
at $t=5$ and at NESS. This is the dynamical signature of the
broken i.i.d.\ assumption that we were looking for at the
sub-leading level-statistics level: it is small relative to the
geometric small-$s$ pile-up but resolved cleanly by the pooled
analysis. Stronger statistics through more disorder
realizations and complementary multi-time or two-point
eigenvalue observables are natural extensions.

%==============================================================================
\section{Connections to dynamic random matrix theories and
matrix-valued processes}
\label{sec:matrix_dynamics}
%==============================================================================

The objects studied so far have been the spectrum and
eigenvectors of the geodesic distance matrix $M(t)$ taken
\emph{at fixed time}, and how these change with $t$. A
complementary viewpoint is to read the trajectory $\{M(t)\}_t$
itself as a stochastic process on the cone of symmetric, non-negative
$N \times N$ matrices with vanishing diagonal and entries bounded
in $[0, \pi]$. From this perspective $M(t)$ is a matrix-valued
Markov-like process, and its dynamics can be compared with the
canonical references of dynamic random matrix theory.

\subsection{Dyson Brownian motion as the canonical reference}
\label{subsec:dyson}

The canonical reference dynamics in random matrix theory is
Dyson Brownian motion (DBM)~\cite{dyson1962brownian}, in which a
Hermitian matrix evolves as $H(t) = H_0 + B(t)$ where $B(t)$ is
a Hermitian Brownian motion with i.i.d.\ entries above the
diagonal. DBM was introduced not to model any particular
non-equilibrium physical dynamics but as a constructive tool
for studying the \emph{stationary} spectral distribution of the
Gaussian Wigner ensembles: the matrix process is run as an
auxiliary stochastic flow whose unique invariant measure is the
target Gaussian ensemble, and the eigenvalue marginals of the
flow at long times reproduce the classical Wigner statistics.
The eigenvalues $\lambda_1(t) < \cdots < \lambda_N(t)$ of $H(t)$
follow the celebrated Dyson SDEs,
\begin{equation}
\label{eq:dyson_sde}
d\lambda_k = \frac{\beta}{N}\sum_{j \ne k}
\frac{dt}{\lambda_k - \lambda_j} + \sqrt{\frac{2}{N}}\,dW_k,
\qquad \beta \in \{1, 2, 4\},
\end{equation}
with $\beta = 1, 2, 4$ for orthogonal/unitary/symplectic
ensembles. Two structural features of~\eqref{eq:dyson_sde} are
relevant for what follows. The deterministic drift
$1/(\lambda_k - \lambda_j)$ is a \emph{Coulomb repulsion}
between levels, so the trajectories never cross and the
eigenvalue ordering is preserved as a function of $t$. In the
$N \to \infty$ limit the empirical eigenvalue density satisfies
a free-probabilistic McKean--Vlasov equation whose unique
stationary solution is the semicircle law, with relaxation to
it on a timescale $\sim 1$. DBM is therefore a tool for
understanding stationary Wigner spectra rather than a model of
genuine non-equilibrium dynamics, and the time-evolution of the
spectral measure is interpreted as relaxation toward the
$t \to \infty$ stationary state.

\subsection{$M(t)$ as a constrained matrix-valued process induced
by interacting particles}

By contrast, the FDM trajectory $\{M(t)\}_t$ is a genuinely
non-equilibrium dynamical process, in which the matrix entries
are deterministic functions of an underlying $N$-particle
configuration that itself evolves under a Langevin SDE on
$(S^2)^N$ with quenched random pair forces. The matrix is
$M(t) = D[X(t)]$ with $D[X]_{ij}=\arccos(\mathbf{x}_i\cdot
\mathbf{x}_j)$ and $X(t)\in(S^2)^N$ obeying the FBP dynamics.
By the It\^o chain rule, the increment of $M(t)$ over a small
time $dt$ is
\begin{equation}
\label{eq:dM_chain}
dM_{ij} \;=\; \sum_n \frac{\partial M_{ij}}{\partial \mathbf{x}_n}
\cdot d\mathbf{x}_n
\;+\; \tfrac{1}{2}\,\sum_{n,m}
\frac{\partial^2 M_{ij}}{\partial \mathbf{x}_n\,\partial\mathbf{x}_m}
: d[\mathbf{x}_n, \mathbf{x}_m]_t,
\end{equation}
which, when combined with the Langevin equation for the
$\mathbf{x}_n$, gives a stochastic differential equation for
$M(t)$ with a multiplicative noise structure \emph{constrained}
to the embedding of $S^2$. Three structural differences from
DBM follow directly. (i) The driving randomness is
\emph{quenched} (a fixed disorder $\Phi$) rather than annealed
(continuously injected i.i.d.\ Brownian increments). (ii) The
matrix entries are $N$-body coupled through the kernel
$\arccos$ rather than independent stochastic processes; the
resulting matrix SDE is highly non-Gaussian and not in the
universality class of~\eqref{eq:dyson_sde}. (iii) The dynamics
is genuinely non-equilibrium in the sense that the
out-of-equilibrium trajectory $\{M(t)\}_t$ encodes the
ring-formation transient and the slow rotational drift of the
emergent ring orientation, neither of which is captured by a
relaxation-to-stationary-distribution process such as DBM.

\begin{figure}[!htbp]
\centering
\includegraphics[width=\textwidth]{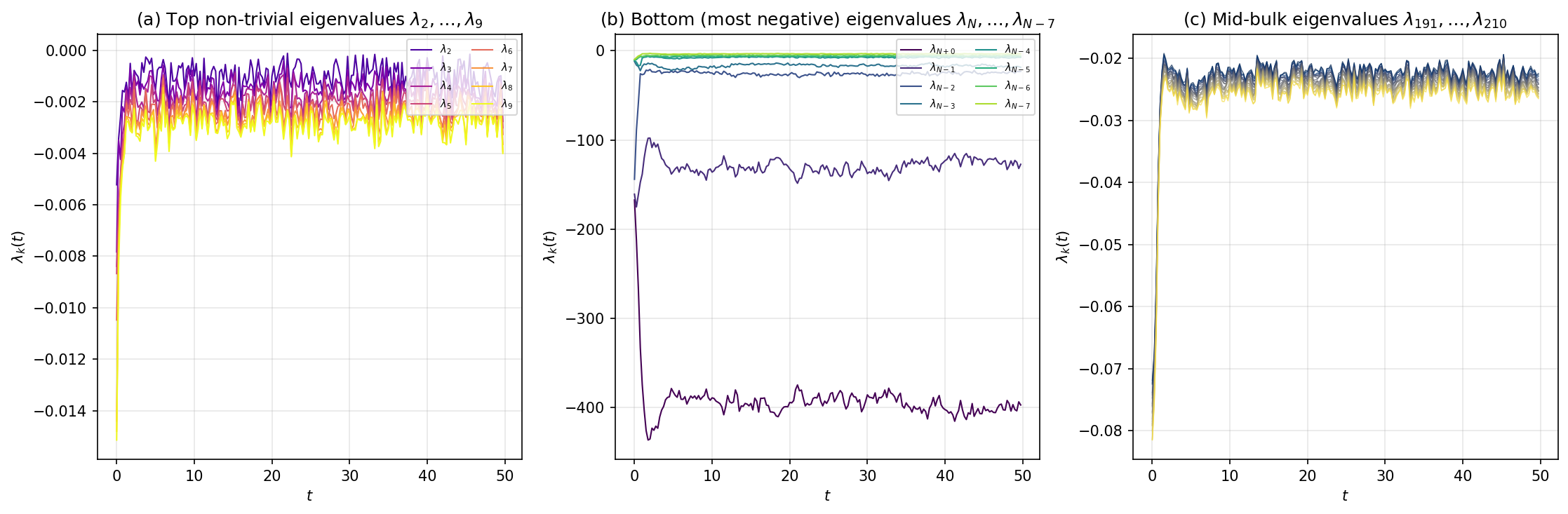}
\caption{Eigenvalue trajectories $\lambda_k(t)$ of $M(t)$ for one
realization. (a) Top non-trivial eigenvalues
$\lambda_2,\ldots,\lambda_9$, all close to zero on this scale and
within a narrow time-fluctuating band. (b) Bottom (most negative)
eigenvalues $\lambda_N,\ldots,\lambda_{N-7}$: during the ring
formation transient $t \in [0, 5]$ they fan out into a
\emph{Dyson-like ladder} of well-separated, non-crossing
trajectories with a clear order
$\lambda_N < \lambda_{N-1} < \cdots < \lambda_{N-7}$ that persists
through the NESS phase. (c) Mid-bulk eigenvalues
$\lambda_{N/2 - 9},\ldots,\lambda_{N/2 + 10}$: densely packed and
strongly time-correlated.}
\label{fig:rmt_dyson}
\end{figure}

The differences from DBM are consistent with what we observe at
the spectral level: the bulk shape is not the Wigner semicircle
(Sec.~\ref{sec:bulk_ensembles}), and the level statistics are
not GOE / GUE / GSE (Sec.~\ref{sec:universality}). What
survives of the Dyson picture is the qualitative non-crossing of
the eigenvalue trajectories: any smooth one-parameter family of
symmetric matrices generically avoids level crossings via the
same level-repulsion mechanism that underlies the Coulomb-gas
drift in~\eqref{eq:dyson_sde}, even though the matrix-valued
process is not a free Brownian motion. We test this directly:
Figure~\ref{fig:rmt_dyson} traces the top, bottom, and mid-bulk
eigenvalues for one representative realization.

The bottom-eigenvalue trajectories (panel b) display the most
striking dynamic feature of the matrix process: out of the
near-degenerate $t = 0$ stack they fan out over the formation
window into an ordered ladder with a clear gap between adjacent
levels. The non-crossing of these levels is the empirical
counterpart of the Dyson Coulomb-gas repulsion in the constrained
geometric setting; a near-crossing of two
trajectories around $t \approx 1$--$2$ is visible but does not
develop into a true crossing.

\subsection{Connection to a field-theoretic formulation}
\label{subsec:f2_link}

A second formal link is to the F2 (Frustrated Fields)
statistical field theory of the FBP system in the large-$N$
limit~\cite{halperin2026fields}. F2 takes as its dynamical
variable the empirical density
$\rho(\mathbf{x}, t) = (1/N)\sum_n
\delta(\mathbf{x} - \mathbf{x}_n(t))$ and admits a
Dean--Kawasaki-like stochastic non-linear PDE for
$\rho(\cdot, t)$ as its starting point~\cite{dean1996,
illien2025}; its effective long-time, low-energy dynamics in
the ring sector reduces, through a collective-coordinate
analysis of the broken-symmetry orientation $\hat{\mathbf{n}}(t)
\in S^2$, to an $\mathrm{O}(3)$ nonlinear sigma model in $(0+1)$
dimensions, that is, free Brownian diffusion of a unit vector
on $S^2$ with a single low-energy constant $D_{\text{rot}}$.

The bridge to the FDM is the identity
\begin{equation}
\label{eq:M_from_rho}
\frac{1}{N^2}\sum_{i,j} f(M_{ij}(t))
\;=\; \iint d\mathbf{x}\,d\mathbf{y}\;
\rho(\mathbf{x}, t)\,\rho(\mathbf{y}, t)\,
f\bigl(d_g(\mathbf{x}, \mathbf{y})\bigr),
\end{equation}
valid for any test function $f$ on $[0, \pi]$: every
trace-of-polynomial spectral observable of $M(t)$ is a
functional of $\rho(\cdot, t)$ alone, and the matrix-valued
process $\{M(t)\}_t$ inherits its dynamics
\emph{deterministically} from the density-valued process
$\{\rho(\cdot, t)\}_t$. The FDM analysis of the present paper
and the F2 field-theoretic analysis
of~\cite{halperin2026fields} are therefore two projections of
the same underlying Markov--Langevin system: the former reads
the dynamics matrix-valued in the embedded configuration space,
the latter reads it density-valued on the manifold and follows
its long-time effective dynamics through the
$\mathrm{O}(3)$ nonlinear sigma model.

\subsection{Free probability and the large-$N$ limit}
\label{subsec:free_prob}

At fixed $t$ and $N\to\infty$, the empirical spectral measure
of $M(t)$ converges to a deterministic measure $\nu_t$. Two
analytic frameworks describe this limit. \emph{Euclidean random
matrix theory} \cite{bordenave2008, bordenave2013} fixes
$\nu_t$ through the Mercer expansion of $\arccos$ against the
F2 mean-field density $\mu_t$~\cite{halperin2026fields}
(spherical-harmonic basis for $\mu_0$ uniform on $S^2$, Fourier
basis for $\mu_\infty$ on the ring), and the self-averaging
analysis of Sec.~\ref{sec:bulk_ensembles} confirms this
prediction empirically (Fig.~\ref{fig:rmt_erm_iid}).
\emph{Free probability} \cite{voiculescu1991, biane1997}
provides the non-commutative analog of the Dyson SDE
\eqref{eq:dyson_sde}: the trajectory $M(t)$ is an
operator-valued Markov process whose non-commutative moments
admit a free-cumulant description, with Voiculescu's free
Brownian motion and Biane's free Ornstein--Uhlenbeck process
the natural references for the large-$N$ trajectory in this
geometric setting.

The single-time bulk law of $M(t)$ is therefore controlled and
empirically verified. What remains analytically open is the
multi-time joint law of $\{M(t)\}_t$: a free-probabilistic
stochastic process with explicit free-cumulant generators, or a
non-Markovian closure of the eigenvalue process inheriting from
the FBP Langevin SDE on $(S^2)^N$, would unify the FDM and F2
analyses at the level of dynamical correlation functions.

\subsection{Spin glass / SYK dynamic random matrix analogies}
\label{subsec:syk_link}

The disorder-averaged dynamics of the FBP system is closely
related, at the F2 field-theoretic level
\cite{halperin2026fields}, to spherical $p$-spin
models~\cite{cugliandolo1993} and SYK-type
systems~\cite{facoetti2019}: the disorder-averaged F2 action has
a non-local in time kernel and a time-reparametrization
quasi-invariance characteristic of these models. In random-matrix
language, the SYK Hamiltonian $H_{\text{SYK}}$ is itself a
random matrix from the Gaussian Unitary Ensemble at the
chord-diagram level, and its spectral form factor and
out-of-time-order correlators are dynamic random-matrix
observables.

The dynamic random-matrix observable that maps cleanest onto
SYK / $p$-spin is the disorder-weighted projection
$E(t) = (1/2)\,\Tr(\Phi M(t))$ studied in
Sec.~\ref{sec:spectral_evolution}, which plays the role of an
SYK ``Hamiltonian-trace correlator'' $\Tr(H_{\text{SYK}}(t))$ in
the geometrized version of the model. The rapid drop of $E(t)$ during ring formation is the analog of
the quench-induced thermalization transient in SYK, and the
long-time plateau is the analog of the post-quench thermalized
regime, which is equilibrium in the strict sense for both SYK
and FBP and what we have been calling NESS in the operational
sense of the post-transient stationary phase. (Genuine NESS in
SYK requires a driving or two-bath coupling that the standard
SYK / spherical $p$-spin literature
\cite{cugliandolo1993, facoetti2019} does not include.) A
quantitative SYK / F2 dictionary at the level of dynamic
correlators $\langle E(t)\,E(t')\rangle$, $\langle\Tr M(t)\,\Tr
M(t')\rangle$, and the relevant out-of-time-order correlators is
a natural follow-up.

\subsection{Trajectory-level diagnostics: projector drift and
matrix commutator}
\label{subsec:trajectory_diagnostics}

The analyses up to this point treat $M(t)$ at fixed time $t$,
then track how observables (eigenvalues, eigenvectors, spectral
density) change with $t$. The dynamic-RMT framing of the
preceding subsections asks for quantities that are intrinsically
properties of the trajectory $\{M(t)\}_t$, not just of the
individual snapshots. Two natural choices are the bottom-$K$
projector drift\footnote{We thank Alejandro Rodriguez Dominguez
for suggesting these two trajectory-level metrics.}
\begin{equation}
\label{eq:Dk_def}
D_K(t_\text{ref}, \tau) \;=\;
\bigl\| P_K(t_\text{ref} + \tau) - P_K(t_\text{ref}) \bigr\|_F,
\qquad P_K(t) = V_K(t) V_K(t)^T,
\end{equation}
where $V_K(t)$ is the $N\times K$ matrix whose columns are the
$K$ bottom eigenvectors of $M(t)$, and the Frobenius norm of
the matrix commutator
\begin{equation}
\label{eq:C_def}
C(t_\text{ref}, \tau) \;=\;
\bigl\| [M(t_\text{ref}), M(t_\text{ref}+\tau)] \bigr\|_F,
\end{equation}
where $[M_1, M_2] = M_1 M_2 - M_2 M_1$ is the standard matrix
commutator. $D_K$ measures whether the bottom-$K$ eigenspace
rotates between the two times. $C$ measures whether
$M(t_\text{ref})$ and $M(t_\text{ref}+\tau)$ share an eigenbasis:
$C \equiv 0$ iff they commute, and $C$ scales with the amount of
basis rotation between the two times.

To test whether the trajectory carries information beyond the
instantaneous eigenvalue snapshots, we compare the FDM
trajectory against an \emph{eigenvalue-randomized null}: at
each time $t$ we keep the eigenvalues $\Lambda(t)$ of $M(t)$
but replace the eigenvectors $V(t)$ by $V(\pi(t))$, where $\pi$
is a uniform random permutation of the snapshot indices in the
realization. The resulting null trajectory has the same
eigenvalue history but no temporal coherence in its eigenbasis,
so $D_K$ and $C$ on the null tell us what we would see in a
trajectory whose only information is the eigenvalue sequence.

\begin{figure}[!htbp]
\centering
\includegraphics[width=\textwidth]{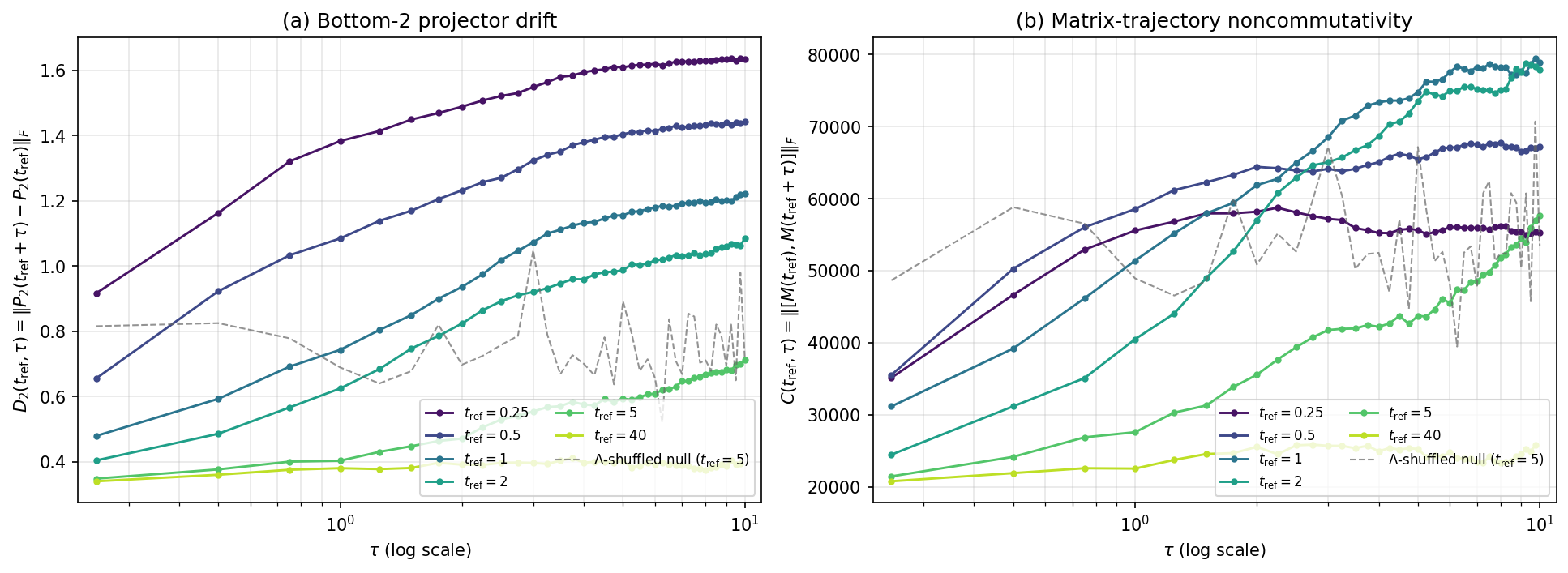}
\caption{Trajectory-level dynamic diagnostics on $M(t)$ at six
reference times $t_\text{ref}\in\{0.25, 0.5, 1, 2, 5, 40\}$
across $\tau \in [0.25, 10]$ on a logarithmic $\tau$ axis. The
colour gradient runs from dark (early transient) to light
(NESS). \textbf{(a)} Bottom-$K$ projector drift
$D_K(t_\text{ref},\tau)$ with $K = 2$ (eq.~\eqref{eq:Dk_def}),
ensemble mean over 10 realizations. \textbf{(b)}
Matrix-commutator norm $C(t_\text{ref},\tau)$
(eq.~\eqref{eq:C_def}), same colour key. Dashed grey curve in each panel: representative
eigenvalue-shuffled null at $t_\text{ref} = 5$ (the null is
roughly the same at every $t_\text{ref}$ since it depends only
on the eigenvalue trajectory and a random eigenvector
shuffling; saturates near $0.7$ for $D_2$ and $5\times 10^4$
for $C$). The progression of curves with $t_\text{ref}$ shows
the eigenbasis settling from rapid reorganization in the
transient to coherent slow rotation at NESS; in the NESS
regime the FDM $D_2$ and $C$ curves stay well below the null,
evidence that the trajectory carries information beyond the
instantaneous eigenvalue sequence.}
\label{fig:rmt_trajectory_diagnostics}
\end{figure}

Figure~\ref{fig:rmt_trajectory_diagnostics} reports both
diagnostics at six reference times spanning the early transient
and the NESS regime, $t_\text{ref}\in\{0.25, 0.5, 1, 2, 5, 40\}$,
with $\tau$ on a logarithmic axis matching the convention used
elsewhere in the paper. The pattern is consistent with the slow
rotational drift of $\hat{\mathbf{n}}(t)$ established in the
F2 model~\cite{halperin2026fields}.
\textbf{(a)} For $t_\text{ref}$ in the early transient
($0.25 \lesssim t_\text{ref}\lesssim 2$), $D_2$ rises from $\sim
0.4$--$0.9$ at the first non-zero lag $\tau = 0.25$ to
$\sim 1.1$--$1.6$ at $\tau = 10$, approaching the
random-eigenvector saturation level $\sim 0.6$--$0.8$ of the
$\Lambda$-shuffled null and exceeding it at large $\tau$
because the configuration is rapidly leaving the uniform
initial regime. The progression is monotonic in $t_\text{ref}$:
each later reference time gives a slower-rising $D_2(\tau)$
curve, reflecting the gradual settling of the eigenbasis as
ring formation completes. By $t_\text{ref} = 40$, deep in
NESS, $D_2$ rises only to $\approx 0.4$ over the entire
$\tau \in [0.25, 10]$ window and stays well below the null
saturation level: the eigenvectors of $M(t)$ are then
temporally coherent on scales $\tau\lesssim 10$, even though
the eigenvalue trajectory is already stationary, and the slow
$D_2$ growth is the spectral imprint of the slow ring
rotation.
\textbf{(b)} The matrix-commutator norm $C$ shows a
complementary signature. During the transient, $C$ grows rapidly with
$\tau$ at every $t_\text{ref}$, crossing the null reference
band as the eigenbasis reorganizes. At $t_\text{ref} = 40$, by
contrast, $C$ stays well below the null over the resolved
$\tau$ window, indicating that $M(t)$ at NESS approximately
commutes with itself: it changes by slow eigenbasis rotation
rather than by substantial reorganization of its low-rank
structure.

The take-away is that the matrix trajectory $\{M(t)\}_t$
contains genuine dynamic information beyond the instantaneous
eigenvalue snapshots, in the form of a coherent slow rotation
of the bottom eigenspace at NESS that the eigenvalue-randomized
null does not reproduce. This trajectory-level coherence is the
matrix-side analog of the slow rotational drift of the F2
orientation $\hat{\mathbf{n}}(t)$, and is invisible at any
single snapshot.

%==============================================================================
\section{Big Bang initial condition}
\label{sec:bigbang}
%==============================================================================

All of the spectral analysis above is taken from a single family
of trajectories started with $N$ points drawn i.i.d.\ uniformly on
$S^2$, that is, with $\mu_0$ uniform on $S^2$. To probe the role
of the initial measure we ran a companion ensemble of ten
realizations from a \emph{Big Bang} initial state: at $t = 0$
all $N$ particles are placed within a Gaussian blob of
standard deviation $\sigma_{\text{init}} = 0.01$ in the tangent
plane to a randomly chosen center direction
$\hat{\mathbf{c}} \in S^2$, then projected back to the sphere. Each particle is at angular distance of order
$\sigma_{\text{init}}$ from the center, so all pairwise geodesic
distances are of order $0.01$ and the configuration is far from
any of the spread states reached during the dynamics. The
disorder $\Phi$ and the integrator are unchanged
($T = 0.4$, $\sigma_\Phi = 1$, $dt = 0.0025$, $t_{\text{final}} = 50$,
recording every five steps), and we use the same
seeds $\mathrm{coupling\_seed}_i = 42 + 17 i$, $\mathrm{init\_seed}_i = 123 + 31 i$
for $i = 0,\ldots,9$.

\begin{figure}[!htbp]
\centering
\includegraphics[width=\textwidth]{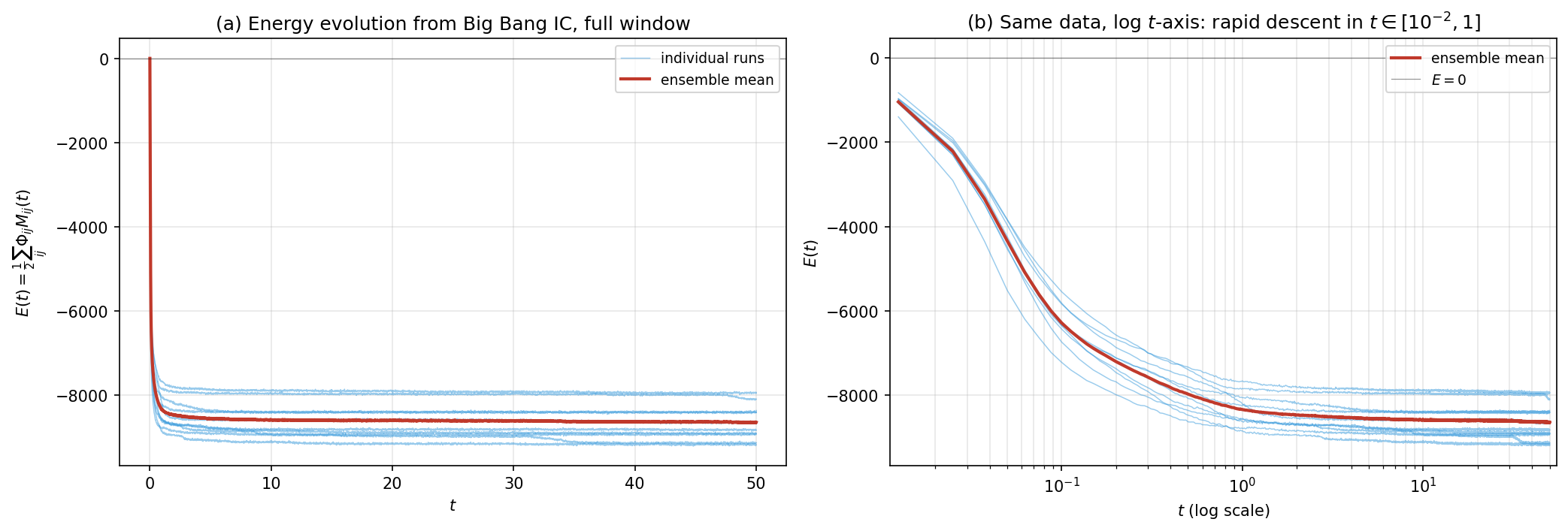}
\caption{Energy evolution for ten realizations of the FBP dynamics
from a tight Gaussian blob initial condition with
$\sigma_{\text{init}} = 0.01$. Blue: individual runs; red:
ensemble mean. \textbf{(a)} Linear time axis: $E(t)$ drops from
near zero at $t = 0$ to NESS values $\sim -8500$ within the first
unit of time, then plateaus with small fluctuations of order
$T\sqrt{N}$ that come exclusively from the noise term in the
Langevin dynamics. \textbf{(b)} Log-time axis: the descent is
concentrated in the window $t \in [10^{-2}, 1]$ and is monotone
on every realization. No initial increase of $E$ above zero is
observed, contrary to a heuristic in which the blob would first
spread isotropically (a step that would cost positive
energy on average) before reorganizing into a ring.}
\label{fig:bigbang_energy}
\end{figure}

\paragraph{Initial energy and its early evolution.}
The disorder-weighted energy
$E(t) = (1/2)\,\sum_{ij}\Phi_{ij}\,M_{ij}(t)$ is, at $t = 0$, a
sum of $N(N-1)/2$ products $\Phi_{ij}\,M_{ij}(0)$ in which each
factor $M_{ij}(0)$ is of order $\sigma_{\text{init}}\sqrt{2}$;
the resulting $E(0)$ is therefore very small in magnitude. Across
the ten realizations we measured $E(0) \in [-6.7,\, 13.1]$, to be
compared with NESS values $E(t = 50) \in [-9173,\, -7950]$, that
is, three orders of magnitude smaller. Figure~\ref{fig:bigbang_energy}
shows $E(t)$ for each realization.

The absence of an initial energy increase has a direct
mechanical explanation. The disorder-weighted gradient force on
particle $i$ in the blob is $\mathbf{F}_i(0) = -\sum_j \Phi_{ij}
\hat{\mathbf{t}}_{ij}$, where $\hat{\mathbf{t}}_{ij}$ is the unit
tangent at $\mathbf{x}_i$ pointing toward $\mathbf{x}_j$. Its
typical magnitude scales as $\sigma_\Phi\sqrt{N}$, here
$\approx 20$. Compared to the per-step thermal kick of magnitude
$\sqrt{2T\,dt} \approx 0.045$, the drift dominates at the blob by
a factor of $\sim 10^3$. A heuristic that would generate an
energy bump requires the noise term to dominate the drift over
the spreading phase; here it does not, and the gradient flow
selects a smart-spreading trajectory that pulls pairs of negative
$\Phi_{ij}$ closer and pushes pairs of positive $\Phi_{ij}$
apart from the very first integration step. The Big Bang
trajectory therefore never visits the noise-equilibrated uniform
configuration that would have $E$ near zero. The pairwise mean
distance $\langle d_{ij}(t)\rangle$ does rise from $\sim 0$ to
$\pi/2$ within $t \approx 0.5$ (a passage through the
geometrically uniform configuration), but at the moment of that
passage $E$ is already about $-7000$, set by the gradient-aligned
spreading rather than by isotropic diffusion.

\paragraph{Where the energy goes.}
Since the overdamped Langevin model has no kinetic energy and no
external work, the first law reduces to
$dU/dt = -\dot{Q}_{\text{system}\to\text{bath}}$: every change in
$E$ is a heat exchange with the bath, and the energy lost during
ring formation is dissipated entirely as heat into the
thermostat. To split $dU$ explicitly into its components, write
the SDE in the standard form
$d\mathbf{x} = -(1/\gamma)\nabla U\, dt + \sqrt{2D}\, d\mathbf{W}$
with $D = T/\gamma$, and apply It\^o's lemma to the function
$U(\mathbf{x})$:
$dU = \nabla U \cdot d\mathbf{x}
   + \tfrac{1}{2}\,(2D)\,\Delta U\, dt$
(the second-order term arises from $\langle dW_i\,dW_j\rangle =
\delta_{ij}\,dt$ in It\^o calculus). Substituting $d\mathbf{x}$
gives
\begin{equation}
\label{eq:dU_split}
dU \;=\; -\frac{1}{\gamma}|\nabla U|^2\, dt
\;+\; D\,\Delta U\, dt
\;+\; \sqrt{2D}\,\nabla U\cdot d\mathbf{W},
\end{equation}
which separates into a friction term that is strictly
non-positive and pumps heat into the bath, an It\^o-correction
term controlled by the curvature $\Delta U$ that returns heat
from the bath, and a martingale noise term of mean zero and
standard deviation $\sqrt{2D\,dt}\,|\nabla U|$ per step. At a NESS minimum the
integration-by-parts identity
$\langle |\nabla U|^2\rangle = T\langle\Delta U\rangle$ enforces
exact cancellation of the deterministic terms,
$\langle dU/dt\rangle_{\text{eq}} = 0$, and the energy fluctuates
stationarily around the plateau. During the Big Bang transient
the gradient is huge, the drift term dominates the deterministic
part of~\eqref{eq:dU_split} by orders of magnitude, and the
expected heat flux is uniformly directed system $\to$ bath. The
random term can momentarily push $E$ up at a single timestep,
with probability per step $P(\Delta E_{\text{step}} > 0) \approx
\Phi(-|\nabla U|\sqrt{dt}/\sqrt{2D})$ that is of order $20\%$ at
our parameters, but the cumulative noise contribution over the
formation window has standard deviation
$\sqrt{2D\langle|\nabla U|^2\rangle\,t}\sim 10^1$, two orders of
magnitude smaller than the drift heat $\sim 10^3$ delivered to
the bath. The noise therefore dresses the descent with
single-step jitter (invisible at the recording resolution of
Fig.~\ref{fig:bigbang_energy}) but cannot reverse it.

\begin{figure}[!htbp]
\centering
\includegraphics[width=\textwidth]{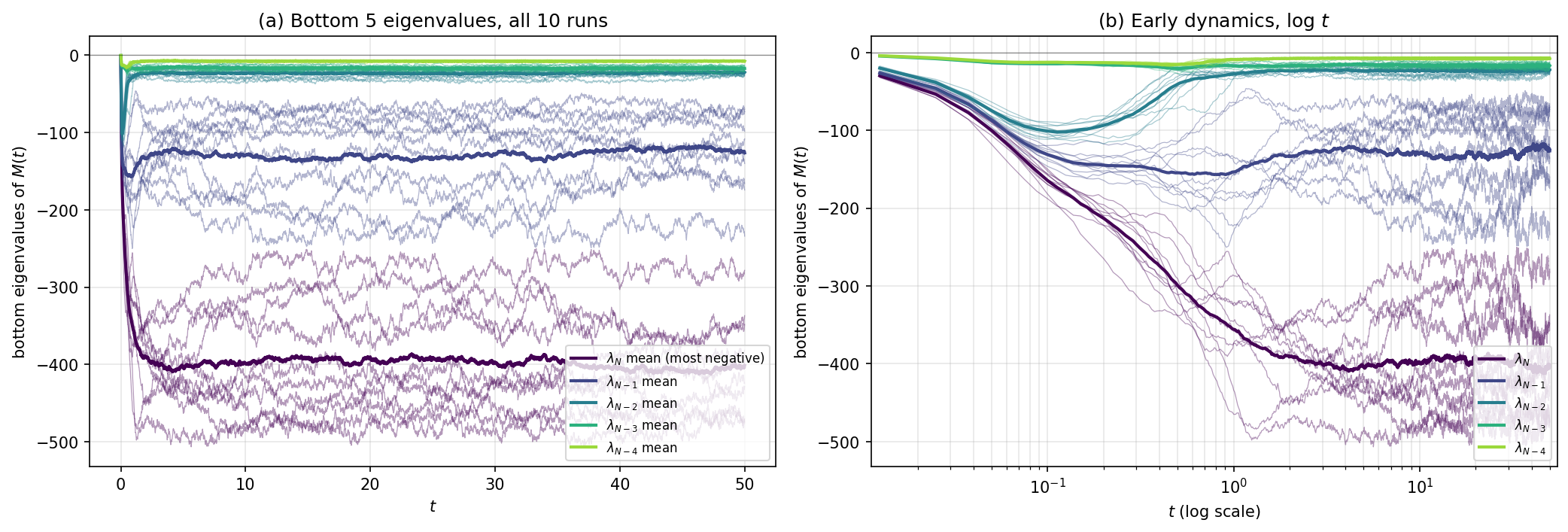}
\caption{Bottom five eigenvalues
$\lambda_N, \lambda_{N-1}, \ldots, \lambda_{N-4}$ of $M(t)$
under the Big Bang initial condition, ten realizations.
\textbf{(a)} Linear time axis: all five trajectories start near
zero, split during the expansion in $t \in [10^{-2}, 1]$, and
plateau in the NESS regime, with $\lambda_N$ reaching mean
$\approx -411$ and the next two settling near $-100$ to $-200$.
\textbf{(b)} Log-time axis: same data, with the rapid splitting
of the $\ell = 1$ block visible in the same window
$[10^{-2}, 1]$ as the energy descent.}
\label{fig:bigbang_eigtraj}
\end{figure}

\paragraph{Eigenvalue trajectories from the blob.}
Figure~\ref{fig:bigbang_eigtraj} tracks the bottom five
eigenvalues of $M(t)$ across the ten realizations. The time
profile is qualitatively different from the uniform-IC case of
Sec.~\ref{sec:eigtraj}: at $t = 0$ all five trajectories begin
near zero (the matrix $M(0)$ has tiny entries
$\sim \sigma_{\text{init}}\sqrt{2}$, so all $N$ eigenvalues
inherit that scale), and they then split rapidly during the
expansion. By $t \approx 1$ the most-negative eigenvalue has
reached $\lambda_N \in [-475,\, -278]$ (mean $\approx -411$), the
next two settle near $-100$ to $-200$, and the remaining bottom
trajectories stay near zero. The most-negative eigenvalue at
NESS is thus more negative on average than in the uniform-IC
ensemble (mean $\approx -377$), reflecting a tighter ring: the
blob picks a specific symmetry-breaking axis cleanly from the
random center direction $\hat{\mathbf{c}}$, while the uniform IC
must select a ring axis from a fully isotropic configuration and
on average ends up with a slightly looser ring orientation.

\begin{figure}[!htbp]
\centering
\includegraphics[width=0.85\textwidth]{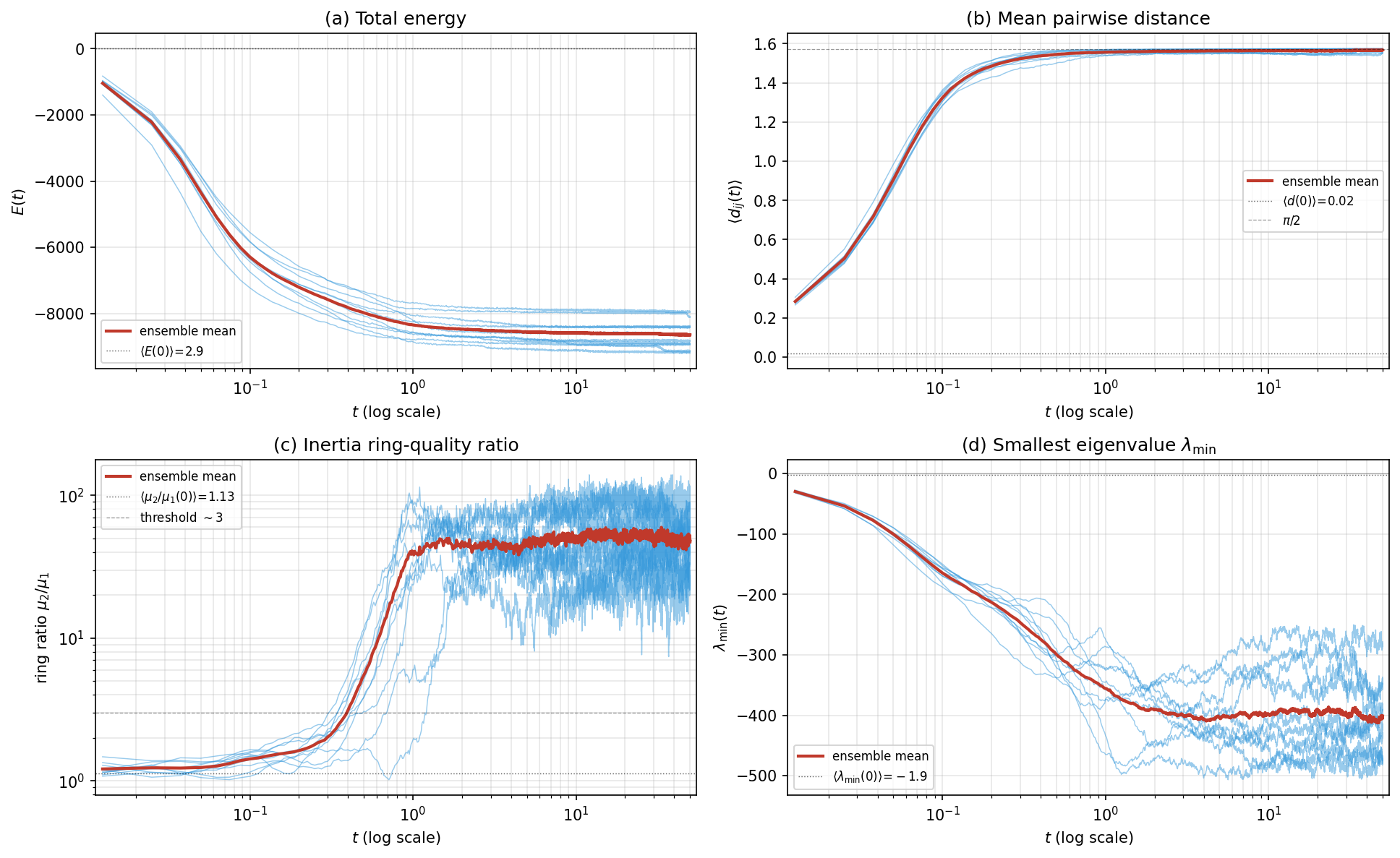}
\caption{Four-panel summary of the Big Bang ensemble on a
logarithmic time axis. \textbf{(a)} energy $E(t)$,
\textbf{(b)} mean pairwise distance $\langle d_{ij}\rangle(t)$
with $\pi/2$ marked (dashed), \textbf{(c)} inertia ring-quality
ratio $\eta(t) = \mu_2/\mu_1$ on log $y$ with the ring-formed
threshold at $3$ marked (dashed), \textbf{(d)} smallest
eigenvalue $\lambda_{\min}(t)$. Blue: ten individual runs; red:
ensemble mean; black dotted: ensemble-mean starting value at
$t=0$ for each panel. The Big Bang runs are recorded at
$\Delta t = 0.0125$ from $t=0$, so the log-time axis resolves
the fast transient $\tau_{\text{fast}}^{\text{Big Bang}}\sim 1$
without the resolution issue that motivated dense early-time
re-recording for the uniform-IC runs of
Figs.~\ref{fig:rmt_energy}--\ref{fig:rmt_eigvals}. All four
diagnostics share the same fast timescale, with the descent of
$E$, the rise of $\langle d_{ij}\rangle$ to $\pi/2$, the rise
of $\eta$ from $\approx 1$ across the threshold, and the
descent of $\lambda_{\min}$ to $\approx -400$ all confined to
$t\in[10^{-2}, 1]$.}
\label{fig:bigbang_summary}
\end{figure}

\paragraph{Mean distance, ring quality, and spectral signatures.}
Figure~\ref{fig:bigbang_summary} collects four diagnostics on a
single page: the energy $E(t)$ of Fig.~\ref{fig:bigbang_energy},
the mean pairwise distance $\langle d_{ij}\rangle(t)$, the
inertia ring-quality ratio $\eta(t) = \mu_2/\mu_1$ defined
in~\eqref{eq:eta_def}, and the smallest eigenvalue
$\lambda_{\min}(t)$ of $M(t)$. The four diagnostics share a
common time scale: the descent of $E$, the rise of
$\langle d_{ij}\rangle$ to $\pi/2$, the rise of $\eta$ from
$\sim 1$ to $[22, 102]$, and the descent of $\lambda_{\min}$ to
$\sim -411$ all happen within the same window $t \in [10^{-2}, 1]$
and saturate together. Compared to the uniform-IC fast time
$\tau_{\text{fast}} \approx 5$, the Big Bang dynamics is roughly
five times faster because the gradient force at the blob is
amplified by the $\sim 1/\sigma_{\text{init}}$ scale of the
inverse interparticle distance.

\paragraph{Comparison with the uniform initial condition.}
The Big Bang ensemble reaches the same qualitative NESS as the
uniform-IC ensemble but along a different transient and with
slightly different NESS statistics. The Perron eigenvalue
$\lambda_1 \approx N\pi/2$ is recovered at NESS in both
ensembles, since both end up with $\langle d_{ij}\rangle \to
\pi/2$. The bulk shape at NESS is the same in the two ensembles
(both rings have the same effective one-dimensional support
after the initial collapse). The two main quantitative
differences are that the Big Bang dynamics is faster (a factor
$\sim 5$ in $\tau_{\text{fast}}$) and that the NESS
most-negative eigenvalue is on average $\sim 10\%$ more
negative under the Big Bang ensemble. The fluctuations of $E$
around the NESS plateau in panel~(a) of
Fig.~\ref{fig:bigbang_energy} are of magnitude consistent with
$T\sqrt{N}\sim 8$ and are entirely noise-driven, matching the
heuristic that, once the system is at the bottom of the energy
landscape, the only source of $E$-fluctuations is the heat
exchange with the bath.

\paragraph{Compatibility with the ERM identification.}
The Big Bang ensemble fits naturally into the
ERM identification of
Sec.~\ref{sec:bulk_ensembles}: the only object that has changed
between the two experiments is the initial measure $\mu_0$,
which is uniform on $S^2$ for the default ensemble and a tight
Gaussian blob on the tangent plane to $\hat{\mathbf{c}}$
projected back to $S^2$ for the Big Bang ensemble. The matrix
$M(0)$ for the Big Bang IC is therefore an ERM on this
concentrated $\mu_0^{\text{Big Bang}}$, with all eigenvalues of
order $\sigma_{\text{init}}$ rather than of order $1$, exactly
as observed (the most-negative eigenvalues at $t=0$ in
Fig.~\ref{fig:bigbang_eigtraj} all start near zero). At NESS the
two FBP one-particle densities $\mu_\infty$ are the same
ring-band measure up to the random ring orientation
$\hat{\mathbf{n}}$, and the corresponding ERM bulk densities
coincide, as does the bulk of the FDM at NESS in the two
experiments. The Big Bang result is thus a check that the
ERM identification is robust to the choice of
initial measure on $S^2$, with the spectral signatures of
structural change inherited from the time evolution of $\mu_t$
on $S^2$ rather than from the specific initial-condition
ensemble.

%==============================================================================
\section{Discussion}
\label{sec:discussion}
%==============================================================================

\paragraph{Position in the random-matrix landscape.}
The Frustrated Distance Matrix (FDM) model proposed here is the
dynamic extension of the static distance-matrix ensemble on
$S^2$ analyzed by BBS~\cite{bogomolny2003}, a canonical
sub-class of the Euclidean random matrix
ensembles~\cite{mezard1999, bordenave2008, bordenave2013,
goetschy2013, ciliberti2004anderson, clapa2012}. The static BBS
analysis is for $N$ i.i.d.\ uniform points on $S^2$; the FDM matrix
$M(t)_{ij}=\arccos(\mathbf{x}_i\cdot\mathbf{x}_j)$ is built on
$N$ Frustrated Brownian Particles (FBP) of
\cite{halperin2026frustrated} whose trajectories evolve by an
overdamped Langevin SDE coupled through the quenched random pair
forces $-\Phi_{ij}$. The FDM is, to our knowledge, the first
random-matrix model that is at once strongly correlated (a
deterministic function of $N$ interacting points on a manifold)
and explicitly dynamic, with both features present in a single
matrix-valued stochastic process. The closest existing analogs
are static (Bordenave Euclidean and sample-covariance ensembles
\cite{bordenave2008, bordenave2013, bouchaudpotters2003,
bunbouchaudpotters2017, baik2005}) or dynamic but with independent
increments (Dyson Brownian motion \cite{dyson1962brownian},
Voiculescu--Biane free Brownian motion \cite{voiculescu1991,
biane1997}); the closest dynamic-and-correlated analog is
SYK / $p$-spin glassy dynamics
\cite{cugliandolo1993, facoetti2019}, but those analyses are
field-theoretic and lack the explicit ERM kernel structure of
the FDM.

The remainder of this section organizes the results around
three pillars: (1) the match between BBS predictions at the
ensemble (disorder-averaged) and individual-trajectory levels,
(2) the spectral signals of structural change extractable from
the time-varying distance matrix, and (3) implications and
extensions to other systems.

\paragraph{(1) Ensemble- and trajectory-level match with BBS.}
The first principal finding is that the static BBS predictions
for distance matrices on $S^2$ are reproduced by the dynamic
FDM at every time along the trajectory, both at the ensemble
(disorder-averaged) and the individual-trajectory (frozen
disorder) levels. \emph{At the ensemble level}, the
disorder-averaged bulk of $M(t)$ coincides quantitatively with
the BBS / ERM density on i.i.d.\ samples from $\mu_t$. The
Perron pattern (one positive eigenvalue, $N{-}1$ non-positive)
holds across the entire dataset of ten realizations $\times$
200 recorded times $\approx 2000$ $M(t)$ configurations, with
no exceptions. The quasi-multiplet ladder of dimensions
$1, 3, 7, 11, 15, \ldots$ matches the BBS prediction
$\Lambda_\ell = N a_\ell/(2\ell+1) =
628.3, -157.1, -9.82, -2.45, -0.96$ for $\ell = 0, 1, 3, 5, 7$
to within $\sim 10\%$. The bulk density follows the BBS
power-law tail $\rho(|\lambda|)\sim|\lambda|^{-5/3}$ in the
delocalized regime $|\lambda|\gtrsim N^{3/4}$ and the linear
vanishing $\rho(|\lambda|)\sim|\lambda|$ in the localized
regime, with a fitted MLE exponent $\alpha\approx 1.73$ at
$t=0$ and $\alpha\approx 1.65$ at NESS, both on the $S^2$ side
of the BBS prediction; the finite-$N$ scan of
\S\ref{sec:bbs_check}\,(vi) shows that the NESS exponent
trajectory converges to $5/3$ from below as $N$ grows rather
than to the ring value $3/2$.

\emph{At the trajectory level}, the same identification holds
through self-averaging. The bulk eigenvalue density of $M(t)$
is an extensive functional whose realization-to-realization
fluctuations at fixed $t$ are $O(1/\sqrt{N})$, so the spectrum
for a typical frozen disorder $\Phi$ at large $N$ coincides
with its disorder average. The disorder average is fixed by
$\mu_t$ alone through the Mercer expansion of $\arccos$
against $\mu_t$ (the F2 mean-field
density~\cite{halperin2026fields}); the non-i.i.d.\
correlations of the FBP joint law contribute only to
sub-leading spectral statistics (level spacings, two-point
eigenvalue correlations, edge fluctuations). The
i.i.d.-resample test (Fig.~\ref{fig:rmt_erm_iid},
Tab.~\ref{tab:rmt_erm_iid}) confirms agreement of the FDM bulk
with the parametric and bootstrap ERM null hypotheses at the
$\sim 2\%$ level on the bulk power-law exponent at $t=0,1,40$,
with the empirical NESS ring band width
$\sigma_\theta\!\approx\!3.9^\circ$ in line with the
$\sim 4.8^\circ$ value reported by the F2
model~\cite{halperin2026fields}. The sub-leading channels are
also tested. The bulk $r$-statistic of the FDM agrees with the
ERM null hypothesis on the FBP one-particle density $\mu_t$ to
within $\sim 1\%$
(Tab.~\ref{tab:rmt_r_erm_null}), and the $P(s)$ shape
overlays the ERM curve at every time
(Fig.~\ref{fig:rmt_p_of_s}), so the sub-GOE Berry--Robnik / BBS
Anderson superposition shape is purely geometric. The
ranked-spectrum envelope test
(Fig.~\ref{fig:rmt_ranked_envelope}) shows the FDM ensemble
mean tracking the bootstrap envelope across the bulk and
deviating only at the deepest negative ranks where the
dynamical $\ell=1$ ring-formation signal sits. The data are
therefore consistent with the picture in which both
the leading bulk density and the sub-leading level statistics
are inherited from the ERM identification on $\mu_t$, with ring
formation entering through the time evolution of $\mu_t$ rather
than through deviations from the i.i.d.\ template. The
finite-$N$ scans of \S\ref{sec:bbs_check} (v) and (vi) make
this picture quantitative: at $t=0$ the bulk exponent decreases
monotonically toward the BBS $d=2$ value $\alpha=5/3$ as $N$
grows, and at NESS the same exponent rises monotonically toward
$\alpha=5/3$ from below, while the most-negative multiplet
position drifts toward its $N\to\infty$ value, consistent with
self-averaging of the bulk under the FBP dynamics.

The same picture extends to the eigenvector level via the
BBS-Anderson correspondence. At the \emph{particle} level, the
FBP dynamics drives the configuration onto a one-dimensional
ring during the fast transient (the \emph{dynamic dimension
reduction} of \cite{halperin2026frustrated}). At the
\emph{matrix} level, the eigenvectors of $M(t)$ are delocalized
at large $|\lambda|$ and exponentially localized at small
$|\lambda|$ on a small subset of particle indices, with
localization length $\xi\propto|\lambda|$ on the emergent 1D
support, recovering the BBS-Anderson scaling
$\mathrm{PR}\approx 4|\lambda|/N$ that is the analog of
Anderson localization in a 1D chain with random hopping
(Sec.~\ref{sec:bbs_check}(iv),
Fig.~\ref{fig:rmt_participation}). The two kinds of
localization are not independent: the FBP transient that
spatially concentrates the particles onto the ring is precisely
what makes the BBS-Anderson correspondence become \emph{exact}
rather than a continuous-approximation analogy. On the
original $d=2$ support the empirical PR sits one to three
orders of magnitude above the 1D scaling, in agreement with
the BBS prediction that the localization volume on a $d > 1$
manifold is set by multipole-moment conditions rather than by
a 1D escape time.

\paragraph{(2) Spectral signals of structural change.}
The second principal finding is a set of concrete spectral
signals in the dynamic distance matrix that flag structural
change in the underlying particle system, computable from
$M(t)$ alone without ever reconstructing the configuration in
$\mathbb{R}^3$. In the FDM these signals correspond to the
ring-formation transient at the fast timescale
$\tau_{\text{fast}}$, in which a uniform configuration on $S^2$
collapses onto an emergent one-dimensional ring with a
subsequent slow Brownian-like drift of the ring orientation
$\hat{\mathbf{n}}(t)$. Both regimes show up in the FDM spectrum
as a redistribution of spectral mass \emph{within} the BBS /
ERM template, identifying ring formation as a structural
reorganization of the matrix rather than a change of its
overall norm. We separate the diagnostics into two groups: the
three structural-change diagnostics that move sharply during
ring formation, and two static cross-checks that hold at every
time and tie the FDM picture to neighbouring constructions.
Both groups are computed from $M(t)$ alone.

\begin{description}
\setlength{\itemsep}{2pt}
\item[\textbf{(I)} Structural-change diagnostics.]
\hfill\\
\textbf{(i)} \emph{$\ell=1$ multiplet rank reduction}, with
the bottom-five non-crossing fan-out as its dynamical
refinement: the lowest BBS multiplet splits from a
near-degenerate 3-cluster into a $1{+}2$ pattern as
$\mathrm{SO}(3)\to\mathrm{SO}(2)$, and the bottom-five
eigenvalues fan out from a near-degenerate stack into a
non-crossing descending ladder.\\
\textbf{(ii)} \emph{Bulk-scale contraction with outlier-count
drop}: $\sigma_{\text{bulk}}(t)$ shrinks by a factor of
$\approx 2.6$ on the fast timescale, and the count of
$|\lambda|>1$ outliers falls from $\approx 30$ to $\approx 17$
in a complementary way.\\
\textbf{(iii)} \emph{Rank-decay exponent shift}: the slope
$\beta$ of $|\lambda_K|$ vs $K$ on the rank window $K\in[2,50]$
shifts from $\beta\approx 1.82$ at $t=0$ to $\beta\approx 2.10$
at NESS, consistent with the BBS predictions $\beta=3/2$ on
$S^2$ and $\beta=2$ on $S^1$: the bottom-$50$ multiplets
reorganize as the effective support of the bottom eigenmodes
collapses from two-dimensional to one-dimensional.
\item[\textbf{(II)} Static cross-checks.]
\hfill\\
\textbf{(iv)} \emph{Eigenvector identity with the F2
orientation estimator}: the bottom-$K$ eigenspace ($K\ge 2$)
of $M(t)$ recovers, through an algebraic identity, the same
ring orientation that the F2 inertia-tensor PCA estimator
provides; the algebraic ``alignment'' between the two
estimators is identically one.\\
\textbf{(v)} \emph{Berry--Robnik bulk level statistics, plus
a sub-leading non-i.i.d.\ footprint at small $s$}:
$P(s)$ shows a small-spacing pile-up with mean $r$-statistic
in $[0.48, 0.51]$, between Poisson and GOE, throughout the
trajectory, set by the spherical-harmonic $\ell$-blocks at
$t=0$ and the Fourier $k\leftrightarrow N-k$ pairs at NESS;
on top of this geometric pile-up a pooled-window
double-difference test detects a sub-leading non-i.i.d.\
contribution to $P(s)$ at small $s$, the spectral footprint
of the attractive pair correlations encoded in $\Phi_{ij}$.
\item[\textbf{(III)} Trajectory-level diagnostics.]
\hfill\\
\textbf{(vi)} \emph{Bottom-$K$ projector drift
$D_K(t_\text{ref}, \tau)$ and Frobenius commutator norm
$C(t_\text{ref}, \tau)$}: at NESS, $D_K$ grows slowly with
$\tau$ and stays well below the eigenvalue-shuffled null
saturation, and $C$ stays well below the null band; the
trajectory $\{M(t)\}_t$ carries information beyond the
instantaneous eigenvalue snapshots in the form of a slow
coherent rotation of the bottom eigenspace, the matrix-side
imprint of the ring-orientation diffusion of
$\hat{\mathbf{n}}(t)$ that the eigenvalue trajectory alone
does not contain.
\end{description}

The sharper version of each entry follows.

\textbf{(i)} The $\ell=1$ multiplet rank-reduces from a
near-degenerate 3-cluster to a $1+2$ split as the rotational
symmetry collapses to $\mathrm{SO}(2)$. The deepest eigenvalue
grows from $\approx -160$ to $\approx -380$ to $-400$ on the
fast timescale, while the second-deepest stays near its $t=0$
value $\approx -150$. The dynamical refinement is the
non-crossing fan-out of Sec.~\ref{sec:eigtraj}: the bottom-five
eigenvalues separate into a non-crossing descending ladder with
persistent inter-level gaps of order $\approx 200$ in NESS,
giving a low-dimensional Dyson-like signature of the geometric
collapse.

\textbf{(ii)} The bulk scale $\sigma_{\text{bulk}}(t)$ contracts
by a factor of $\approx 2.6$ on the same timescale. The absolute
outlier count above $|\lambda|>1$ drops in a complementary way
from $\approx 30$ to $\approx 17$ as the spectrum reorganizes
around fewer, larger outliers.

\textbf{(iii)} The rank-decay exponent $\beta$ of
Sec.~\ref{sec:powerlaw}, fitted on $K\in[2,50]$, moves cleanly
from $\beta\approx 1.82$ at $t=0$ to $\beta\approx 2.10$ at NESS.
The BBS predictions are $\beta=3/2$ on $S^2$ and $\beta=2$ on
$S^1$, so the empirical shift tracks the dimensional collapse
of the lowest few multiplets from a two-dimensional sphere to a
one-dimensional ring. By contrast, the bulk \emph{density}
exponent $\alpha$ from the MLE fit on $|\lambda|>x_{\min}$ is
\emph{not} a structural-change diagnostic: $\alpha$ stays near
the BBS $d=2$ value $5/3$ at all times and at all $N$ tested,
and the finite-$N$ scan at the FBP NESS of
\S\ref{sec:bbs_check}\,(vi) shows that the small
$t=0\to$NESS shift seen at fixed $N=400$ is a finite-$N$
feature that moves \emph{upward} toward $5/3$ as $N$ grows,
not downward toward $3/2$. The ring-formation signal in the
power-law analysis lives in the rank-decay window on the
bottom multiplets, not in the bulk density tail.

\paragraph{Shared characteristics on log time.}
The dense early-time recording and log-time presentation of
Figs.~\ref{fig:rmt_energy}--\ref{fig:rmt_eigvals},
\ref{fig:rmt_powerlaw}, \ref{fig:rmt_outliers},
\ref{fig:rmt_eigtraj} make four properties of the three
change-diagnostics directly readable from the data, none of
them a separate diagnostic but each a useful refinement.
\textbf{First}, $\tau_{\text{fast}}$ becomes a measurable
quantity rather than a verbal label: each diagnostic saturates
at $t \approx 5$ for the uniform initial condition, and at
$t \approx 1$ for the Big Bang initial condition
(Sec.~\ref{sec:bigbang}, Fig.~\ref{fig:bigbang_summary}), a
factor of five faster.
\textbf{Second}, the three diagnostics descend together: their
log-time curves sit on top of each other in shape and
saturation point, so the synchronization on a common
$\tau_{\text{fast}}$ is itself diagnostic of a single
underlying mechanism (the geometric collapse) rather than
several weakly-coupled transitions.
\textbf{Third}, the descent is approximately logarithmic in
$t$ across $t\in[10^{-1}, 5]$, that is, linear on log-$t$
axes; logarithmic relaxation is characteristic of glassy /
activated dynamics in spin-glass-type systems and is
consistent with the SYK / $p$-spin analogies of
Sec.~\ref{subsec:syk_link}.
\textbf{Fourth}, the earliest detectable change occurs at the
first recorded snapshot $t \approx 10^{-2}$, well before
$\tau_{\text{fast}}$, so each of (i)--(iii) doubles as an
early-warning indicator of structural change with a detection
threshold one to two decades below the saturation time.

\textbf{(iv)} The bottom-$K$ eigenspace ($K\ge 2$) of $M(t)$
recovers, through an explicit algebraic identity, the same PCA
estimate of the ring orientation that the F2
model~\cite{halperin2026fields} obtains by diagonalizing the
inertia tensor $C = X^T X/N$
(\S\ref{sec:eigvec_alignment}, within
Sec.~\ref{sec:spectral_evolution}). Both constructions
diagonalize the same $3\times 3$ object via the rank-three
$\ell=1$ Legendre contribution $\propto XX^T$ of the arccos
kernel, and the algebraic alignment between the two estimators
is identically one. This is a static identity, not a
structural-change signal: it holds at every $t$, but it ties
the FDM and F2 descriptions together at the eigenvector level.

\textbf{(v)} The level-spacing statistics of the bulk
(Sec.~\ref{sec:universality}) place $M(t)$ in neither the GOE
nor the Poisson universality class, but in a Berry--Robnik
superposition of symmetry blocks. The blocks are the
spherical-harmonic $\ell$-blocks at $t=0$ and the Fourier
$k\leftrightarrow N-k$ pairs in the ring NESS; both produce a
small-spacing pile-up in $P(s)$ while keeping the mean
$r$-statistic in $[0.48, 0.51]$ throughout. The leading
small-$s$ pile-up is geometric and is reproduced by the
ERM($\mu_t$) null hypothesis. The pooled-window
double-difference test (Fig.~\ref{fig:rmt_p_of_s_residual},
Tab.~\ref{tab:p_of_s_residual}) further detects a sub-leading
non-i.i.d.\ contribution at small $s$ ($3$--$8\sigma$ across
$s\in[0, 0.35]$) which is the spectral footprint of the
attractive pair correlations $\Phi_{ij}$ that ERM resampling on
$\mu_t$ alone cannot capture.

\textbf{(vi)} The trajectory-level projector drift $D_K(t_\text{ref},
\tau)$ and matrix-commutator norm $C(t_\text{ref}, \tau)$ of
Sec.~\ref{subsec:trajectory_diagnostics} probe whether the matrix
trajectory $\{M(t)\}_t$ carries information beyond the
eigenvalue snapshots tracked by (i)-(iii). Compared to an
eigenvalue-randomized null that preserves $\Lambda(t)$ at every
$t$ but shuffles the eigenvectors across the trajectory, the
FDM at NESS has $D_K$ that grows slowly with $\tau$ from $0$ to
$\approx 0.4$ across $\tau\in[0.25, 10]$ and $C$ that stays at
$\sim 2\times 10^4$ throughout, both well below the null
saturation. The dynamical signature there is a slow coherent
rotation of the bottom eigenspace at NESS, the matrix-side
imprint of the rotational diffusion of the ring orientation
$\hat{\mathbf{n}}(t)$ on $S^2$ that the eigenvalue trajectory
alone does not contain. The pattern is mechanism-sensitive in
the same sense as the level-spacing residual of (v): an
alternative dynamics producing the same final ring without the
slow coherent reorientation would not give the same trajectory
diagnostics, even with identical eigenvalue sequences.

The
``Big Bang'' experiment of Sec.~\ref{sec:bigbang}, in which the
particles start in a tiny Gaussian blob rather than uniformly on
$S^2$, reaches the same qualitative NESS through a faster
collapse ($\tau_{\text{fast}}^{\text{Big Bang}}\sim 1$, a factor
of $\sim 5$ shorter) and dissipates the disorder-weighted energy
monotonically into the heat bath without overshoot, confirming
that the FDM spectral signatures of structural change identified
above are robust to the choice of initial measure.

\paragraph{Practical relevance: detecting structural changes
through the distance matrix alone.}
The transferable lesson of these signals is that delocalized
collective states leave a sharp signature in the lowest few
non-Perron eigenvalues of the distance matrix, the bulk scale,
and the rank-decay exponent on the bottom multiplets, with the
rest of the spectrum, including the bulk density tail
exponent, left essentially untouched by the structural change.
The finite-$N$ scan at the FBP NESS of
\S\ref{sec:bbs_check}\,(vi) clarifies which power-law
diagnostic carries the ring-formation signal: the bulk density exponent $\alpha$
reflects the dimension of the embedding $S^2$ and is not a
useful diagnostic, while the rank-decay exponent $\beta$ on
the bottom-$50$ window picks up the dimensional collapse of
the bottom multiplets and shifts cleanly from $\beta\approx
3/2$ to $\beta\approx 2$. The i.i.d.-resample construction of
Sec.~\ref{sec:bulk_ensembles} supplies the natural \emph{null
spectrum} against which an empirical matrix should be compared.
Given a fitted one-particle density $\hat\mu_t$ (analytical
or non-parametric), the i.i.d.\ ERM bulk on $\hat\mu_t$ is the
spectrum that would arise from the marginal alone, with no
inter-sample correlations beyond those induced by the kernel;
time-resolved deviations of the empirical spectrum from this
null hypothesis localize the genuine non-i.i.d.\ structural
content of the underlying system.

\paragraph{What the matrix carries: collective dynamics in the
low eigenspace, mechanism in the sub-leading channels.}
The framing of the spectral diagnostics is set by the
applications in which the underlying configuration $X(t)$ is
not directly observable. In those settings the matrix $M(t)$
is the primary observable, and the question is what
information about the system can be extracted from $M$ alone.
The answer separates into two layers.

The first is a multidimensional-scaling-type
(MDS-type) layer. For an $N\times N$ distance matrix on $N$
points in an ambient space of dimension $D \ll N$, the bottom
$D$ eigenvectors of $M$ encode the collective coordinates of
the underlying configuration up to a global isometry: this is
the standard MDS / kernel-PCA reading of distance matrices,
in which the low eigenspace recovers the embedding without the
embedding being observed. In our setting $D = 3$ (the ambient
$\mathbb{R}^3$ in which $S^2$ sits), and the rank-three $\ell=1$
Legendre block $-(3\pi/8)\,X X^T$ of $M(t)$ is exactly that
low-eigenspace contribution; the bottom-three eigenspace of
$M(t)$ recovers the inertia tensor $X^T X$ algebraically
(Sec.~\ref{sec:eigvec_alignment}). When the configuration
collapses from $D = 3$ to an effective $D - 1 = 2$ during ring
formation (the third inertia moment along $\hat{\mathbf{n}}(t)$
becoming near-zero), the corresponding low eigenspace of $M(t)$
rank-reduces from three to two. This is the MDS-side reading
of the leading change-diagnostics (i)-(iii): the
low-eigenspace structure of $M$ reads off the collective
geometric organization of the underlying configuration, and the
dimensional collapse appears as a rank reduction of that
low-eigenspace, all without observing $X(t)$ in $\mathbb{R}^3$.
This layer is what the spectral approach delivers in the
target applications: the geometric story of the underlying
$N$-particle system, recovered from the matrix alone.

The second layer is the dynamical mechanism. The MDS-type
layer reads off only the geometric arrangement, which is a
property of the marginal one-particle density $\mu_t$ alone.
Any dynamics that produced an FBP-like point cloud (an external
ring-shaped potential on $S^2$, attractive isotropic pair
interactions, or a pre-projected ring with thermal noise) would
generate the same MDS-type signature, and the leading
change-diagnostics (i)-(iii) would fire in qualitatively the
same way. The Big Bang experiment of Sec.~\ref{sec:bigbang}
confirms the corresponding robustness within the FBP dynamics
under a change of $\mu_0$. The spectral channels that go beyond
the MDS-type layer, and that distinguish the FBP mechanism from
disorder-free alternatives producing the same ring, are
sub-leading. They are the small-$s$ non-i.i.d.\ residual in the
level-spacing distribution $P(s)$ (Sec.~\ref{sec:universality},
Fig.~\ref{fig:rmt_p_of_s_residual}), where the $3$-$8\sigma$
excess is the spectral footprint of the attractive pair
correlations encoded in $\Phi_{ij}$ that a disorder-free ring
potential would not produce, and the trajectory-level $D_K$
and $C$ diagnostics
(Sec.~\ref{subsec:trajectory_diagnostics}), where the slow
coherent rotation of the bottom eigenspace is a property of
the ring-orientation diffusion that the FBP dynamics produces,
not of the ring geometry alone. A controlled comparison with
alternative dynamics producing similar ring-like geometry would
test this two-layer stratification more sharply and is a
natural follow-up. The load-bearing claim for the spectral
approach in target applications is the two-layer combination:
the low-eigenspace structure of $M$ reads off the collective
geometric dynamics of the underlying $N$-particle system in an
MDS-type sense, and the sub-leading spectral channels access
the dynamical mechanism that produced that geometry.

%==============================================================================
\section{Summary and outlook}
\label{sec:summary}
%==============================================================================

\paragraph{Brief recap.}
We introduced the Frustrated Distance Matrix (FDM) model as a
dynamic extension of the static distance-matrix ensemble of
Bogomolny, Bohigas, and Schmit on $S^2$, built on $N$ Frustrated
Brownian Particles evolving under quenched random pairwise
couplings linear in the geodesic distance. The static BBS
template (Perron eigenvalue, $(2\ell+1)$ quasi-multiplets,
power-law bulk tail, BBS-Anderson localization crossover) is
preserved at every time along the FDM trajectory, with the
dynamics entering as a redistribution of spectral mass within
that template that flags ring formation through three sharp
diagnostics computable from $M(t)$ alone: $\ell=1$ multiplet
rank reduction with bottom-five fan-out, bulk-scale contraction
with outlier-count drop, and a rank-decay-exponent shift from
the sphere value $3/2$ to the ring value $2$. The bulk density
exponent $\alpha$ stays near the $d=2$ value $5/3$ throughout
and is not a useful ring-formation diagnostic, as the
finite-$N$ scan at the FBP NESS in
\S\ref{sec:bbs_check}\,(vi) showed explicitly. Two static cross-checks tie the picture together:
the bottom-$K$ eigenspace of $M(t)$ recovers the same ring
orientation as the F2 inertia-tensor PCA estimator through the
$\ell=1$ Legendre block of the arccos kernel, and the bulk
level statistics sit in a Berry--Robnik superposition between
Poisson and GOE consistent with the ERM null hypothesis on the
FBP one-particle density $\mu_t$. The pooled-window
double-difference test of \S\ref{sec:universality} also detects
a sub-leading non-i.i.d.\ contribution to the small-$s$
pile-up of $P(s)$ at $3$--$8\,\sigma$ significance: the
attractive pair correlations encoded in the quenched couplings
$\Phi_{ij}$ generate extra small-magnitude entries in $M(t)$
beyond what i.i.d.\ resampling on $\mu_t$ produces, and these
extra near-degeneracies show up where they are most visible,
in the small-$s$ region of the level-spacing density. The
trajectory-level diagnostics
$D_K(t_\text{ref}, \tau) = \|P_K(t_\text{ref}+\tau) -
P_K(t_\text{ref})\|_F$ and $C(t_\text{ref}, \tau) =
\|[M(t_\text{ref}), M(t_\text{ref}+\tau)]\|_F$
of \S\ref{subsec:trajectory_diagnostics} further test whether
the matrix trajectory carries information beyond the eigenvalue
snapshots: at NESS both stay well below the
eigenvalue-shuffled null saturation, evidence that
$\{M(t)\}_t$ has a slow coherent rotation of its bottom
eigenspace, the matrix-side imprint of the rotational diffusion
of $\hat{\mathbf{n}}(t)$ that the eigenvalue trajectory alone
does not contain. We close by sketching how the construction
extends beyond the FBP test bed.

\paragraph{Implications and extensions.}
The framework developed here is a controlled laboratory for the
dynamical inverse problem of detecting structural changes in an
underlying $N$-particle system from the spectrum of a
time-varying distance, similarity, or correlation matrix. The
FBP dynamics on $S^2$ is the simplest setting in which the
underlying structural change is unambiguously visualizable
(ring formation from a uniform configuration), the BBS / ERM
template is exactly known, and the corresponding spectral
signals can be calibrated against ground truth. The same
construction extends to any base space and any
$N$-particle system in which pairwise distances, similarities,
or correlations are observable and the underlying configuration
either is high-dimensional or is not observed directly.

\emph{Empirical settings on physical or quasi-physical
manifolds.} In financial markets, time-varying asset correlation
matrices are analyzed through their spectra to clean estimators
and to detect regime changes~\cite{bouchaudpotters2003,
bunbouchaudpotters2017}. In network science, the eigenvalue
distribution of empirical graph and network adjacency matrices
is sensitive to topology and community
structure~\cite{goh2001, chunglu2003}. Similarity matrices in
molecular dynamics resolve conformational reorganizations
(folding transitions, glass transitions) without an explicit
reaction coordinate. In each case the time-evolving spectrum
plays the role of $M(t)$ in the FDM laboratory, and the
ERM null hypothesis built on the empirical one-particle density
$\hat\mu_t$ is the natural baseline against which
structural-change diagnostics should be calibrated.

\emph{Dynamics in high-dimensional and non-physical (abstract)
spaces.} Several settings of practical interest involve a
``configuration space'' that is high-dimensional and not a
physical manifold but in which pairwise distances or
similarities are nevertheless easily computable. \emph{Neural
networks} are a clean example: pairwise distances or
similarities between weight vectors at successive training
steps, between hidden-state representations across a dataset,
or between learned features at different layers, define a
time-varying distance matrix on an abstract high-dimensional
parameter or representation manifold. Structural transitions in
the learning dynamics (onset of feature emergence, phase
transitions between memorization and generalization, mode
collapse, grokking) should manifest as time-resolved
redistributions of the spectrum within the ERM reference
distribution built on the empirical one-particle density
$\hat\mu_t$, calibratable against the ground truth available
in controlled experimental settings (toy datasets,
small-scale models). \emph{Brain-network connectivity} from
fMRI / EEG time series is another natural example: the
underlying neural state is high-dimensional, the connectivity
matrix between regions or sensors is the primary observable,
and structural transitions (state changes, task transitions,
seizure onset) should leave time-resolved spectral signatures.
Other examples include time-varying gene-expression
similarity matrices in molecular and developmental biology,
where the underlying regulatory network is non-observable;
spatial-correlation matrices of climate or ecological time
series; and configuration-space distance matrices of
high-dimensional dynamical systems (granular media, plasma,
turbulent flow) where direct visualization of the configuration
is impractical. In each case the FDM construction supplies a
controlled microscopic test bed for the spectral detection
strategy, while the ERM identification on the time-evolving
$\mu_t$ supplies the geometry-aware null hypothesis.

\paragraph{Relation to the F2 model.}
The picture that emerges here is geometric and complementary to
the F2 (Frustrated Fields) model developed in
Ref.~\cite{halperin2026fields}, where the same FBP dynamics is
reduced to an $\mathrm{O}(3)$ nonlinear sigma model in $(0+1)$
dimensions for the orientation $\hat{\mathbf{n}}(t)$ with a
single low-energy constant $D_{\text{rot}}$. The FDM and F2
descriptions agree on the existence of a sharp ring-formation
transient at $\tau_{\text{fast}}$, on the post-formation slow
diffusion of $\hat{\mathbf{n}}(t)$, and on the disorder-induced
sample-to-sample variability of $D_{\text{rot}}$; the FDM view
in addition exposes a hierarchy of large-eigenvalue and
large-eigenvector diagnostics that bypass the F2 reduction and
are computed directly from the matrix.

\paragraph{Future directions.}
Three immediate extensions suggest themselves. First, building
on the self-averaging identification of
Sec.~\ref{sec:bulk_ensembles} of the FDM bulk with the ERM
density on $\mu_t$, a quantitative analytical computation of
the latter through a self-consistent Mezard--Parisi--Zee
\cite{mezard1999} or Goetschy--Skipetrov \cite{goetschy2013}
closure on the F2 mean-field density
$\mu_t$~\cite{halperin2026fields}. The same line should
extend to the sub-leading non-i.i.d.\ corrections detected
empirically in \S\ref{sec:universality}: the small-$s$
double-difference signal in $P(s)$ has a clear physical origin
in the attractive pair correlations $\Phi_{ij}$ but no
quantitative theory yet, and complementary observables
(two-point eigenvalue correlations, edge fluctuations,
multi-time correlators of the matrix trajectory) would sharpen
the picture and allow a direct comparison with field-theoretic
predictions for sub-leading corrections to the ERM ansatz. Second, the design of a universal disorder-blind detector
of ring formation that combines the bulk-scale, the bottom-block
level gaps, the eigenvector alignment, and the
participation-ratio crossover position into a single
dimensionless statistic, ready for use on observed distance,
similarity, or correlation matrices where the underlying
configuration is unknown. Third, the extension of the analysis
to other base manifolds (torus, cylinder, deformed spheres),
where the BBS quasi-multiplet multiplicities and the bulk
power-law exponent change in predictable ways, and to other
kernels (Coulomb, Riesz), where the singularity structure shifts
the BBS exponent through the same Mercer-expansion machinery
used in Sec.~\ref{sec:bbs_check}. A fourth direction, suggested
by the geometry-versus-mechanism distinction discussed in
Sec.~\ref{sec:discussion}, is a controlled comparison of the
FBP spectral trajectory with alternative dynamics that produce
similar ring-like geometries from different mechanisms (an
external ring-shaped potential on $S^2$, attractive isotropic
pair interactions, a pre-projected ring with thermal noise).
The leading change-diagnostics (i)-(iii) should fire in
qualitatively the same way for all these mechanisms; the
mechanism-sensitive channels (the level-spacing residual at
small $s$ and the trajectory-level $D_K$ and $C$ of
Sec.~\ref{subsec:trajectory_diagnostics}) should distinguish
them, and quantifying that distinction is the natural test of
how much mechanism information the matrix trajectory carries on
top of the geometry it records.

\end{document}